\documentclass[12pt,twoside]{report}

\usepackage{vmargin}
\usepackage{bin}
\usepackage[dvips]{graphicx}
\usepackage{amsmath,amssymb,latexsym}
\usepackage{beton,euler}
\setpapersize{A4}
\setmargins{4.0cm}{1.5cm}{14cm}{22cm}{0cm}{1.2cm}{0cm}{1.5cm} 
\begin{document}
\begin{titlepage}
\vspace{7cm}
\begin{center}
{\Huge Two-dimensional Quantum Gravity}\\[4cm]
{\Large Juri Rolf \\
Niels Bohr Institute\\University of Copenhagen}\\[2cm]
\begin{figure}[h]
\begin{center}
\includegraphics[width=6cm]{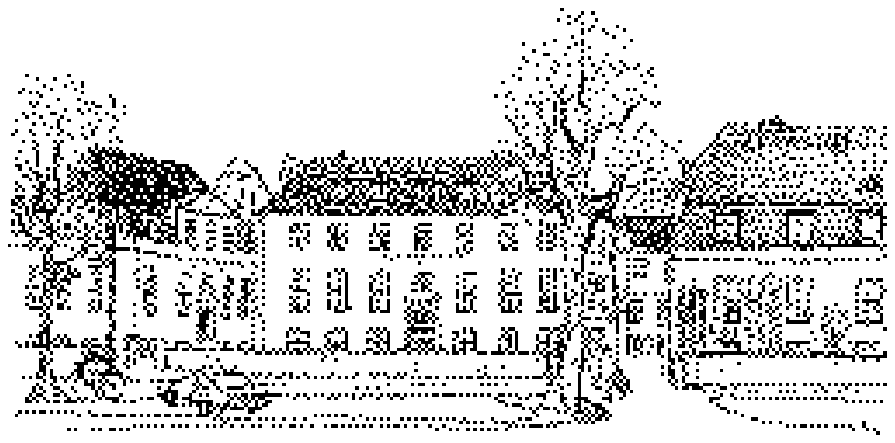}
\end{center}
\end{figure}
Thesis submitted for the Ph.D. degree in physics \\
at the Faculty of Science, University of Copenhagen.\\
{May 1998}
\end{center}
\end{titlepage}
\newpage\thispagestyle{empty} ~ \newpage
\pagenumbering{roman}   
\tableofcontents
\chapter*{Acknowledgement}
\addtocontents{toc}{\protect\contentsline {chapter}{\protect\numberline {\ }Acknowledgement}{iii}}   

I want to thank all the people who have directly or indirectly contributed
to this work and made my stay in Copenhagen possible, in particular:
\begin{itemize}
\item my supervisor Jan Ambj{\o}rn for the interesting project, for many discussions, for a good collaboration
  and for partial financial support of my visit in Santa Fe at the workshop
  New Directions in Simplicial Quantum Gravity,
\item Jakob L.~Nielsen for many discussions and for a good collaboration,
\item my collaborators Dimitrij Boulatov, George Savvidy and Yoshiyuki Watabiki for the good work done together,
\item Martin Harris and Jakob L. Nielsen for proofreading the thesis,
\item my parents and my wife for their support and encouragement,
\item Konstantinos N.~Anagnostopoulos, Martin Harris, Lars Jensen, Jakob L.~Nielsen,
  Kaspar Olsen, J{\o}rgen Rasmussen and Morten Weis for some discussions and for 
  creating the nice atmosphere in our office,
\item the Niels Bohr Institute for warm hospitality,
\item the Studienstiftung des deutschen Volkes for financial support.
\end{itemize}
\newpage\thispagestyle{empty} ~ \newpage
\setcounter{chapter}{0}
\chapter*{Introduction}
\addtocontents{toc}{\protect\contentsline {chapter}{\protect\numberline {\ }Introduction}{1}}   

This Ph.D. thesis pursues two goals: The study of the geometrical
structure of two-dimensional quantum gravity and in particular
its fractal nature. To address these questions we review the continuum
formalism of quantum gravity with special focus on the scaling properties
of the theory. We discuss several concepts of fractal dimensions which
characterize the extrinsic and intrinsic geometry of quantum gravity.
This work is partly based on work done in collaboration with
Jan Ambj{\o}rn, Dimitrij Boulatov, Jakob L. Nielsen and Yoshiyuki Watabiki \cite{Ambjorn:1997jf}.

The other goal is the discussion of the discretization of quantum gravity and
to address the so called quantum failure of Regge calculus. We review
dynamical triangulations and show that it agrees with the continuum theory
in two dimensions. Then we discuss Regge calculus and prove that
a continuum limit cannot be taken in a sensible way and that it does
not reproduce continuum results. This work is partly based on work done
in collaboration with Jan Ambj{\o}rn, Jakob L.~Nielsen and George Savvidy \cite{Ambjorn:1997ub}.

In chapter \sref{chap1} we introduce the main ingredients for the formulation
of two-dimensional quantum gravity as an Euclidean functional integral over geometries.
It contains a brief reminder of Liouville theory and the technical issues in
the continuum formalism. We use these techniques to discuss the extrinsic and intrinsic
Hausdorff dimension and the spectral dimension of two-dimensional quantum gravity.

Chapter \sref{disc} is a review of dynamical triangulation in two dimensions. We begin with
an introduction of the main ideas of how to discretize two-dimensional quantum geometries.
The scaling properties are illustrated by means of the two-point function of pure
gravity and of branched polymers.

In chapter \sref{regge} we discuss quantum Regge calculus which has been suggested
as an alternative method to discretize quantum geometries.
We prove by a simple scaling argument that a sensible continuum limit of this theory
cannot be taken and that it disagrees with continuum results. 
\pagenumbering{arabic}   
\setcounter{chapter}{0}
\chapter{Two-dimensional quantum gravity}
\label{sec:chap1}

Euclidean quantum gravity is an attempt to quantize general relativity based on Feynman's functional
integral and on the Einstein-Hilbert action principle. One integrates over all Riemannian metrics
on a $d$-dimensional manifold $M$. It is based on the hope that one can recover the Lorentzian signature
after performing the integration analogously to the Wick rotation in Euclidean quantum field theory.
For a general discussion of further problems and for motivation of a theory of quantum gravity we refer to
\cite{Hawking:1979zw,Hawking:1980gf,Isham:1995wr}.

General relativity is a reparametrization invariant theory which can be formulated with no
reference to coordinates at all. This diffeomorphism invariance is a central issue in the quantum
theory. Its importance is most apparent in two dimensions, since the Einstein-Hilbert action
is trivial and consists only of a topological term and the cosmological constant coupled to the volume of
the spacetime. All the non-trivial dynamics of the two-dimensional theory of quantum gravity thus
come from gauge fixing the diffeomorphisms while keeping the geometry exactly fixed. This is
the famous representation of the functional integral over geometries as a Liouville field theory
by Polyakov \cite{Polyakov:1981rd}. Based on this formulation the scaling exponents can be obtained
\cite{Knizhnik:1988ak,David:1988hj,Distler:1989jt}.

Any theory of quantum gravity must aim at answering questions about the geometrical structure
of quantum spacetime. The interplay between matter and geometry is well known from general relativity.
The quantum average over all geometries changes the dynamics of this interaction. It turns
out that the quantum spacetime has a fractal nature and even its dimension is a dynamical quantity.
The characterization of the fractal nature of the quantum spacetime in two-dimensional quantum gravity
is one of the central themes of this work.

In section \sref{sqg} the main concepts in the continuum formalism are introduced. In the following section
we review technical details about the factorization of the diffeomorphisms from the functional integral.
In section \sref{fd} we introduce fractal dimensions to characterize the fractal nature of quantum spacetime.
That section, and in particular section \sref{erg:sd} are partly based on work presented
in \cite{Ambjorn:1997jf,Ambjorn:1997pr}.

Reviews about the continuum approach to two-dimensional quantum gravity
can be found in \cite{Ginsparg:1993is} and the references therein. For an introduction
to conformal field theory see \cite{Belavin:1984vu,Cardy:1987,Cardy:1988}.

\section{Scaling in quantum gravity}
\label{sec:sqg}

In this section we introduce the basic concepts in two-dimensional quantum gravity.
We begin with the partition function and the Hartle-Hawking wavefunctionals. The most
natural object to address questions about the scaling behaviour of the theory
is the geodesic two-point function. We discuss its scaling behaviour in detail and
introduce the concept of the intrinsic fractal dimension which illustrates
the fractal nature of the intrinsic geometry of the two-dimensional quantum space-time.

\subsection{Partition function}

Let $M$ be a two-dimensional, closed, compact, connected, orientable
manifold of genus $h$.
Then the partition function for two-dimensional quantum gravity
can formally be written as the functional integral
\begin{equation}
  \label{e1}
  Z(G,\Lambda) =
  \sum_{h\geq 0}
  \int\! {\cal D}[{\gmn}]\ 
  e^{-S_{\text{EH}}(g;G,\Lambda)}
  \int {\cal D}_gX\ e^{-S_{\text{matter}}(g, X)}.
\end{equation}
Here
\begin{equation}
  \label{e2}
  S_{\text{EH}}(g;G,\Lambda) =
  \Lambda \int_{M}\! d^2\xi \sqrt{g}
  -\frac{1}{4\pi G} \int_{M}\! d^2\xi \sqrt{g}\ {\cal R}(\xi)
\end{equation}
is the classical reparametrization invariant
Einstein-Hilbert action \cite{Einstein:1916,Hilbert:1915} with the gravitational
coupling constant $G$, the cosmological constant $\Lambda$ and 
the curvature scalar $\cal R$.
According to the Gauss--Bonnet theorem, 
the last term in (\ref{e2}) is a topological invariant,
called the Euler characteristic $\chi$ of $M$:
\begin{equation}
  \label{e3}
  \chi(h) = \frac{1}{4\pi} \int_{M}\! d^2\xi \sqrt{g}\ {\cal R}(\xi)  = 2-2h,
\end{equation}
while the first term is the volume $V_g$ of $M$ equipped with the
metric ${\gmn}$:
\begin{equation}
  \label{e4}
  V_g = \int_{M}\! d^2\xi \sqrt{g}.
\end{equation}
Therefore, \eref{e1} can be rewritten as
\begin{equation}
  \label{e5}
  Z(G,\Lambda) = \sum_{h\geq 0} e^{\frac{\chi(h)}{G}} Z(\Lambda),
\end{equation}
where $Z(\Lambda)$ is defined as
\begin{equation}
  \label{e5a}
  Z(\Lambda) = \int\! {\cal D}[{\gmn}]\ e^{-S({g}, \La)}
  \int\! {\cD}_gX\ e^{-S_{\text{matter}}(g, X)},
\end{equation}
with $S({g}, \La) = \La V_g$.

$S_{\text{matter}}(g, X)$ in \eref{e1} denotes 
any reparametrization invariant action for
conformal matter fields $X$ with central charge $D$ coupled to gravity.
A typical example is the coupling of
$D$ free scalar matter fields $X^1, \ldots, X^D$ to gravity. In this case
\begin{equation}
  \label{e2a}
  S_{\text{matter}}(g, X) = \frac{1}{8\pi} \int_M\! d^2\xi\sqrt{g}\
  g^{\mu\nu} \partial_{\mu}X^a\partial_{\nu}X^a,
\end{equation}
which is diffeomorphism invariant and
invariant under Weyl rescalings of the metric:
\begin{equation}
  \label{e5b}
  S_{\text{matter}}(e^{\phi}g, X) = S_{\text{matter}}(g, X).
\end{equation}
Note that \eref{e2a} can also be interpreted
as an embedding of $M$ in a $D$-dimensional Euclidean
space, thus leading to an interpretation of \eref{e1}
as bosonic string theory in $D$ dimensions \cite{Polyakov:1981rd}.

The functional integration $\int\!\cD[\gmn]$ in \eref{e1} is an integration
over all geometries, that means all diffeomorphism classes $[\gmn]$ of metrics $\gmn$ on
the manifold $M$. This is often denoted formally as
\begin{equation}
  \label{add1}
  \int\!\frac{\cD\gmn}{\text{Vol(Diff)}},
\end{equation}
where $\text{Vol(Diff)}$ is the volume of the group of diffeomorphisms on $M$.
Since this group is not compact the quotient \eref{add1} does not make sense
beyond a formal level. We will derive expressions for the measures
$\cD\gmn$, $\cD[\gmn]$ and $\cD_gX$ in section \sref{cc}.

In \eref{e1} the sum goes over all topologies of two-dimensional manifolds,
that means over all genera $h$. It is presently unknown how to define such a sum over
topologies different from a perturbative expansion in $h$. It comes
as a bitter aspect of the theory that we only know how to perform the functional
integration in \eref{e1} for a given manifold and thus only for a given fixed topology,
while a summation over topologies has to be performed by hand.
In four dimensions it is known that the topologies even cannot be classified.
Therefore the role of topological fluctuations is still unclear in two and
in higher dimensions. In the remainder of this work we will always fix the topology
and thereby disregard the sum over the genera in \eref{e1}.

Sometimes it is useful to define the partition function
in an ensemble of universes with fixed volume $V$:
\begin{eqnarray}
  \label{e3a}
  Z(V) &=& \int\! {\cal D}[{\gmn}]\ \de(V-V_g)
  \int {\cal D}_gX\ e^{-S_{\text{matter}}(g, X)}\nonumber\\
  &\equiv& \int\! {\cal D}[{\gmn}]_V\ \int {\cal D}_gX\ e^{-S_{\text{matter}}(g, X)}.
\end{eqnarray}
$Z(\La)$ is the Laplace transform of $Z(V)$
\footnote{That is also the reason why we denote
  both with the same symbol and distinguish between them
  by the names of their arguments.}, that means:
\begin{eqnarray}
  \label{e3b}
  Z(\La) &=& \int_0^{\infty}\! dV\ e^{-\La V} Z(V),\\
  Z(V) &=& \int_{\sigma-i\infty}^{\sigma+i\infty}
  \frac{d\La}{2\pi i}\ e^{\La V} Z(\La),
\end{eqnarray}
where $\sigma$ is a real constant which exceeds the
real part of all the singularities of $Z(\La)$.
Formula \eref{e3a} shows that although the action of
pure two-dimensional quantum gravity is trivial and does not
contain any propagating degrees of freedom, quantum gravity
in two dimensions is a true quantum problem, since each
equivalence class of metrics
is counted once with the same weight. That means that there is no classical spacetime
around which we expand and the theory is as ``quantum-like'' as it can get. Therefore
it might contain important information about quantum gravity in higher dimensions.

As we show in section \sref{cc} $Z(V)$ scales as
\begin{equation}
  \label{e3c}
  Z(V) \propto V^{\ga-3}, \quad\text{or~}
  Z(\La) \propto \La^{2-\ga},
\end{equation}
which defines a scaling exponent $\ga$
that depends on the matter coupled to gravity and on the
topology of the universe. The computation of $\ga$ gives the
result \cite{Knizhnik:1988ak,David:1988hj,Distler:1989jt}
\begin{equation}
  \label{e3d}
  \ga = 2 + \frac{1-h}{12}\left(D-25 - \sqrt{(25-D)(1-D)}\right).
\end{equation}
This formula is often called KPZ-formula.
For pure gravity ($D=0$) $\ga$ equals $-1/2$, for the Ising model coupled to
gravity ($D=1/2$) we have $\ga = -1/3$ and at $D=1$, $\ga$ equals $0$ for spherical
topology.
Clearly this formula breaks down for conformal charges $D>1$, where $\ga$ assumes
complex values. This has sometimes been called the $D=1$ barrier.
There is now some evidence for the fact that two-dimensional quantum gravity
coupled to matter with central charge $D>1$ is in a branched polymer phase~\cite{Harris:1996hk,David:1996vp}
in which the surfaces collapse to tree-like objects.

The expectation value of some observable $\cO(g,X)$ is defined as
\begin{equation}
  \label{add2}
  \bra \cO(g,X)\ket_{\La} = \frac{1}{Z(\La)} \int\! {\cal D}[{\gmn}]\ e^{-\La V_g}
  \int\! {\cD}_gX\ e^{-S_{\text{matter}}(g, X)} \cO(g,X),
\end{equation}
or for fixed volume $V$ as 
\begin{equation}
  \label{add3}
  \bra \cO(g,X)\ket_{V} = \frac{1}{Z(V)} \int\! {\cal D}[{\gmn}]_V
  \int\! {\cD}_gX\ e^{-S_{\text{matter}}(g, X)} \cO(g,X).
\end{equation}

\subsection{Hartle-Hawking wavefunctionals}
\label{sec:hh}

Let us in the remainder of this section ignore the coupling of possible matter
fields to quantum gravity. The definitions below can easily be expanded to
the general case. Furthermore we concentrate on spherical topology only. Again
the generalization is straightforward.

Typical observables in two-dimensional quantum gravity are
loop amplitudes, that means amplitudes for one-dimensional universes.
Let $M$ be topologically equivalent to a sphere
with $b$ holes. The induced gravity on the boundary is covered
by the modified action
\begin{equation}
  \label{e7}
  S(g,\La,Z_1, \ldots,Z_b) = \La V_g + \sum_{i=1}^b Z_i L_{g,i},
\end{equation}
where $L_{g,i}$ denotes the length of the $i$'th boundary component in the
metric ${\gmn}$.
In this case the partition function is given by
\begin{equation}
  \label{e8}
  W(\La, Z_1, \ldots, Z_b) = \int\! {\cal D}[{\gmn}]\
  e^{-S(g, \La, Z_1, \ldots, Z_b)}.
\end{equation}
This can also be interpreted as the amplitude for $b$
one-dimensional universes of arbitrary lengths.
Since the lengths of the boundary components are invariant
under diffeomorphisms it makes sense to fix them to prescribed
values $L_1, \ldots, L_b$. This yields the definition
of the Hartle-Hawking wave functionals \cite{Hartle:1983ai}:
\begin{equation}
  \label{e6}
  W(\La, L_1, \ldots, L_b) = \int\! {\cal D}[{\gmn}]\
  e^{-S(g, \La)}  \prod_{i=1}^b \delta(L_i-L_{g,i}).
\end{equation}
We note that (\ref{e8}) is the Laplace transform of
(\ref{e6}), that means:
\begin{equation}
  \label{e9}
  W(\La, Z_1, \ldots, Z_b) = \int_{0}^{\infty} \prod_{i=1}^b dL_i
  e^{-Z_iL_i}\  W(\La, L_1, \ldots, L_b).
\end{equation}
In the ensemble of universes with fixed volume $V$
the wave functionals are given by:
\begin{equation}
  \label{e12}
  W(V,L_1, \ldots, L_b) = \int\! {\cal D}[{\gmn}]\
  \delta(V-V_g)\prod_{i=1}^b\delta(L_i-L_{g,i}).
\end{equation}
The cosmological constant $\La$ and the volume $V$ are
conjugate variables, that means:
\begin{equation}
  \label{e13}
  W(\La, L_1, \ldots, L_b) = \int_0^{\infty}\! dV\
  e^{-\La V} W(V, L_1, \ldots, L_b).
\end{equation}
In the case $b=0$ one recovers the partition functions
\eref{e5a} and \eref{e3a}.

\subsection{The two-point function}
\label{sec:obs}


To define reparametrization invariant 
correlation functions let $d_g(\xi,\xi')$ be the geodesic
distance between two points $\xi$ and $\xi'$ with respect to the metric ${\gmn}$.
Then the invariant two-point function is defined as
\begin{equation}
  \label{e10}
  G(\La,R) = \int\! {\cal D}[{\gmn}]\ e^{-S(g,\La)}
  \int_M\! d^2\xi \sqrt{g(\xi)} \int_M\! d^2\xi'\sqrt{g(\xi')}\ \delta(R-d_g(\xi,\xi')).
\end{equation}
$G(\La,R)$ can also be interpreted as the partition function for universes with
two marked points separated by a geodesic distance $R$.
The integrated two-point function, called the susceptibility $\chi$
has the following behaviour:
\begin{equation}
  \label{e10a}
  \chi(\La) = \int_0^{\infty}\! dR\ G(\La, R) =
  \frac{\partial^2 Z(\La)}{\partial \La^2}
  \sim \La^{-\ga}.
\end{equation}
Therefore $\ga$ is often called the susceptibility exponent.
Yet another characterization of the exponent $\ga$ from the branching ratio
of the two-dimensional universes will be explained in chapter \sref{disc}.
For fixed volume of spacetime the definition of the two-point function is
\begin{equation}
  \label{e16}
  G(V, R) = \int\! {\cal D}[{\gmn}]_V\ \int_M\! d^2\xi\sqrt{g(\xi)}
  \int_M\! d^2\xi'\sqrt{g(\xi')}\ \delta(R-d_g(\xi, \xi')),
\end{equation}
which is related to formula (\ref{e10}) via
\begin{equation}
  \label{e17}
  G(\La,R) = \int_0^{\infty}\! dV\ e^{-\La V} G(V, R).
\end{equation}
Similar to (\ref{e10}) a whole
set of invariant correlation functions which depend on
the geodesic distance can be defined by multiplying
the local measures  $d^2\xi\sqrt{g(\xi)}$
and $d^2\xi'\sqrt{g(\xi')}$ by powers of the invariant
curvature scalars ${\cal R}(\xi)$ and ${\cal R}(\xi')$.

Note that by (\ref{e10}) or \eref{e16} the concept of geodesic distance
becomes meaningful even in quantum gravity. The two-point
function can be interpreted as the partition function of the
ensemble of universes which have two marked points separated by 
a geodesic distance $R$. Both, its short distance behaviour and
its long distance behaviour reveal the fractal structure of the
most important metrics that contribute to the functional
integral (\ref{e1}).

An important property of the two-point function \eref{e10} is the
inequality
\begin{equation}
  \label{e16a}
  G(\La, R_1+R_2) \geq \text{const}\times G(\La, R_1)\
  G(\La, R_2),
\end{equation}
which follows from a simple gluing argument which is explained in
chapter \sref{disc}. Up to a constant,
\eref{e16a} is equivalent to the subadditivity of $-\log{G(\La, R)}$,
from which the existence of the limit
\begin{equation}
  \label{e16b}
  \lim_{R\rightarrow\infty} \frac{-\log{G(\La, R)}}{R} =
  M(\La) 
\end{equation}
follows from general arguments (see for example \cite{Ruelle:1977}).
Furthermore one can deduce that $M(\La)$ is an increasing function
of $\La$ and $M(\La)>0$ for $\La>0$. At this stage one cannot prove
that $M(\La)$ scales to zero as $\La$ goes to zero. However let us
assume that this is the case and verify it later by an explicit calculation
of the two-point function in section \sref{fs}:
\begin{equation}
  \label{e16c}
  M(\La) = c\ \La^{\nu},
\end{equation}
with a dimensionless constant $c$. 
That means that for large $R\gg M(\La)^{-1}$ the two point function
falls off exponentially with a subleading correction:
\begin{equation}
  \label{e16d}
  G(\La, R) \sim \La^{\nu-\ga}\ e^{-c\La^{\nu}R},\text{~for~}
  R\gg\frac{1}{M(\La)}.
\end{equation}
The power correction can be found by applying dimensional arguments
to \eref{e10a}.
We conclude that the average volume of two-dimensional universes with
two marked points separated by a geodesic distance $R$ is proportional
to $R$ if $R$ is large enough:
\begin{equation}
  \label{e16e}
  \bra V\ket_R \equiv - \frac{\partial \log{G(\La, R)}}{\partial \La}
  \underset{R\gg M(\La)^{-1}}{\sim} \La^{\nu - 1} R.
\end{equation}
That means, for large $R$ typical universes have the shape of long thin tubes.
On the other hand, for $R\sim\La^{-\nu}$ the exponential decay turns
over into some power law and we get
\begin{equation}
  \label{e16f}
  \bra V\ket_R \sim R^{\frac{1}{\nu}}, \text{~for~}R\sim\La^{-\nu},
\end{equation}
by inserting into \eref{e16e}.
Per definition, the exponent of $R$ in this equation equals the
(grand canonical) intrinsic Hausdorff dimension $d_H=\frac{1}{\nu}$.

The large $R$ behaviour of $G(V, R)$ can be computed from \eref{e16d} by a saddle-point
calculation. The result up to power corrections is:
\begin{equation}
  \label{e16g}
  {G(V,R)} \sim e^{-\tilde{c}\left( \frac{R}{V^{\nu}}\right)^{\frac{1}{1-\nu}}},
  \text{~for~}\frac{R}{V^{\nu}}\gg 1,
\end{equation}
where $\tilde{c}=\frac{1-\nu}{\nu}(\nu c)^{\frac{1}{1-\nu}}$.

Another concept of fractal dimension can be applied in the ensemble
of surfaces with fixed volume $V$ for small distances $R$.
Let us define with
\begin{equation}
  \label{e17a}
  l(g,R) = \frac{1}{V_g} \int_{M}\! d^2\xi\sqrt{g(\xi)}
  \int_{M}\! d^2\xi'\sqrt{g(\xi')}\
  \de(R-d_g(\xi,\xi'))
\end{equation}
the average length of boundaries on a manifold $M$ with the metric $\gmn$,
which have a geodesic distance $R$ from a marked point. Then
$l(g,R)dR$ is the average volume of a spherical shell
of radius $R$ and thickness $dR$. The quantum expectation value
in the ensemble of universes with fixed volume is given by
\begin{equation}
  \label{e17b}
  \bra l(g,R)\ket_V = \frac{1}{Z(V)} \int\! \cD[\gmn]_V\ l(g,R) = \frac{G(V,R)}{VZ(V)}.
\end{equation}
Now we conclude from
\begin{equation}
  \label{e17ca}
  \int_0^{\infty}\! dR\ \bra l(g,R)\ket_V = V
\end{equation}
that
\begin{equation}
  \label{add4}
  \text{dim}[V] = \text{dim}[R] + \text{dim}[\bra l(g,R)\ket_V].
\end{equation}
Thus $\bra l(g,R)\ket_V$ has the scaling behaviour
\begin{equation}
  \label{add5}
  \bra l(g,R)\ket_V=V^{1-\nu}\ F\left(\frac{R}{V^{\nu}}\right),
\end{equation}
where $F(x)$ is a function which falls off exponentially for large $x$, see \eref{e16g}.
The (canonical) intrinsic Hausdorff dimension $d_h$ is now defined
by the scaling of $\bra l(g,R)\ket_V\sim R^{d_h-1}$ for
small $R$. To be precise we expand \eref{add5} around $R=0$ and get:
\begin{equation}
  \label{e17e}
  \bra l(g,R)\ket_V \sim V^{1-\nu d_h}\ R^{d_h-1}, \text{~for~}
  R\ll V^{\frac{1}{d_h}}.
\end{equation}
For smooth $d$-dimensional manifolds we have $d_H=d_h=d$. If $\bra l(g,R)\ket_V$
stays nonzero and finite for $V\rightarrow \infty$ we have
\begin{equation}
  \label{e17f}
  \nu d_h=1, \text{~that means $d_H=d_h$}.
\end{equation}
This requirement is called the smooth fractal condition. It means that
the average circumference of circles with a small geodesic radius $R$
does not depend on the global volume of the universe.
However it is well known that the smooth fractal condition is not fulfilled
for the model of multicritical branched polymers \cite{Ambjorn:1990wp}, compare section \sref{bpo}.
Therefore its validity is non-trivial and should not be taken for granted uncritically.
It turns out that in two-dimensional quantum gravity the smooth fractal condition
is fulfilled. That means that there is only one
fractal dimension for short and for long distances together.

For pure gravity, in the case of spherical topology, the two-point function
$G(\La, R)$ can be calculated exactly with a transfer-matrix method \cite{Kawai:1993cj}
or alternatively by a peeling method \cite{Watabiki:1995ym}. The result 
\begin{equation}
  \label{e17g}
  G(\La, R) \propto \La^{3/4} \frac{\cosh{c\La^{1/4}R}}{\sinh^3{c\La^{1/4}R}}
\end{equation}
fulfills the standard scaling relations, but with non-standard exponents.
$G(\La, R)$ falls off as $e^{-2c\La^{1/4}R}$, from which we read
off $\nu=1/4$, so that the Hausdorff dimension of pure two-dimensional quantum gravity equals {\em four}.

Another scaling exponent, though not independent of $\ga$ and $\nu$, can be defined by:
\begin{equation}
  \label{e17h}
  G(\La, R) \sim R^{1-\eta}, \text{~for $1\ll R\ll \La^{-\nu}$}.
\end{equation}
$\eta$ is called the anomalous scaling dimension. By expanding
\eref{e17g} for small $R$ we get
\begin{equation}
  \label{e17j}
  G(\La, R) = \frac{1}{R^{3}} - \frac{1}{15}\La R + O(R^3),
\end{equation}
and thus $\eta=4$. This is a
notable result, since in ordinary statistical systems $\eta$ is
always smaller than $2$.

The so-called Fisher scaling relation
\begin{equation}
  \label{e17i}
  \ga = \nu(2-\eta)
\end{equation}
relates the exponents defined above. It can be 
derived by applying dimensional arguments to \eref{e10a}.

\section{Liouville theory: A brief reminder}
\label{sec:cc}

A central problem in the continuum approach to two-dimensional quantum gravity
is posed by the diffeomorphism invariance of the theory.
It is comparatively easy \cite{Mottola:1995sj} to derive an expression for the formal integration $\cD\gmn$
since one can define a natural scalar product on the cotangent space to the
space of all metrics which defines a volume form in the same way as in finite dimensional
Riemannian geometry. 

However, since the measure and the action are diffeomorphism invariant, these expressions are ill-defined.
Therefore a gauge fixing and the factorization of the diffeomorphisms from the measure is
required. This has been performed in \cite{Polyakov:1981rd} where the functional integration
over geometries is expressed as a Liouville field theory.
This theory has been developed and explained in \cite{Alvarez:1983zi,Moore:1986}.
The measure for the Liouville mode is very
complicated. Therefore two-dimensional quantum gravity has strictly speaking not been solved yet
in the continuum approach.

However, the critical scaling exponents could be obtained in the light-cone gauge \cite{Knizhnik:1988ak}
and later in the conformal gauge by consistent scaling assumptions \cite{David:1988hj,Distler:1989jt}.

\subsection{Functional measures}
\label{sec:fm}

Let $R(M)$ be the space of all positive definite Riemannian metrics
on a $d$-dimensional manifold ${M}$
and let $T_gR(M)$ be its cotangent space at a point $\gmn\in R(M)$.

To define the functional measure
$\cD\gmn$ one makes use  of the fact, that the cotangent space $T_gR(M)$ can
be naturally equipped with a diffeomorphism invariant scalar product
$\bra\cdot,\cdot\ket_T$. This is used to define a measure $\cD_g\de\gmn$
on $T_gR(M)$ which descends to a measure on $R(M)$. The line element $ds^2$
for fluctuations $\de\gmn\in T_gR(M)$ of the metric can be written as
\begin{equation}
  \label{meas1.1}
  ds^2 = \bra\de g,\de g\ket_T\equiv
  \int_{{M}}\! d^d\xi \sqrt{g}\ \dgmn \GMNAB \dgab,
\end{equation}
with the DeWitt (super)metric \cite{DeWitt:1967}
\begin{equation}
  \label{meas1.3}
  \GMNAB = \frac{1}{2} ( g^{\mu\al} g^{\nu\be} + g^{\mu\be} g^{\nu\al}
  + C \gMN\gAB).
\end{equation}
Up to an overall normalization constant, \eref{meas1.1} is the only
diffeomorphism invariant, ultralocal, that means dynamically neutral
distance on $T_gR(M)$. The constant $C$ in \eref{meas1.3} takes the value
$-2$ in canonical quantum gravity \cite{DeWitt:1967}.
In our framework $C$ cannot be computed. The choice of $C$ determines
the signature of the metric $\bra\cdot,\cdot\ket_T$ as can be seen by splitting
the fluctuation $\de\gmn$ into its trace part $\gmn\de c$ and its
tracefree part $\de h_{\mu\nu}$. This decomposition is orthogonal
with respect to $\bra\cdot,\cdot\ket_T$. On the tracefree subspace
$\GMNAB$ has eigenvalue $1$, while it has eigenvalue $1+\frac{Cd}{2}$
on the trace sector. Thus, the DeWitt metric is positive definite for
$C>-\frac{2}{d}$ and negative definite for $C<-\frac{2}{d}$. For
$C=-\frac{2}{d}$, $\GMNAB$ is the orthogonal projection on the tracefree part
of $\de\gmn$.
The inverse of the DeWitt metric, which satisfies
\begin{equation}
  \label{meas1.10}
  \Gmnab G^{\al\be,\rho\sig} = \frac{1}{2} \left( \de_{\mu}^{\rho}
  \de_{\nu}^{\sig} + \de_{\mu}^{\sig}  \de_{\nu}^{\rho} \right),
\end{equation}
is given by
\begin{equation}
  \label{meas1.11}
  \Gmnab = \frac{1}{2} \left( g_{\mu\al} g_{\nu\be} + g_{\mu\be}
  g_{\nu\al} -\frac{C}{1+\frac{Cd}{2}}{\gmn} g_{\al\be} \right),
\end{equation}
as can be verified by multiplication with \eref{meas1.3}.

The line element \eref{meas1.1} can be rewritten as
\begin{equation}
  \label{meas1.7}
  ds^2 = \int_{{M}}\! d^d\xi \int_{{M}}\! d^d\xi'\
  \sum_{\mu\leq\nu}\sum_{\al\leq\be}
  \dgmn(\xi)
  \Om^{\mu\nu,\al\be}(\xi,\xi') \dgab(\xi'),
\end{equation}
with $\Om^{\mu\nu,\al\be}(\xi,\xi')=  \sqrt{g(\xi)}\ \Ga^{\mu\nu,\al\be}(\xi)\ \de(\xi-\xi')$,
and the $\frac{d(d+1)}{2}\times\frac{d(d+1)}{2}$ matrix $\Ga^{\mu\nu,\al\be}(\xi)$ is
defined as
\begin{equation}
  \label{meas1.7a}
  \Ga^{\mu\nu,\al\be}(\xi) = \left\{
    \begin{array}{ll}
      4\GMNAB(\xi), & \text{if $\mu<\nu$ and $\al<\be$},\\
      2\GMNAB(\xi), & \text{if $\mu<\nu$ and $\al=\be$, or $\mu=\nu$ and $\al<\be$},\\
      \GMNAB(\xi), & \text{if $\mu=\nu$ and $\al=\be$}.
    \end{array}
    \right.
\end{equation}
This induces a volume form $\cD\dgmn$ on the cotangent space, which descends
to a volume form $\cD\gmn$ on $R(M)$:
\begin{eqnarray}
  \label{meas1.9}
  \cD\dgmn &=& \sqrt{\det \Om^{\mu\nu,\al\be}(\xi,\xi')}
  \prod_{\xi\in{M},\
    \mu\leq\nu}\! d\dgmn(\xi),\\
  \label{meas1.9a}
  \cD\gmn &=& \sqrt{\det \Om^{\mu\nu,\al\be}(\xi,\xi')}
  \prod_{\xi\in{M},\
  \mu\leq\nu}\! d\gmn(\xi).
\end{eqnarray}
We first note that due
to the $\de$-function the determinant of $\Om^{\mu\nu,\al\be}(\xi,\xi')$
factorizes into a product over all spacetime points $\xi\in{M}$. The local determinant of the
$\frac{d(d+1)}{2}\times\frac{d(d+1)}{2}$ matrix $\sqrt{g(\xi)}\ \Ga^{\mu\nu,\al\be}(\xi)$
can be evaluated by variation with respect to $\de\log g=-\gMN\dgmn$:
\begin{eqnarray}
  \label{meas1.12}
  \de \log\det\left(\sqrt{g}\ \Ga^{\mu\nu,\al\be}\right) &=&
  \tr\ \de\log\left(\Ga^{\mu\nu,\al\be}\right) +
  \tr\ \de\log\left(\sqrt{g}\ \id^{\mu\nu,\al\be}\right) \nonumber\\
  &=& \Gmnab\de G^{\al\be,\mu\nu} + \frac{d(d+1)}{4} \de\log g
  \nonumber \\
  &=& \frac{1}{4} (d+1)(d-4)\ \de\log g,
\end{eqnarray}
where it has been used that the inverse of $\Ga^{\mu\nu,\al\be}$
equals $\Gmnab$ restricted to all indices with $\mu\leq\nu$ and
$\al\leq\be$.
Therefore it follows that
\begin{equation}
  \label{meas1.13}
  \det\left(\sqrt{g}\ \Ga^{\mu\nu,\al\be}\right) = \kappa
  g^{\frac{(d+1)(d-4)}{4}},
\end{equation}
where the constant $\kappa$ can be determined by specializing to
$\gmn=\de_{\mu\nu}$, which gives
$\ka=\left( 1+\frac{Cd}{2} \right)/2^{\frac{d(d-1)}{2}}$.
Thus the measures \eref{meas1.9} and \eref{meas1.9a} equal
\begin{eqnarray}
  \label{meas1.15}
  {\cal D}\dgmn &=& \text{const}\times \prod_{\xi\in{M}}
  \prod_{\mu\leq\nu} g(\xi)^{\frac{d-4}{4d}} d\dgmn(\xi),\\
  \label{meas1.15a}
  {\cal D}\gmn &=& \text{const}\times \prod_{\xi\in{M}}
  \prod_{\mu\leq\nu} g(\xi)^{\frac{d-4}{4d}}  d\gmn(\xi),
\end{eqnarray}
which we normalize such that
\begin{equation}
  \label{meas1.15b}
  \int\! \cD\dgmn\ e^{-\frac{1}{2} \bra\de g,\de g\ket_T} = 1.
\end{equation}
Note that $\sig = \frac{d-4}{4d}$ is the only exponent of $g(\xi)$ in
\eref{meas1.15} or \eref{meas1.15a} such that these measures are
diffeomorphism invariant. With the measure
\begin{equation}
  \label{meas1.15d}
  \prod_{\xi\in{M}}\prod_{\mu\leq\nu} g(\xi)^{\sig}d\dgmn(\xi),
\end{equation}
the Gaussian integral \eref{meas1.15b} is resolved by computing the
determinant of $g(\xi)^{\frac{1}{2}-2\sig}\Ga^{\mu\nu,\al\be}$. This can be done
along the lines of \eref{meas1.12}. The result is
\begin{equation}
  \label{meas1.15c}
  \det\left( g(\xi)^{\frac{1}{2}-2\sig}\Ga^{\mu\nu,\al\be}\right)
  = \text{const}\times g(\xi)^{\frac{(d+1)((1-4\sig)d-4)}{4}},
\end{equation}
which is a diffeomorphism invariant constant if and only if
$\sig = \frac{d-4}{4d}$.
The same result can be derived more rigorously
by using the BRS-symmetry associated
with general coordinate transformations 
\cite{Fujikawa:1983im,Fujikawa:1984qk}.

Analogously, the measures for scalar and vector fields can be derived. The
corresponding metrics $\bra\cdot,\cdot\ket_S$ and $\bra\cdot,\cdot\ket_V$ are defined as
\begin{eqnarray}
  \label{add10}
  \bra\phi,\phi\ket_S &=& \int_M\! d^d\xi \sqrt{g}\ \phi(\xi) \phi(\xi),\\
  \bra \psi,\psi\ket_V &=& \int_M\! d^d\xi \sqrt{g}\ \psi_{\mu}(\xi) \gMN \psi_{\nu}(\xi).
\end{eqnarray}
The measures $\cD_g\phi$ and $\cD_g\psi_{\mu}$ turn out as:
\begin{eqnarray}
  \label{add11}
  \cD_g\phi &=& \prod_{\xi\in M} g(\xi)^{\frac{1}{4}}d\phi(\xi),\\
  \label{add11a}
  \cD_g\psi_{\mu} &=& \prod_{\xi\in M} \prod_{\mu=1}^d g(\xi)^{\frac{d-2}{4d}}d\psi_{\mu}(\xi). 
\end{eqnarray}

\subsection{Factorization of the diffeomorphisms}
\label{sec:liouville}

By construction, the functional measure and the action
in \eref{e1} and \eref{e3a} are invariant under diffeomorphisms.
Therefore, the measure overcounts physically equivalent configurations
related by the group of diffeomorphisms. To avoid this
overcounting of gauge equivalent metrics, the diffeomorphisms
have to be factored from the measure. In two dimensions, this
can be done in a covariant way in the conformal gauge.

Let us first recall that the space of metrics on
$M$ modulo diffeomorphisms and Weyl transformations is a finite
dimensional compact
space, called the moduli space $\cM$ of $M$,
which has dimension $0$, $2$ and $6h-6$ for genus $0$, $1$
and $h\geq 2$ respectively. It is
parametrized by the Teichm\"uller parameters $\tau=(\tau_1,\ldots,\tau_N)$.
That means that if for each moduli $\tau\in\cM$ we choose a fixed
background metric $\hat{g}_{\mu\nu}(\tau)$, all other metrics
$\gmn$ are contained
in the orbits under diffeomorphisms and Weyl transformations:
\begin{equation}
  \label{e18}
  \gmn = f^{\star}\left(e^{\phi} \hat{g}_{\mu\nu}(\tau)\right),
\end{equation}
where $f^{\star}$ denotes the action of the diffeomorphism
$f:{M}\rightarrow {M}$.
(Actually the background metrics $\hat{g}_{\mu\nu}$ can be chosen
such that they all have constant curvature.)

Any infinitesimal change $\de\gmn$ in the metric can be decomposed
into an infinitesimal Weyl transformation $\de\phi$, the effect of an infinitesimal
diffeomorphism $\xi$ and the effect of varying the Teichm\"uller parameters:
\begin{equation}
  \label{z1}
  \de \gmn = f^{\star}\de\phi\ \gmn +
  (\nabla_{\mu}\xi_{\nu}+\nabla_{\nu}\xi_{\mu}) +
  \frac{\partial \gmn}{\partial\tau_i}\de\tau_i.
\end{equation}
Here $\nabla_{\mu}$ denotes the covariant derivative. We want to
orthogonalize this decomposition with respect to the
scalar product $\bra\cdot,\cdot\ket_T$.
First note that the tracefree part
of the effect of the diffeomorphism $\xi$ is given by the conformal
Killing form
\begin{equation}
  \label{z2}
  (P\xi)_{\mu\nu} = \nabla_{\mu}\xi_{\nu}+\nabla_{\nu}\xi_{\mu}
  -\gmn \nabla_{\al}\xi^{\al},
\end{equation}
which maps vector fields into traceless symmetric tensors.
The zero modes of $P$ are called conformal Killing vectors. For spherical
topology there are six linearly independent conformal Killing vectors, for
torus topology there are two while there are none for higher genus. The
conformal Killing vectors are important because they induce a variation
$\de\gmn$ in the direction of $\gmn$ which can thus be compensated by a
Weyl rescaling. In other words, the decomposition \eref{z1}
is in general not unique.
To make it unique one chooses a diffeomorphism
$\tilde{\xi}\in (\text{ker} P)^{\bot}$
orthogonal to the zero modes of $P$.
That means the gauge will only be fixed up to the conformal Killing vectors
and we expect a remaining, at most six-dimensional symmetry in the gauge-fixed
expressions. It follows that all variations $\de\gmn\in T_gR(M)$ of the
metric can be uniquely decomposed into the variation of the conformal
factor, the action of a diffeomorphism orthogonal to the conformal Killing
vectors and the variation of the Teichm\"uller parameters. This decomposition
can be written as a mutually orthogonal sum of a trace part, a tracefree part
orthogonal to the moduli deformations and these moduli deformations:
\begin{eqnarray}
  \label{z3}
  \de\gmn &=& \lbrace \tilde{f}^{\star}\de\phi +
  \nabla_{\al}\tilde{\xi}^{\al}
  +\frac{1}{2}\gAB\frac{\partial \gab}{\partial\tau_i}\de\tau_i\rbrace\gmn
  +\lbrace P\tilde{\xi} + P(P^+P)^{-1}P^+k^i_{\mu\nu}\de\tau_i\rbrace\nonumber\\
  &&+\left(1-P(P^+P)^{-1}P^+\right)k_{\mu\nu}^i\de\tau_i,
\end{eqnarray}
where
\begin{equation}
  \label{z4}
  k_{\mu\nu}^i = \frac{\partial\gmn}{\partial\tau_i}
  - \frac{1}{2}\gmn \gAB \frac{\partial\gab}{\partial\tau_i}.
\end{equation}
The adjoint $P^+$ of $P$ is defined by the relation
$\bra h,P\xi\ket_T = \bra P^+h,\xi\ket_V$.
To prove that the decomposition \eref{z3} is orthogonal just note
that $1-P(P^+P)^{-1}P^+$ is the projector on the zero modes $\psi_l$ of ${P^+}$.
The tracefree part is orthogonal to the trace part by definition of the
scalar product $\bra\cdot,\cdot\ket_T$.

The change of variables $\gmn\rightarrow(\phi,\tilde{f},\tau)$
involves a Jacobian $J(\phi,\tau)$:
\begin{equation}
  \label{z5}
  \int\!\cD[\gmn]\ \cF(g) = \int\!\cD f\int\!\frac{\cD\tilde{f}}{\cD f}\cD_{e^{\phi}\hat{g}}\phi
  d\tau\ J(\phi,\tau)\ \cF(e^{\phi}\hat{g}(\tau)).
\end{equation}
Here one integrates some reparametrization invariant functional $\cF$. To find $J$
one substitutes the orthogonal decomposition \eref{z3} into the normalization condition
\begin{equation}
  \label{z6}
  1=\int\!\cD\de\gmn\ e^{-\frac{1}{2}\bra\de g,\de g\ket_T}
\end{equation}
and performs the change of variables on the tangent space. The result of these
Gaussian integrations is
\begin{equation}
  \label{z7}
  1=J(\phi,\tau) \bigg[\frac{\det\bra\psi_k,\psi_l\ket_T}{\det'(P^+P)}\bigg]^{\frac{1}{2}}\
  \bigg[\det\bra\psi_m,\frac{\partial g}{\partial\tau_n}\ket_T\bigg]^{-1}.
\end{equation}
At last, the change from the diffeomorphisms
$f$ to $\tilde{f}$ and the conformal Killing
vectors $\om_a$, which fulfill $P(e^{\phi}\om_a)=0$,
has to be computed. The final result
is
\begin{eqnarray}
  \label{z8}
  \int\!\cD[\gmn]\ &&\hspace{-0.7cm}\cF(g) =
  \int\!\cD f_{\mu}\int\!\frac{d\tau}{v(\tau)}
  \det\bra\psi_m,\frac{\partial g}{\partial\tau_n}\ket_T\times
  \nonumber\\
  &&
  \int\!\cD_{e^{\phi}\hat{g}}\phi
  \bigg[\frac{\det'(P^+P)}{\det\bra\psi_k,\psi_l\ket_T
    \det\bra\om_a,\om_b\ket_V}\bigg]^{\frac{1}{2}}\
  \cF(e^{\phi}\hat{g}(\tau)),
\end{eqnarray}
where $v(\tau)$ is the volume of the group
generated by the conformal Killing vectors.
Now the diffeomorphisms can be factored out.
Since it turns out that
$\frac{d\tau}{v(\tau)}
  \det\bra\psi_m,\frac{\partial g}{\partial\tau_n}\ket_T$
depends only on the Teichm\"uller parameters and not on $\phi$ we denote this
as $\cD\tau$. The square root is called the
Faddeev-Popov determinant and denoted as $Z_{\text{FP}}(e^{\phi}\hat{g})$.

\subsection{Gravitational dressing of scaling exponents}
\label{sec:gd}

After factoring the diffeomorphisms from the measure the gauge-fixed
expression for the partition function \eref{e3a} is:
\begin{equation}
  \label{e18b}
  Z(V) = \int\! \cD\tau\!\int\! {\cal D}_{e^{\phi}\hat{g}}\phi\
  \de(V-V_{e^{\phi}\hat{g}})\ Z_{\text{FP}}(e^{\phi}\hat{g})
  Z_{\text{mat}}(e^{\phi}\hat{g}),
\end{equation}
where $Z_{\text{mat}}$ is defined as:
\begin{equation}
  \label{add20}
  Z_{\text{mat}}(e^{\phi}\hat{g}) = \int\! {\cal D}_{e^{\phi}\hat{g}}X\
  e^{-S_{\text{matter}}(e^{\phi}\hat{g}, X)},
\end{equation}
with some matter action for matter with central charge $D$.
The measure for the Liouville mode $\phi$ is defined by the line element
\begin{equation}
  \label{e18c}
  d^2s = \bra\de\phi,\de\phi\ket_S = \int_{M}\! d^2\xi\sqrt{\hat{g}}\ e^{\phi}\
  \de\phi(\xi) \de\phi(\xi),
\end{equation}
which depends on $\phi$ itself. Therefore
the corresponding measure $\cD_{e^{\phi}\hat{g}}\phi$ is very complicated and it is
unknown how to perform the functional integration over $\phi$ with this measure.
The idea used to overcome these problems is to go over to the translational invariant
measure $\cD_{\hat{g}}\phi$ and to shift all dependence on $\phi$
in \eref{e18b} into the action. 

For this transformation we use the relations
\begin{eqnarray}
  \label{e19}
  Z_{\text{mat}}(e^{\phi}\hat{g}) &=& Z_{\text{mat}}(\hat{g}) e^{\frac{D}{48\pi}S_{\text{L}}(\hat{g}, \phi)},\\
  \label{e18a}
  Z_{\text{FP}}(e^{\phi}\hat{g}) &=& Z_{\text{FP}}(\hat{g}) e^{-\frac{26}{48\pi}S_{\text{L}}(\hat{g}, \phi)},\\
  \label{e21}
  {\cal D}_{e^{\phi}\hat{g}}\phi &=& {\cal D}_{\hat{g}}\phi\ e^{\frac{1}{48\pi}S_{\text{L}}(\hat{g}, \phi)},  
\end{eqnarray}
where $S_{\text{L}}(\hat{g}, \phi)$ is the Liouville action:
\begin{equation}
  \label{e20}
  S_{\text{L}}(\hat{g}, \phi) = \int_{M}\! d^2\xi\sqrt{\hat{g}}\
  \left(\frac{1}{2}\hat{g}^{\mu\nu}\partial_{\mu}\phi\partial_{\nu}\phi
  + \hat{\cal R}\phi\right).
\end{equation}
Here $\hat{\cal R}$ is the curvature scalar in the metric $\hat{g}_{\mu\nu}$.

The relation \eref{e19} for the matter partition function can be derived by
a couple of methods such as the heat kernel regularization \cite{Brown:1977sj} or the
zeta function regularization \cite{Hawking:1977ja}.
Since the matter action is invariant under Weyl transformations we can actually
say that the whole contribution comes from the transformation of the measure
$\cD_{e^{\phi}\hat{g}}X$ under conformal rescalings of the metric.
This is also the reason for the trace anomaly of the energy momentum tensor.

The transformation property \eref{e18a} for the Faddeev-Popov determinant can be
derived in a similar way. However, the involved operators are much more complicated,
compare \eref{z8}.

The transformation \eref{e21} could not be derived rigorously by these or other
methods. David \cite{David:1988hj} and Distler and Kawai \cite{Distler:1989jt} use
a selfconsistent bootstrap method to find this relation. 
Under the assumption that
the factor takes the form of an exponential of a
Liouville action with arbitrary
coefficients the coefficients are determined from general invariance considerations.

In this calculation the Liouville mode $\phi$ is rescaled by a factor $\al$, which
has to be determined. Taking all factors together, the partition function \eref{e18b}
can be rewritten as:
\begin{equation}
  \label{e22}
  Z(V) = \int\! \cD\tau\! \int\! {\cal D}_{\hat{g}}\phi\
  Z_{\text{FP}}(\hat{g})\
  Z_{\text{mat}}(\hat{g})\  
  e^{\frac{D-25}{48\pi}S_{\text{L}}(\hat{g}, \phi)}\
  \de(V-V_{e^{\alpha \phi}\hat{g}}),
\end{equation}
where the measures depend only on the background metric $\hat{g}$.
Thus the partition function scales under a rescaling of the volume as:
\begin{equation}
  \label{e33}
  Z(\la V) = \la^{\frac{D-25}{12\al}\chi - 1} Z(V),
\end{equation}
where $\chi$ is the Euler characteristic of the manifold $M$. To find
\eref{e33} we have shifted $\phi$ by $\frac{1}{\al}\log\la$ and used
the Gauss-Bonnet theorem \eref{e3}.

Now we have to compute the dressing exponent $\al$. It turns out that more generally
we can compute the anomalous scaling dimensions for a whole family of operators.
Consider the observable
\begin{equation}
  \label{add30}
  \cO(g) = \int_{M}\! d^2\xi\sqrt{g}\ \Phi(g),
\end{equation}
where $\Phi$ is a spinless primary field of dimension $\De_0$.
Typical simple examples are the volume $V_g=\int_M\! d^2\xi\sqrt{g}$ with $\De_0=0$
and the identity $1=\int_M\! d^2\xi\sqrt{g}\ \frac{1}{\sqrt{g}}\de(\xi)$ with $\De_0=1$.
Then $\cO(g)$ transforms under a rescaling $\gmn\rightarrow\la\gmn$ of the metric as
\begin{equation}
  \label{add31}
  \cO(\la g) = \la^{1-\De_0}\cO(g),
\end{equation}
if the geometry is kept fixed. Under the average over all geometries
the scaling behaviour exponent $\De_0$ is in general changed to some other value $\De$:
\begin{equation}
  \label{add32}
  \bra \cO(g)\ket_{\la V} = \la^{1-\De}\bra \cO(g)\ket_V.
\end{equation}
The coupling to gravity dresses the scaling behaviour of the field $\Phi$.
Therefore we make the ansatz
\begin{equation}
  \label{add33}
  \cO(\hat{g},\phi) = \int_M\! d^2\xi \sqrt{g}\ e^{\be\phi}\Phi(\hat{g})
\end{equation}
for the observable after the transition to the translational invariant measure.
Note that $\cO(\hat{g},\phi)$ is diffeomorphism invariant and fulfills $\cO(\hat{g},0) = \cO(\hat{g})$.
Now observe that \eref{e19}, \eref{e18a}, \eref{e21} and
$\cO(e^{\phi}\hat{g})$ are invariant under the transformation
\begin{equation}
  \label{e23}
  \hat{g}_{\mu\nu}(\xi)\rightarrow e^{\sigma(\xi)}\hat{g}_{\mu\nu}(\xi),
  \quad \phi(\xi) \rightarrow\phi(\xi) - \sigma(\xi).
\end{equation}
Therefore also $\cO(\hat{g}, \phi)$ has to be invariant.
This requirement leads to an expression for the gravitational dressing exponent
$\be$ as follows. Under \eref{e23} the product $\sqrt{\hat{g}}\ \Phi(\hat{g})$ transforms as
\begin{equation}
  \label{add34}
  \sqrt{\hat{g}(\xi)}\Phi(\hat{g}) \rightarrow e^{(1-\De_0)\sig(\xi)}\sqrt{\hat{g}(\xi)}\Phi(\hat{g}).
\end{equation}
Thus $e^{\be\phi(\xi)}$ has to transform as
\begin{equation}
  \label{add35}
  e^{\be\phi(\xi)} \rightarrow e^{(\De_0-1)\sig(\xi)}e^{\be\phi(\xi)}.
\end{equation}
On the other hand we can determine the transformation of $e^{\be\phi(\xi)}$ by
computing the Gaussian integral
\begin{equation}
  \label{add36}
  F_{\phi}(\hat{g},\be) = \int\!\cD_{\hat{g}}\phi\
  e^{\frac{D-25}{48\pi}S_{\text{L}}(\hat{g},\phi) + \be\phi(\xi)},
\end{equation}
where $\be\phi(\xi)$ is interpreted as an additional $\de$-function source. If we define
\begin{equation}
  \label{add37}
  \hat{J}(x) = \hat{\cR}(x) + \frac{48\pi\be}{D-25}\frac{1}{\sqrt{\hat{g}(x)}}\de(x-\xi),
\end{equation}
we get:
\begin{eqnarray}
  \label{add38}
  \lefteqn{
  F_{\phi}(\hat{g},\be) = \exp{\left[\frac{D-25}{96\pi}\int_M\! d^2x\sqrt{g}\
    \hat{J}(x)(\De_{\hat{g}}^{-1}\hat{J})(x)\right]}}\\
  &=& F_{\phi}(\hat{g},0)\ \exp{\left[\be\int_M\! d^2x\sqrt{\hat{g}}\
    \De_{\hat{g}}^{-1}(\xi,x) \hat{\cR}(x)\
  +\ \frac{24\pi\be^2}{D-25} \De_{\hat{g}}^{-1}(\xi,\xi)\right]}.\nonumber
\end{eqnarray}
Under the transformation \eref{e23} the first term in the exponent gives
an extra $-\be\sig(\xi)$. The second term is an evaluation of the propagator at
coinciding points which naively seen is singular. Thus we have to renormalize this
expression \cite{Polchinski:1986zf}. 
The renormalized propagator is defined as 
\begin{equation}
  \label{e30}
  \De_{\hat{g},\text{R}}^{-1}(\xi,\xi) = \lim_{x\rightarrow \xi}
  \left( \De_{\hat{g}}^{-1}(x,\xi) - \frac{1}{4\pi}\log{d_{\hat{g}}(x,\xi)^2}\right),
\end{equation}
where $d_{\hat{g}}(x,\xi)$ is the geodesic distance between $x$ and $\xi$.
Therefore the second term in the exponent gains an extra $\frac{6\be^2}{25-D}\sig(\xi)$
under \eref{e23}. Taking both contributions together we conclude:
\begin{equation}
  \label{e31}
  e^{\be\phi(x)} \rightarrow
  e^{-\be\sig(x) + \frac{6\be^2}{25-c}\sig(x)}\ e^{\be\phi(x)}
  \overset{!}{=} e^{(\De_0-1)\sig(x)}\ e^{\be\phi(x)}.
\end{equation}
The solution to this quadratic equation determines $\be$ and thus also $\al$:
\begin{eqnarray}
  \label{e32}
  \be &=& \frac{1}{12}\left(25-D - \sqrt{(25-D)(1-D+24\De_0)}\right),\\
  \al &=& \frac{1}{12}\left(25-D - \sqrt{(25-D)(1-D)}\right).
\end{eqnarray}
The sign of the square root is determined by requesting that in
the classical limit $c\rightarrow -\infty$ the scaling of observables is the same
as in a fixed geometry. 
With $\al$ inserted into \eref{e33} we can now verify the form \eref{e3d} for the susceptibility
exponent $\ga$.

For the expectation values of observables the dependence on the topology cancels , and we get:
\begin{equation}
  \label{e35}
  \bra\cO(g)\ket_{\la V} = \la^{\frac{\be}{\al}} \bra\cO(g)\ket_V = \la^{1-\De} \bra\cO(g)\ket_V.
\end{equation}
$\De$ evaluates as
\begin{equation}
  \label{add39}
  \De = \frac{\sqrt{1-D+24\De_0}-\sqrt{1-D}}{\sqrt{25-D}-\sqrt{1-D}}.
\end{equation}
The expectation value of products of observables scales with the sum of their scaling exponents.

\section{Fractal dimensions}
\label{sec:fd}

Quantum gravity as defined above is a functional integration over geometries.
Therefore we have to ask which geometrical concepts survive the quantum average
and how geometry is changed by it. In any theory of quantum gravity
the structure of spacetime is of primary interest and even its
dimensionality is a {\em dynamical} quantity. The transfer matrix method
in \cite{Kawai:1993cj} applied to pure two-dimensional quantum gravity revealed
that the spacetime has a self-similar structure on all scales with a fractal dimension
{\em four}. An understanding of the quantum spacetime in two-dimensional quantum gravity
might help to understand similar dynamical changes of the geometry in higher dimensions.

In this section we introduce three different concepts of dimensionality to
characterize the fractal structure of spacetime, each of which might describe different
aspects of the geometry.
The intrinsic Hausdorff dimension or fractal dimension $d_H$ has already been defined above
by the scaling of observables such as the average circumference of circles with their
``geodesic'' radii. These measurements are performed within the intrinsic geometry of the
universes. They are analogous to the usual ``clock and rod'' measurements in classical
gravity. General results for $d_H$ from Liouville theory
have been given in \cite{Watabiki:1993fk}.

We can also characterize the fractal structure of spacetime by embedding it into a
$D$-dimensional, typically Euclidean flat space. Measurements of, for example, the
mean square extent of the universes are then performed with the metric of the embedding
space. The finite size scaling of the mean square extent of the surfaces defines the
extrinsic Hausdorff dimension $D_H$. Intuitively the embedding gives us an idea what
typical universes would ``look like''. Indeed it is known that the extrinsic Hausdorff dimension
of two-dimensional quantum gravity is infinity \cite{Distler:1990jv,Kawai:1991qv}.
That means that the surfaces are crumpled together in a complicated way. 

Instead of determining the metrical structure of the spacetime directly by putting out
clocks and rods, one can also choose to observe the propagation of test particles.
From such observations one can in the classical case determine the propagator and
via an inverse Laplace transformation the associated heat kernel. Since all this can
be done in a reparametrization invariant way, we can define corresponding quantum
observables. This leads to the definition of the spectral dimension $d_s$ as
the scaling exponent of the short time return probability distribution of diffusing matter. 
The first general analytical result for the spectral dimension of two-dimensional quantum
gravity coupled to conformal matter with central charge $D$ was given in \cite{Ambjorn:1997jf}.

\subsection{Extrinsic Hausdorff dimension}
\label{sec:eHd}

The extrinsic Hausdorff dimension $D_H$ of a fractal object embedded
in a $D$-dimensional space
is a measure for the extent of the object measured with the metric
of the embedding space. $D_H$ can be defined for mathematical fractals
as well as for random surfaces and for quantum gravity. Let us focus
on quantum gravity. The partition function for two-dimensional quantum
gravity with fixed volume $V$ of spacetime coupled to $D$ Gaussian matter fields is
\begin{equation}
  \label{eh1}
  Z(V) = \int\!\cD[\gmn]_V \int\!\cD_gX_{\mu}\big\vert_{\text{cm}}
  e^{-\int_M\! d^2\xi\sqrt{g}\ \gAB\partial_{\al}X_{\mu}\partial_{\be}X_{\mu}}.
\end{equation}
Here $\cD_gX_{\mu}\big\vert_{\text{cm}}$ denotes the functional integration over
the $D$ matter fields such that the centre of mass is fixed to zero,
which amounts in dropping the zero modes of the Laplacian when integrating
over the matter fields. This
is a slight variation of \eref{e3a} which allows us to define the extrinsic
Hausdorff dimension as
\begin{equation}
  \label{eh2}
  \bra X^2\ket_V \sim V^{\frac{2}{D_H}}\text{,~for $V\rightarrow\infty$},
\end{equation}
with the definition
\begin{eqnarray}
  \label{eh3a}
  \bra X^2\ket_V &\equiv&
  \frac{1}{Z(V)} \int\!\cD[\gmn]_V\int\!\cD_gX_{\mu}\big\vert_{\text{cm}}
  e^{-\int_M\! d^2\xi\sqrt{g}\ \gAB\partial_{\al}X_{\mu}\partial_{\be}X_{\mu}}
  \nonumber \\
  &&\times \frac{1}{DV}\int_M\!d^2\xi \sqrt{g}\ X_{\mu}^2(\xi),
\end{eqnarray}
which is equivalent to
\begin{eqnarray}
  \label{eh3}
  \bra X^2\ket_V &\equiv&
  \frac{1}{Z(V)} \int\!\cD[\gmn]_V\int\!\cD_gX_{\mu}\big\vert_{\text{cm}}
  e^{-\int_M\! d^2\xi\sqrt{g}\ \gAB\partial_{\al}X_{\mu}\partial_{\be}X_{\mu}}
  \nonumber \\
  &&\times \frac{1}{2DV^2}\int_M\!d^2\xi_1 \sqrt{g} \int_M\!d^2\xi_2 \sqrt{g}\
  (X(\xi_1)-X(\xi_2))^2.
\end{eqnarray}
To understand the definition of $D_H$ note that the fields $X_{\mu}$ define an embedding
of the manifold $M$ in a $D$-dimensional Euclidean space.
That means that we measure the Euclidean mean-square extent of spacetime
embedded in $D$ dimensions.
Mathematically
more conservatively the extrinsic Hausdorff dimension is defined by
constructing a covering of the geometrical object embedded in some $D$-dimensional
space by a union of $n$ small balls. $n$ depends on the typical macroscopic
scale $r$ of the system and scales with the dimension $D_H$ and is related
to the mean square extent of the system as
\begin{equation}
  \label{eh4}
  n\sim r^{D_H}\sim \bra X^2\ket^{\frac{D_H}{2}}.
\end{equation}

The extrinsic Hausdorff dimension for two-dimensional
quantum gravity has been computed in \cite{Distler:1990jv,Kawai:1991qv}. Note
that the treatment in \cite{Distler:1990jv} is based on a more general scaling
assumption than is needed in two-dimensional quantum gravity. Their general result
was specialized to quantum gravity in \cite{Kawai:1991qv}. 
In the following we briefly summarize the derivation.

If we define
the two-point function by
\begin{equation}
  \label{eh5}
  G(p) = \bra \int_M\!d^2\xi_1\sqrt{g}\int_M\!d^2\xi_2\sqrt{g}\
  e^{ip(X(\xi_1)-X(\xi_2))}\ket_V,
\end{equation}
it follows that
\begin{equation}
  \label{eh6}
  \bra X^2\ket_V = -\frac{1}{2DV^2}\frac{\partial^2}{\partial p^2}G(p)\big\vert_{p=0}.
\end{equation}
In flat space the two-point function behaves as $V^{2-\De_0(p)}$ with $\De_0\sim p^2$.
Coupled to gravity this acquires a gravitational dressing which can be computed
to be:
\begin{equation}
  \label{eh7}
  \De(p) = \frac{\sqrt{1-D+24\De_0(p)}-\sqrt{1-D}}{\sqrt{25-D}-\sqrt{1-D}}.
\end{equation}
If we differentiate $G(p)\underset{V\rightarrow\infty}{\sim} V^{2-\De(p)}$ two times with respect to $p$ at $p=0$ we
get
\begin{eqnarray}
  \label{eh8}
  \bra X^2\ket_V&\sim&\log V,\text{~for $D<1$},\\
  \bra X^2\ket_V&\sim&\log^2 V,\text{~for $D=1$},
\end{eqnarray}
for large volumes $V$. In both cases $\bra X^2\ket_V$ grows slower than any power of
$V$. Thus we conclude that $D_H=\infty$.
For spherical topology this has already been found for $D\rightarrow -\infty$ and to one-loop order
in \cite{Jurkiewicz:1984pq}.

\subsection{Spectral dimension}
\label{sec:erg:sd}

To probe the structure of classical spacetime one can study the properties of
propagating particles, typically of free particles or diffusing matter.
Let $\Psi(\xi,T)$ be the wave function for diffusion on the
$d$-dimensional compact manifold $M$. $\Psi$ depends on the time-parameter $T$ and is a function
of the points in $M$ which fulfills the diffusion equation
\begin{equation}
  \label{erg:sd1}
  \frac{\partial}{\partial T} \Psi(\xi,T) = \De_g\Psi(\xi,T)
\end{equation}
for the initial condition
\begin{equation}
  \label{erg:sd2}
  \Psi(\xi,0) = \frac{1}{\sqrt{g(\xi)}}\de(\xi_0-\xi).
\end{equation}
Here $\De_g$ is the Laplace-Beltrami operator corresponding to the metric
$\gmn$ of $M$.
The solution of the diffusion equation can be written as
\begin{equation}
  \label{erg:sd2a}
  \Psi(\xi,T) = e^{T\De_g'}\Psi(\xi,0) =
  \int_M\! d^d \xi'\sqrt{g}\ K_g'(\xi,\xi';T) \Psi(\xi',0),
\end{equation}
which defines the probability distribution (or heat kernel) $K_g'(\xi,\xi';T)$ for diffusion
on a compact manifold $M$ with metric $\gmn$. 
Here and in the following we take into account, that the Laplace operator $\De_g$
on compact surfaces has zero modes which should be projected out. This is indicated
with a prime. 
$K_g'(\xi,\xi';T)$ is related to the massless scalar propagator $(-\De_g)^{-1}$ by
\begin{equation}
  \label{erg:sd2b}
  \bra \xi'\vert(-\De_g)^{-1}\vert\xi\ket' = \int_{0}^{\infty}\! dT\ K_g'(\xi,\xi';T).
\end{equation}
It is known that the average return probability distribution
\begin{equation}
  \label{erg:sd2ba}
  RP_g'(T) \equiv \frac{1}{V_g}\int_M\! d^d\xi\sqrt{g}\ K_g'(\xi,\xi;T)
\end{equation}
at time $T$ for diffusion on a $d$-dimensional manifold
with a smooth geometry $[\gmn]$ admits the following asymptotic expansion
for small times $T$:
\begin{equation}
  \label{erg:sd2c}
  RP_g'(T) \sim \frac{1}{T^{d/2}}\sum_{r=0}^{\infty} a_n T^n
  = \frac{1}{T^{d/2}}(1+O(T)),
\end{equation}
where the coefficients $a_n$ are {\em diffeomorphism invariant} integrals of polynomials
in the metric and the curvature and in their covariant derivatives,
see for example \cite{DeWitt:1979book}. This asymptotic expansion
breaks down when $T$ is of the order $V^{\frac{2}{d}}$, when the exponential decay of
the heat kernel becomes dominant. 

Since $RP_g'(T)$ is reparametrization invariant, we can define its quantum average
over geometries for fixed volume $V$ of spacetime as:
\begin{equation}
  \label{erg:sd5}
  RP_V'(T) \equiv \frac{1}{Z(V)}\int\!\cD[\gmn]_V\
  e^{-S_{\text{eff}}(g)} RP_g'(T).
\end{equation}
Here $S_{\text{eff}}(g)$ denotes the effective action of quantum gravity
after integrating out possible matter fields.


The spectral dimension
$d_s$ is now defined by the short time asymptotic behaviour of $RP_V'(T)$:
\begin{equation}
  \label{erg:sd6}
  RP_V'(T) \sim \frac{1}{T^{d_s/2}}(1+O(T)).
\end{equation}
It is natural to expect that under the average \eref{erg:sd5} over all geometries the only
remaining geometric invariant is the fixed volume $V$. Thus we expect that
$RP_V'(T)$ has the form
\begin{equation}
  \label{erg:sd7}
  RP_V'(T) = \frac{1}{T^{d_s/2}} F\left(\frac{T}{V^{2/d_s}}\right),
\end{equation}
where $F(0)>0$ and $F(x)$ falls off exponentially for $x\rightarrow\infty$.
This scaling assumption is the main input into our derivation of the spectral dimension
of two-dimensional quantum gravity. It is very well documented in numerical
simulations \cite{Ambjorn1998ab,Ambjorn:1995rg}.

For a fixed smooth geometry the spectral dimension is by definition equal to
the dimension $d$ of the manifold. However, the quantum average can a priori
change this behaviour. Actually we know that the intrinsic Hausdorff dimension of
two-dimensional quantum gravity is different from two. In this sense generic geometries
are fractal with probability one. For diffusion on fixed (often embedded) fractal structures it is
well known, that the spectral dimension can be different from the embedding dimension as
well as from the fractal dimension.
The diffusion law, measured in the embedding space, becomes anomalous:
\begin{equation}
  \label{erg:sd7a}
  \bra r^2(T)\ket \sim T^{2/d_w},
\end{equation}
with a gap exponent $d_w>2$. This slowing down of the transport is caused by the fractal ramification
of the system. The gap exponent $d_w$ is related to the spectral dimension $d_s$ and the
intrinsic Hausdorff dimension $d_h$ by
\begin{equation}
  \label{erg:sd7b}
  d_s = \frac{2d_h}{d_w}.
\end{equation}
If the diffusion law is not anomalous, we have $d_w=2$ and $d_s=d_h$, analogous to
$d_s=d$ for diffusion on fixed smooth $d$-dimensional geometries. For a review
of diffusion on fractals see \cite{Havlin:1987}.

While the above concepts are in principal valid for Euclidean quantum gravity
in any dimension $d$, let us now specialize to two dimensions.
The starting point is again the
partition function \eref{eh1} for two-dimensional quantum gravity coupled to $D$
Gaussian matter fields $X_{\mu}$ for fixed volume $V$ of spacetime. We use \eref{eh3a} to
define $\bra X^2\ket_V$. The Gaussian action implies that:
\begin{eqnarray}
  \label{erg:sd8}
  \bra X^2\ket_V &=& \frac{1}{DV} \frac{\partial}{\partial\om}
  \bra e^{\om\int_M\!d^2\xi\sqrt{g}\ X_{\mu}^2(\xi)}\ket_V\Big\vert_{\om=0}\nonumber\\
  &=& \frac{1}{DVZ(V)}\frac{\partial}{\partial\om}
  \int\!\cD[\gmn]_V\ \big({\det}'(-\De_g-\om)\big)^{-D/2}\Big\vert_{\om=0}\nonumber\\
  &=& \frac{1}{2VZ(V)}\int\!\cD[\gmn]_V\ \big({\det}'(-\De_g)\big)^{-D/2}
  \tr'\left[\frac{1}{-\De_g}\right]\nonumber\\
  &=& \frac{1}{2V}\bra \tr'\left[\frac{1}{-\De_g}\right]\ket_V.
\end{eqnarray}
By inserting \eref{erg:sd2b} and \eref{erg:sd7} into this formula we get:
\begin{eqnarray}
  \label{erg:sd9}
  \bra X^2\ket_V &=&
  \frac{1}{2V} \bra\int_0^{\infty}\! dT\ \int_M\! d^2\xi\sqrt{g}\ K_g'(\xi,\xi;T)\ket_V\nonumber\\
  &=& \frac{1}{2}\int_{0}^{\infty}\! dT\ RP_V'(T) =
  \frac{1}{2}\int_{0}^{\infty}\! dT\ \frac{1}{T^{d_s/2}} F(\frac{T}{V^{2/d_s}})\nonumber\\
  &\sim& V^{\frac{2}{d_s}-1},
\end{eqnarray}
for $V\rightarrow\infty$. Comparing this with \eref{eh2} we conclude:
\begin{equation}
  \label{erg:sd10}
  \frac{1}{d_s} = \frac{1}{D_H} + \frac{1}{2}.
\end{equation}
Using the result $D_H=\infty$ from the preceeding section we arrive at:
\begin{equation}
  \label{erg:sd11}
  d_s = 2,\text{~for all $D\leq 1$.}
\end{equation}
Strictly speaking, this assumes $d_s\leq 2$. However, 
at $D=-\infty$ a fixed geometry implies $d_s=2$ and $D_H=\infty$, in agreement with \eref{erg:sd10},
and we expect that a saddle point calculation around $D=-\infty$ is reliable, compare
\cite{Jurkiewicz:1984pq}. Therefore $d_s$ equals $2$ in a neighbourhood of $D=-\infty$.
The average over all geometries in \eref{erg:sd5} includes many degenerate geometries, for
which $D_H<\infty$, for example the branched polymer-like geometries which are discussed
in section \sref{bpo}. Thus, a priori we would expect that under the average over fluctuating geometries
the spectral dimension decreases. Therefore it is
reasonable to assume that $d_s\leq 2$.
\footnote{If $d_s>2$, the integral in \eref{erg:sd9} diverges at $0$, and we have to
  introduce a cut-off $\eps$ for small times $T$. Then it is convenient to
  consider $\bra (X^2)^n\ket_V\sim V^{2n/D_H}$ instead, with $n=[d_s/2]+1$
  for non-integer $d_s$. Then the leading large $V$ behaviour will be $V^{2n/d_s-1}$,
  and we get $\frac{1}{d_s} = \frac{1}{D_H} + \frac{1}{2n}$.}

Thus we have shown, that the spectral dimension of two-dimensional quantum gravity
coupled to $D$ Gaussian fields equals 
two for $D\leq 1$ \cite{Ambjorn:1997jf}. Our derivation relies on
the scaling form \eref{erg:sd7} of the averaged return probability $RP_V'(T)$ and on properties
of Gaussian matter fields. To corroborate the result \eref{erg:sd11}, let us present another derivation
completely within Liouville theory\footnote{Thanks to Jakob L.~Nielsen for showing me this argument.}.
Let us define an observable
\begin{equation}
  \label{erg:sd12}
  \cO(g) = \int_M\! d^2\xi_0\sqrt{g}\ \frac{1}{-\De_g}
  \frac{1}{\sqrt{g(\xi)}}\de(\xi_0-\xi)\Big\vert_{\xi=\xi_0}.
\end{equation}
Under a rescaling of the metric, $\cO$ behaves like $\cO(\la g) = \la\cO(g)$, that
means it is a correlator with conformal weight $(-1,-1)$. Therefore we deduce
from \eref{e35} that
\begin{equation}
  \label{erg:sd13}
  \bra\cO(g)\ket_{\la V} = \la \bra\cO(g)\ket_V,
\end{equation}
or equivalently:
\begin{equation}
  \label{erg:sd14}
  \bra\frac{1}{V}\cO(g)\ket_{\la V} = \bra\cO(g)\ket_V.
\end{equation}
On the other hand we can write:
\begin{eqnarray}
  \label{erg:sd15}
  \bra\frac{1}{V}\cO(g)\ket_{V} &=&
  \bra \frac{1}{V}\int_M\!d^2\xi_0\sqrt{g(\xi_0)}\
  \frac{1}{-\De_g} \frac{1}{\sqrt{g(\xi)}}\de(\xi_0-\xi)\Big\vert_{\xi=\xi_0} \ket_V\nonumber\\
  &=& \bra \frac{1}{V}\int_M\!d^2\xi_0\sqrt{g(\xi_0)}\
  \int_0^{\infty}\! dT\ e^{T\De_g}\frac{1}{\sqrt{g(\xi)}}\de(\xi_0-\xi)\Big\vert_{\xi=\xi_0} \ket_V \nonumber\\
  &=&\int_0^{\infty}\! dT\ RP_V'(T)\sim V^{\frac{2}{d_s}-1},
\end{eqnarray}
using the scaling assumption \eref{erg:sd7}.
From \eref{erg:sd14} and \eref{erg:sd15} it follows that
$d_s=2$ in Liouville theory for all types of matter coupled to gravity
with central charge $D\leq 1$, at which point the continuum calculations break down.

This situation is very remarkable: A generic geometry in the functional integral \eref{eh1}
is a typical fractal, when looked at in the usual way by computing the fractal dimension
$d_h$ from measurements of the circumference $\bra l(g,R)\ket_V$ of circles versus the
geodesic distance $R$. Also the gap exponent $d_w$ is anomalous and larger than two.
It is exactly equal to the intrinsic Hausdorff dimension and consequently, the spectral dimension
$d_s$ equals two. It takes the same value as for smooth two-dimensional geometries.

The spectral dimension has received considerable attention also in numerical studies
of the fractal structure of two-dimensional quantum gravity. In \cite{Ambjorn:1995rg}
the return probability is computed for pure gravity and for gravity coupled to matter
with central charge $\frac{1}{2}$ and $1$. They find that their results are consistent with
$d_s=2$. In \cite{Ambjorn1998ab} $d_s$ is measured for central charge $-2$, $0$,
$\frac{1}{2}$ and $\frac{4}{5}$.
They get the results $d_s=2.00(3)$, $d_s=1.991(6)$, $d_s=1.989(5)$ and $d_s=1.991(5)$,
respectively\footnote{Thanks to Konstantinos N.~Anagnostopoulos for showing me these data
  prior to the publication of \cite{Ambjorn1998ab}}.

For $D>1$ it is generally believed that two-dimensional quantum gravity is in a branched polymer
phase, see \cite{Harris:1996hk,David:1996vp} for recent analytical and \cite{Thorleifsson:1997ac}
for some numerical evidence. The Gaussian fields define an embedding of these polymers
in $\mathbb{R}^D$. The extrinsic Hausdorff dimension of generic branched polymers is
$D_H=4$, thus we conclude from \eref{erg:sd10} that the spectral dimension equals $\frac{4}{3}$,
the famous Alexander-Orbach value \cite{Alexander:1982}. The value $d_s=\frac{4}{3}$ and
formula \eref{erg:sd10} for branched polymers have been derived in \cite{Cates:1984} by different methods,
see also \cite{Jonsson:1997gk} for a recent complete proof of $d_s=\frac{4}{3}$. Furthermore,
for the multicritical branched polymers \cite{Ambjorn:1990wp} it is known that
$D_H=\frac{2m}{m-1}, m=2,3,\ldots$, where $m=2$ for the ordinary branched polymers.
Thus we obtain $d_s = \frac{2m}{2m-1}$, in agreement with the analysis in \cite{Correia:1997gf}.
For $m\rightarrow\infty$ the multicritical branched polymers approach ordinary random walks,
for which $D_H=2$ and $d_s=1$.

Let us in the end comment on a subtlety in the definition \eref{erg:sd5} of the averaged return
probability $RP_V'(T)$. We have defined it as the quantum average of the return
probability for fixed geometry over fluctuating geometries. Instead, we could have defined
it as the limit for $R\rightarrow 0$ of
\begin{eqnarray}
  \label{erg:sd100}
  K_V'(R,T) &=& \frac{1}{G(V,R)}\int\!\cD[\gmn]_V\
  e^{-S_{\text{eff}}(g)}
  \\
 &&\times \int_M\! d^2\xi_1\sqrt{g} \int_M\! d^2\xi_2\sqrt{g}\ \de(R-d_g(\xi_1,\xi_2))K_g'(\xi_1,\xi_2;t),\nonumber
\end{eqnarray}
where $G(V,R)=V\bra l(g,R)\ket_VZ(V)$ is the partition function for universes with two marked
points separated by a geodesic distance $R$ (``two-point function'')
as defined (up to matter fields) in \eref{e16},
see also \eref{e17b}. $K_V'(R,T)$ is the average probability distribution for diffusing
a geodesic distance $R$ in the time $T$. It is natural to identify $RP_V'(T)$ with
$K_V'(0,T)$. However it is unknown whether the limit $R\rightarrow 0$ of \eref{erg:sd100}
commutes with the functional integration over geometries. Actually, for the
two-point function $G(V,R)$ it is easy to see that this is not the case.
If the limits do not commute there are two inequivalent definitions of the
return probability $RP_V'(T)$. However, the scaling form of $K_V'(0,T)$ is
the same as the scaling form \eref{erg:sd7} of $RP_V'(T)$ \cite{Watabiki:1996ja}.
Indeed, we have
\begin{equation}
  \label{erg:sd101}
  \int_0^{\infty}\! dR\bra l(g,R)\ket_V\ K_V'(R,T) = 1,
\end{equation}
and using $\text{dim}[R]=\text{dim}[V^{\nu}]$ we arrive at the scaling form
\begin{equation}
  \label{erg:sd102}
  K_V'(R,T) = \frac{1}{V}P\left(\frac{R}{V^{\nu}},\frac{T}{V^{\la}}\right),
\end{equation}
where $\la$ is chosen such that $\frac{T}{V^{\la}}$ is dimensionless.
Expanding the return probability $K_V'(0,T)$ around $T=0$ we get
\begin{equation}
  \label{erg:sd103}
  K_V'(0,T) \sim \frac{V^{\frac{\la d_s}{2}-1}}{T^{d_s/2}},\text{~for $T\sim 0$,}
\end{equation}
by definition of $d_s$. If this short time asymptotic stays nonzero and finite
for infinite volume $V$ we get $\la=\frac{2}{d_s}$, and $K_V'(0,T)$ scales as
\begin{equation}
  \label{erg:sd104}
  K_V'(0,T) = \frac{1}{T^{d_s/2}} \tilde{F}\left(\frac{T}{V^{2/d_s}}\right),
\end{equation}
which has the same form as the scaling behaviour of $RP_V'(T)$. 

\subsection{Intrinsic Hausdorff dimension}
\label{sec:erg:ihd}

The intrinsic Hausdorff dimension $d_h$ of two-dimensional quantum gravity
has been defined in section \sref{obs}. One way is to measure the average
circumference $\bra l(g,R)\ket_V$ of circles with a small radius $R$, which scales
as $\bra l(g,R)\ket_V\sim R^{d_h-1}$. This is analogous to the classical ``clock and rod'' procedure
to study the geometrical properties of spacetime. Alternatively one can define
the fractal dimension $d_H$ as the scaling exponent of the average volume
of the universes: $\bra V\ket_R \sim R^{d_H}$, compare \eref{e16f}.
In two-dimensional quantum gravity the first ``local'' and the second ``global'' concept
of the intrinsic Hausdorff dimension agree. For pure gravity, we have $d_H=4$, see \eref{e17g}.

The intrinsic Hausdorff dimension has been computed in \cite{Watabiki:1993fk} by studying the
diffusion equation in Liouville theory. Although the mathematical status of this derivation is
unclear and although it is not consistent with the derivation of the spectral dimension in
section~\sref{erg:sd}, the result is valid for a number of special cases and agrees with
many numerical simulations.

In~\cite{Watabiki:1993fk} the mean square distance
$\bra R^2(T)\ket_V$ of diffusion in the fluctuating space time is defined as:
\begin{eqnarray}
  \label{erg:id1}
  \bra R^2(T)\ket_V &\equiv&
  \frac{1}{VZ(V)} \int\!\cD[\gmn]_V\ e^{-S_{\text{eff}}(g)}
  \int_M\! d^2\xi_1\sqrt{g}\int_M\! d^2\xi_2\sqrt{g}\ 
  \nonumber\\ && \times
  d_g^2(\xi_1,\xi_2)K_g'(\xi_1,\xi_2;T).
\end{eqnarray}
$\bra R^2(T)\ket_V$ scales with the volume as 
\begin{equation}
  \label{erg:id1suppl}
  \bra R^2(T)\ket_V \sim V^{\frac{2}{d_h}},
\end{equation}
which yields yet another way to define the intrinsic Hausdorff dimension $d_h$.
On the other hand, the mean 
square distance for diffusion on a fixed smooth geometry $[\gmn]$
with the initial condition \eref{erg:sd2} can be expanded for small times $T$ as \cite{DeWitt:1965}
\begin{equation}
  \label{erg:id4}
  \int_M\! d^2\xi\sqrt{g}\ d_g^2(\xi_0,\xi) K_g'(\xi_0,\xi;T) = 4T - \frac{2}{3}T^2\cR(\xi_0) + O(T^3),
\end{equation}
with the curvature scalar $\cR$. If we assume that this expansion commutes with the
functional average over geometries in \eref{erg:id1}, we get
\begin{equation}
  \label{erg32}
  \bra R^2(T)\ket_V \sim \text{const}\ \cdot T, \text{~for small times $T$.}
\end{equation}
The dimension of $T$ can be computed from Liouville theory by expanding
the (unnormalized) return probability for small times to observe how
this scales under the transformation $V\rightarrow\la V$:
\begin{eqnarray}
  \label{erg:id5}
  \lefteqn{\!\!\!\!\!\!\!\!\!\!\!\!\bra\int_M\!d^2\xi_0\sqrt{g}\ K_g'(\xi_0,\xi_0;T)\ket_{\la V}
   =\ \bra\int_M\!d^2\xi_0\sqrt{g}\ e^{T\De_g}\Psi(\xi,0)\Big\vert_{\xi=\xi_0}\ket_{\la V} }
  \nonumber \\
  &=&  \bra\int_M\!d^2\xi_0\sqrt{g}\ \left[
      \Psi(\xi_0,0) + T\De_g\Psi(\xi,0)\Big\vert_{\xi=\xi_0} + O(T^2)\right]\ket_{\la V}\nonumber\\
  &=&  1 + T\bra\int_M\! d^2\xi_0\sqrt{g}\ \De_g \frac{1}{\sqrt{g(\xi)}}\de(\xi-\xi_0)
  \Big\vert_{\xi=\xi_0}\ket_{\la V} + O(T^2)\nonumber\\
  &\overset{\eref{e35}}{=}& 1 + T\ \la^{\frac{\be}{\al}} + O(T^2),
\end{eqnarray}
where $\be$ is computed for $\De_0=2$.
This calculation is based on the assumption that the expansion of the exponential commutes with the
functional integration over geometries. 
Then we have:
\begin{equation}
  \label{erg:id7}
  \text{dim}[T] = \text{dim}[V^{2/d_H}] = \text{dim}[V^{-\be/\al}], 
\end{equation}
and thus
\begin{equation}
  \label{erg:id8}
  d_H = -\frac{2\al}{\be} = 2\ \frac{\sqrt{25-D}+\sqrt{49-D}}{\sqrt{25-D}+\sqrt{1-D}},
\end{equation}
for matter with central charge $D$ coupled to two-dimensional gravity. For $D=0$,
that means in the case of pure gravity, \eref{erg:id8} predicts $d_H=4$, in agreement with
\eref{e17g}. In the semiclassical limit $D\rightarrow -\infty$ we expect $d_H=2$,
which is also correctly predicted by \eref{erg:id8}.

The validity of \eref{erg:id8} has been widely discussed.
The main assumption of the derivation sketched above is that
one is allowed to interchange the expansion of $e^{T\De_g}$ with the functional integration
in \eref{erg:id5}. Further input is the scaling assumption for $\bra R^2(T)\ket_V$.
In principle the same result for the intrinsic Hausdorff dimension should be obtained
if one uses some power of the Laplace operator. However this is not true and results obtained
from higher orders of the expansion of $e^{T\De_g}$ contradict the result from the first order. 

An alternative theoretical prediction based on the transfer matrix method has been given
in \cite{Distler:1990jv,Ishibashi:1994sv,Ishibashi:1995in}. They get:
\begin{equation}
  \label{erg:id9}
  d_H = -\frac{2}{\ga} = \frac{24}{1-D+\sqrt{(25-D)(1-D)}}. 
\end{equation}
For $D=0$ this yields again $d_H=4$, while for $D\rightarrow-\infty$ \eref{erg:id9}
predicts $d_H=0$.

To settle this question, the authors of \cite{Ambjorn:1997kb,Ambjorn:1997sy}
performed a high precision Monte Carlo analysis for $D=-2$. Their result is
$d_H=3.574(8)$, in agreement with the value $d_H=3.562$ from \eref{erg:id8} and in
clear disagreement to $d_H=2$ from \eref{erg:id9}.
This clearly rules out the validity of \eref{erg:id9} for matter with $D<0$ coupled
to gravity. 

However, for unitary matter with $D>0$ coupled to gravity the situation is less
clear. Most simulations report values $d_H\approx 4$ \cite{Bowick:1995,Ambjorn:1995rg,Ambjorn:1997wc},
and a back-reaction of the matter on the structure of the quantum spacetime
cannot be observed. 
However, values extracted from the scaling of the matter correlation functions are higher,
though less reliable. All data are consistent with the conjecture \cite{Bowick:1995,Ambjorn:1995rg}:
\begin{equation}
  \label{erg:id10}
  d_H=4,\text{~for $0\leq D\leq 1$,}
\end{equation}
although the validity of \eref{erg:id8} for $D>0$ is not completely ruled out.
\begin{table}[tb]
\renewcommand{\baselinestretch}{1.2}
  \normalsize
  \begin{center}
    \begin{tabular}{lllllll}
      \hline\hline\hline
      $D=-5$  &  $D=-2$  &  $D=0$  &  $D=\frac{1}{2}$  &  $D=\frac{4}{5}$  &  $D=1$  &  ref.\\ \hline
      3.236  &  3.562  &  4  &  4.212  &  4.421  &  4.828  &  \eref{erg:id8}\\
      1.236  &  2  &  4  &  6  &  10  &  $\infty$  &  \eref{erg:id9}\\ \hline
      3.36(4)  &  3.574(8)  &  4.05(15)  &  4.11(10)  &  4.01(9)  &  3.8--4.0  &  \cite{Anagnostopoulos:1998ab,Ambjorn:1997nf}\\ \hline      
        &    &    &  3.96--4.38  &  3.97--4.39  &    &  \cite{Ambjorn:1997nf}\\
      \hline\hline\hline
    \end{tabular}
    \renewcommand{\baselinestretch}{1.0}
    \normalsize
    \parbox[t]{\textwidth}
    {
      \caption[bv1]
      {\label{tab:erg:id1}
        \small
        Current status of numerical values for the intrinsic Hausdorff dimension
        $d_H$ compared to the two different theoretical predictions \eref{erg:id8}
        and \eref{erg:id9}. The values in the first part of the table are the theoretical
        predictions. In the second part of the table we summarize results from the scaling
        of the two-point function for the random surfaces, while in the last part the
        Hausdorff dimension is determined from the scaling of the matter correlation
        functions. See the references for technical details about the
        simulations and the data analysis.
        }
      }
  \end{center}
\end{table}
The current status of the numerical simulations is summarized in table \tref{erg:id1}.


\chapter{Discretization of $2d$ quantum gravity}
\label{sec:disc}

The formulation of quantum field theory in terms of renormalized Euclidean
functional integrals leads to an identification of quantum field theory
with statistical mechanics. A key ingredient for this identification is the discretization
of spacetime. Then the powerful techniques from the theory of critical phenomena
get available and have proved to be invaluable for the understanding
of renormalization and of non-perturbative phenomena.

Therefore it is natural to attempt a discretization of the continuum theory of
quantum gravity. However, spacetime no longer plays the role of a mere
background but is the dynamical variable itself. This, combined with the
diffeomorphism invariance of the continuum theory poses a problem for the discretization
which has successfully been solved by the method of dynamical
triangulations~\cite{Ambjorn:1985az,David:1985tx,Kazakov:1985ea}.
The main idea is to discretize geometry directly, with no reference
to coordinate parametrizations. That means that the functional
integral over equivalence classes of metrics on a manifold $M$ is
replaced by a finite sum over piecewise linear spaces which are constructed
by successively gluing $d$-simplices together. The fixed edge length of the
simplices introduces a reparametrization invariant cutoff into the theory.
In this way quantum gravity can be formulated as an ensemble of
discrete random manifolds and the machinery of critical phenomena
can be applied. The continuum theory is recovered at a critical
point of this ensemble. Masses and continuum coupling constants are
defined by the approach to the critical point in the scaling limit. 

Research in the formulation of classical gravity in purely geometrical
terms was initiated in~1961 by T.~Regge~\cite{Regge:1961}, while the first
attempt at a discretization of quantum gravity in the sense of this chapter
has been done by Weingarten in~1980~\cite{Weingarten:1980,Weingarten:1982}.
A good review with some comments about the historical development of the
subject is~\cite{Ambjorn1997}. Other reviews can be found in~\cite{David:1992,Ambjorn:1994,Ambjorn:1995aw}.
Much of the material presented in this chapter is taken from these articles.
In~\cite{DiFrancesco:1995nw} the relation between the matrix-model
technique and dynamical triangulation is reviewed.

In the first section of this chapter we introduce the method of dynamical triangulations
and discuss the scaling properties of the theory. Branched polymers, which have been
mentioned before, are discussed in some detail. In section~\sref{mm} we give a brief review
of the matrix-model techniques which have been applied to solve two-dimensional quantum gravity.
For more details and further issues we refer to the review articles given above. The explicit solution
of discretized two-dimensional quantum gravity shows, that the continuum theory is
the scaling limit of dynamical triangulation. Therefore both theories can be identified.

In section~\sref{fs} we outline how the fractal structure and the scaling properties
of pure two-dimensional quantum gravity can be obtained.

An alternative method for the discretization of quantum gravity,
known as quantum Regge calculus, has been suggested. This is discussed in chapter~\sref{regge}
of this thesis. 

\section{Dynamical triangulation}

Dynamical triangulation is a discretization of quantum geometries. No reference to
parametrizations has to be made. At the same time it provides a regularization
of the continuum theory by introducing an explicit cutoff. In this section we can only
outline the beauty and power of dynamical triangulations in two dimensions.
We begin with a short introduction into discretization of continuum manifolds and geometries.
Then we define the partition function of two-dimensional dynamical triangulation and discuss
its scaling properties. The susceptibility exponent $\ga$ is characterized through the fractal
structure of the random spacetimes. We end this section with a discussion of the scaling
properties of the two-point function which are illustrated by means of
the model of branched polymers.

\subsection{Discretization of geometry}
\label{sec:dg}

Let us define an oriented $n$-simplex as an $(n+1)$-tuple
of points modulo even permutations. Thus a $0$-simplex is a
point, a $1$-simplex is a pair of points which can be identified with a
line, a $2$-simplex is a triangle, a $3$-simplex a tetrahedron etc.
A subsimplex $\sig'$ of a simplex $\sig$ is defined as a subset of
$\sig$: $\sig'\subseteq\sig$. By gluing simplices together one gets a simplicial complex.
Simplicial complexes have
no fixed dimension, that means they are not manifolds.
However, their structure is sufficient to define exterior calculus
on them, which physically means that scalar fields, gauge fields, antisymmetric
tensor fields etc can be defined on them.

A $d$-dimensional simplicial manifold is a simplicial complex such
that the neighbourhood of each point in the complex is homeomorphic to
a $d$-dimensional ball. In one and two dimensions each simplicial
manifold can be obtained by gluing pairs of $d$-simplices along some
of their $(d-1)$-edges. In higher dimensions this successive gluing
yields in some cases only pseudo-manifolds, in which a neighbourhood
of a point can also be homeomorphic to a topologically more complicated object.

The structure of simplicial manifolds corresponds to a discretization
of the continuous manifold structure. To define geometrical concepts we
have to introduce a metric. A $d$-dimensional simplicial manifold can be equipped canonically
with a Riemannian metric by demanding that the metric is flat inside
each $d$-simplex, continuous when the $(d-1)$-subsimplices are crossed 
and that each $(d-1)$-subsimplex is a linear flat subspace of this simplex.
A simplicial manifold equipped with such a metric is called piecewise
linear space.

The canonical metric can be defined by assigning lengths to the $N_l$ links
($1$-subsimplices) of the simplicial manifold. 
First note that any $d$-dimensional simplicial manifold $M$ can be
covered with charts $(U,\phi)$ such that each $d$-simplex is 
parametrized by barycentric coordinates. Here
\begin{equation}
  \label{r0}
  U=\lbrace\xi\in \mathbb{R}^d_+\vert\xi_1+\cdots +\xi_d<1\rbrace,
\end{equation}
and $\phi: U\rightarrow M$ is given by
\begin{equation}
  \label{r0a}
  \phi(\xi) = \xi_1y_1 + \ldots + \xi_dy_d + (1-\xi_1-\ldots-\xi_d)y_{d+1},
\end{equation}
where $y_1, \ldots, y_{d+1}$ are the coordinates of the vertices of the simplicial complex
which live in some ambient space $\mathbb{R}^n$.
Then on each chart
the canonical metric is defined by
\begin{equation}
  \label{r0b}
  \gmn(\xi) = \frac{\partial\phi}{\partial\xi^{\mu}}\cdot
              \frac{\partial\phi}{\partial\xi^{\nu}}.
\end{equation}
This metric is Euclidean inside each $d$-simplex and continuous if a $(d-1)$-face
is crossed. It is compatible with the manifold structure and can be expressed solely
in terms of the link lengths $l_1, \ldots, l_{N_l}$.
For a triangle~\eref{r0b} can be written as 
\begin{equation}
  \label{r0cc}
  \gmn = \left(
  \begin{array}[c]{cc}
    l_1^2 & \frac{1}{2}(l_1^2+l_2^2-l_3^2) \\
    \frac{1}{2}(l_1^2+l_2^2-l_3^2) & l_2^2
  \end{array}
  \right).
\end{equation}

The intrinsic curvature of a two-dimensional piecewise linear space is located at the
vertices, that means its $0$-subsimplices. The curvature can clearly not reside
in the triangles, since they are flat. Furthermore the metric is continuous
when crossing the edges of the triangles.
Since we define curvature as an
intrinsic concept of the piecewise linear space it should be independent of
the embedding. We can bend the surface around an edge without
changing the geometry -- an observation which the German mathematician
C.~F.~Gauss made in his famous paper {\em Disquisitiones generales circa superficies curva} (1828)
and which he called {\em theorema egregrium}.
Therefore it is intuitive that the curvature cannot reside
in the edges. Rather, it is concentrated at the vertices or in general at the $(d-2)$-hinges of the
piecewise linear space. To each vertex $v$ in the surface
we can assign a deficit angle $\de_v$ as the difference between $2\pi$ and the
sum of the angles meeting at $v$:
\begin{equation}
  \label{di1}
  \de_v = 2\pi - \sum_{t:v\in t}\al_{v,t},
\end{equation}
where the sum goes over all triangles $t$ containing $v$. $\al_{v,t}$ is the angle
at $v$ inside $t$. The scalar curvature $\cR_v$ at $v$ is defined as
\begin{equation}
  \label{meas2.5}
  \cR_v = 2\frac{\de_v}{dA_v},~dA_v = \frac{1}{3}\sum_{t:\, v\in t} A_t,
\end{equation}
where $A_t$ is the area of the triangle $t$.
The scalar curvature $\cR$ in two dimensions equals two times the Gaussian curvature.
With these definitions we finally have:
\begin{equation}
  \label{di5}
  \text{total area of the surface}=\sum_{v=1}^{N_v}dA_v,
\end{equation}
and the Gauss-Bonnet theorem is now valid in the form
\begin{equation}
  \label{di4}
  \sum_{v=1}^{N_v} \cR_vdA_v = 4\pi\chi,
\end{equation}
which is equivalent to the polyeder formula $N_t-N_l+N_v=\chi$ by Descartes and Euler.

\subsection{Definition of the model}

The original proposal of Regge~\cite{Regge:1961} consists in constructing a
sequence of piecewise linear spaces to approximate a given smooth manifold $M$ such that
for any sufficiently smooth function $f$
\begin{equation}
  \label{di2}
  \sum_{v}dA_v f_v\rightarrow \int_M\! dA(\xi)\ f(\xi), 
\end{equation}
where $f_v$ is the value of the function $f$ at the vertex $v$. The applicability of
this point of view in the quantum case will be discussed in chapter~\sref{regge}.
Here we take a somewhat different point of view. Our aim is to integrate over all
equivalence classes of metrics on a given manifold $M$. We introduce a reparametrization
invariant cutoff by assigning the length $a$ to each edge of piecewise linear spaces.
We construct all possible piecewise linear spaces by gluing equilateral triangles together.
In this case the formulas above simplify considerably. 
Let $n_v$ be the order of the vertex $v$. Then we have:
\begin{equation}
  \label{di6}
  dA_v = \frac{\sqrt{3}}{12}a^2n_v,\quad \cR_vdA_v = \frac{2\pi}{3}(6-n_v).
\end{equation}
For simplicity we rescale $a^2$ by $\frac{\sqrt{3}}{4}$ and then set $a=1$.
\eref{di6} shows that two triangulations which cannot be mapped onto each other
by a relabelling of the vertices lead to different local curvature assignments.
Thus they define different metric structures. The set of combinatorially
non-equivalent triangulations defines a grid in the space of diffeomorphism classes
of metrics. The proposal described in this chapter relies on the hope that this grid
gets uniformly dense in the limit $a\rightarrow 0$.

To define a regularized theory of quantum gravity  we replace
the action $S({g}, \La) = \La V_g$ in~\eref{e5a} by
\begin{equation}
  \label{di7}
  S_T(\mu) = \mu N_t,
\end{equation}
where $\mu$ is the bare cosmological constant. The integration over equivalence classes
of metrics on $M$ is replaced by a summation over all non-isomorphic equilateral
triangulations $T$ with the topology of $M$.
In the case of matter coupled to quantum gravity, the matter fields $X$ can be defined
on the vertices, the links or on the triangles. In general the action $S_{\text{matter}}$
and the functional integration over the matter fields 
will depend on the triangulation $T$.
The discretized version of the partition function can thus be written as
\begin{equation}
  \label{di8}
  Z(\mu) = \sum_{T\in \cT} \frac{1}{C_T} e^{-S_T(\mu)}\int\! \cD_T[X]\ e^{-S_{\text{matter}}(T,X)},
\end{equation}
where $C_T$ denotes the symmetry factor of the triangulation $T$ which is equal to the
order of the automorphism group of $T$. The summation goes over a suitable class
$\cT$ of abstract triangulations $T$ defined by their vertices and a connectivity matrix.
Appealing to universality details of the chosen class
of triangulations such as whether closed two-loops are allowed or not should not be important
for the theory. This has been verified a posteriori.

For simplicity we will ignore any possible matter fields coupled to gravity in the following
equations.

Then we can write:
\begin{equation}
  \label{di9}
  Z(\mu) = \sum_{N=1}^{\infty} e^{-\mu N} Z(N),\quad Z(N) = \sum_{T\in \cT_N} \frac{1}{C_T},
\end{equation}
where $\cT_N$ denotes the subset of triangulations with $N$ triangles of $\cT$.
Expectation values of observables $\cO$ are defined as:
\begin{eqnarray}
  \label{di9a}
  \bra\cO\ket_{\mu} &=& \frac{1}{Z(\mu)} \sum_{T\in \cT} \frac{1}{C_T} e^{-S_T(\mu)} \cO(T),\\
  \bra\cO\ket_{N} &=& \frac{1}{Z(N)}\sum_{T\in \cT_N} \frac{1}{C_T} \cO(T),
\end{eqnarray}
in the ``grand-canonical'' ensemble with fixed cosmological constant and in the
``canonical'' ensemble with fixed volume respectively.
$\mu$ can be understood as the chemical potential for adding new triangles to the surfaces.

$Z(N)$ can be interpreted as the number of triangulations in $\cT_N$. This number is exponentially
bounded:
\begin{equation}
  \label{di10}
  Z(N) = e^{\mu_c N}N^{\ga-3}\left(1+O(1/N)\right).
\end{equation}
The proof for spherical topology has been given in~\cite{Tutte:1962}.
For a general proof by purely combinatorial methods in the spirit of Tutte see~\cite{Ambjorn1997}.
It can be proved that the critical point $\mu_c$ does not depend on the topology.

That means that the statistical ensemble defined by~\eref{di8} has a critical point $\mu_c$.
$Z(\mu)$ is analytical for $\mu>\mu_c$ and contains non-analytical parts at the critical point.
The latter have our biggest interest since they are the universal parts given by the large $N$
behaviour of $Z(N)$. The continuum limit should be defined as $\mu\rightarrow\mu_c$.
Close to the critical point we have:
\begin{equation}
  \label{di11}
  Z(\mu) = (\mu-\mu_c)^{2-\ga} + \text{less singular terms,}
\end{equation}
from a discrete Laplace transformation of~\eref{di10}.
Many analytical and numerical investigations have revealed that $\ga$ assumes the same values
as in the continuum theory of two-dimensional quantum gravity and is given by~\eref{e3d}.
Thus the susceptibility exponent $\ga$ can also be characterized as the subleading power
correction of the exponentially growing number of triangulations for a fixed topology.
Furthermore, $\ga$ can be characterized by the branching ratio into minimal bottleneck
baby universes~\cite{Jain:1992bs}, that means parts of the universe which are connected to the rest by
a three-loop of links which do not form a triangle, see figure \fref{di1}.
\begin{figure}[htbp]
  \begin{center}
    \includegraphics[width=0.6\linewidth]{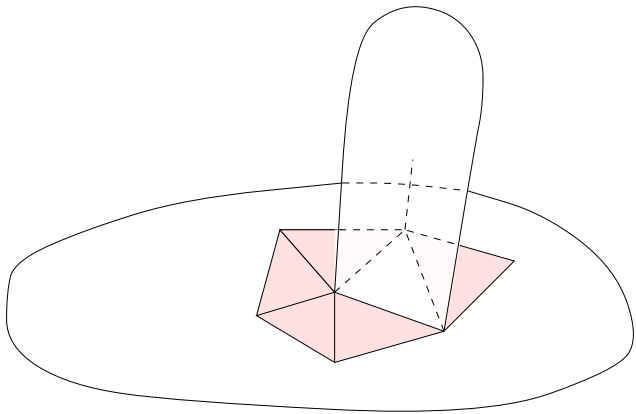}
    \parbox[t]{.85\textwidth}
    {
      \caption[babyfig]
      {
        \label{fig:di1}
        \small
        Branching of a minimal bottleneck baby universe from the larger parent.
        }
      }    
  \end{center}
\end{figure}
The smaller part is then called
minimal bottleneck baby universe and the larger part is called parent.
In the ensemble of universes with volume $N$ the average number $\bra\cN(V)\ket_N$ of baby universes
with volume $V$ is given by
\begin{equation}
  \label{di12}
  \bra\cN(V)\ket_N \sim \frac{3!}{Z(N)} V Z(V)\ (N-V)Z(N-V).
\end{equation}
The factors $Z(V)$ and $Z(N-V)$ are the weights of the baby universe and of the parent respectively.
The volume factors $V$ and $N-V$ reflect the fact that the baby universe can be attached at any
triangle. Finally, $3!$ is the number of ways the two boundaries of the parent and of the
minimal bottleneck baby universe can be glued together.~\eref{di12} is valid 
in the generic case where there are no additional symmetries.
Assuming that $Z(N)$ is given by~\eref{di10} we get:
\begin{equation}
  \label{di13}
  \bra\cN(V)\ket_N \sim  N V^{\ga-2} \left(1-\frac{V}{N}\right)^{\ga-2}.
\end{equation}
For $V\ll N$ this reduces to $\bra\cN(V)\ket_N \sim  N V^{\ga-2}$.
It is interesting to note that $\ga$ is a function of the matter coupled to
gravity. Thus it describes an aspect of the fractal structure of quantum spacetime
that originates from the back-reaction of the matter on gravity. 


\subsection{The two-point function}
\label{sec:tpf}

In analogy to the continuum formalism in chapter~\sref{chap1} we can define a two-point function
which in a natural way contains information about the fractal structure of the random surfaces.

In principle, we could define a geodesic distance by using the canonical metric described in section~\sref{dg}.
Alternatively, the use of simplified definitions has been suggested. The geodesic distance
between two links $l_1$ and $l_2$ is defined as the length of the shortest path of triangles
connecting $l_1$ and $l_2$. The geodesic distance between a link $l_1$ and a set of links $\cL$ is defined
as the minimum of the geodesic distances between $l_1$ and the elements of $\cL$.
Furthermore we define the geodesic distance between a loop $\cL_1$ and a loop $\cL_2$ to be $r$ if all
links of $\cL_1$ lie a geodesic distance $r$ from the loop $\cL_2$. Note that this definition
is not symmetric in $\cL_1$ and $\cL_2$. 
Similarly we could have defined the geodesic distance as the distance between vertices or between triangles.

Then the two-point function is defined as:
\begin{eqnarray}
  \label{di14}
  G(\mu,r) &=& \sum_{T\in \cT(2,r)} e^{-\mu N_t} = \sum_{N=1}^{\infty} e^{-\mu N} G(N,r),\\
  G(N,r) &=& \sum_{T\in \cT_N(2,r)} 1,
\end{eqnarray}
where $\cT(2,r)$ denotes the class of triangulations $T$
with two marked links separated a geodesic distance $r$, and $\cT_N(2,r)$
the subclass where all triangulations consist of $N$ triangles. Note that $r$ is an integer in units of the
lattice spacing. $G(\mu,r)$ is the discretized form of $G(\La,R)$ in~\eref{e10}, while $G(N,r)$ is the
discretized form of $G(V,R)$ in~\eref{e16}.

\begin{figure}[htbp]
  \begin{center}
    \includegraphics[width=0.75\linewidth]{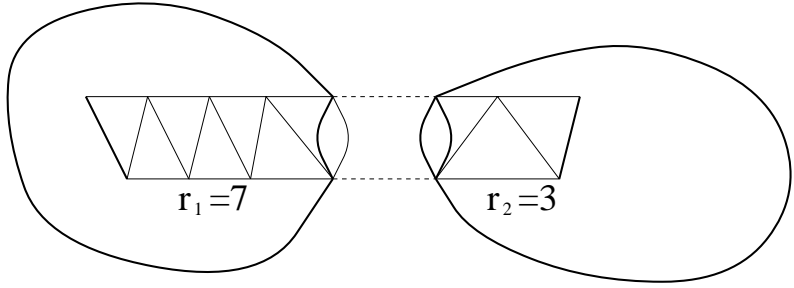}
    \parbox[t]{.85\textwidth}
    {
      \caption[gluefig]
      {
        \label{fig:di2}
        \small
        Two surfaces with two marked links separated a geodesic distance $r_1$ and $r_2$
        respectively can be glued together after cutting open one of the marked links each
        to form a surface with two marked links separated a geodesic distance $r_1+r_2$.
        }
      }    
  \end{center}
\end{figure}
An important property of the two-point function is:
\begin{equation}
  \label{di15}
  G(\mu,r_1+r_2)\geq \text{const}\times G(\mu,r_1)\ G(\mu,r_1),
\end{equation}
which follows from the fact that each term on the right hand side uniquely defines a term on the left
hand side.  This is demonstrated in figure \fref{di2}.
From this one can conclude that the limit
\begin{equation}
  \label{di16}
  \lim_{r\rightarrow\infty}\frac{-\log G(\mu,r)}{r} = m(\mu)
\end{equation}
exists with $m(\mu)\geq 0$ and $m'(\mu)>0$, for $\mu>\mu_c$.
Thus the two-point function falls off exponentially for $r\rightarrow\infty$.
These relations are by now
well known. However, the complete proofs require some technical arguments, compare~\cite{Ambjorn1997}.

Below we will show that a continuum limit of the discretized theory
can only exist if $m(\mu)$ scales to zero as $\mu\rightarrow\mu_c$. Let us
assume that this is the case:
\begin{equation}
  \label{di17a}
  m(\mu) \sim (\mu-\mu_c)^{\nu},~\text{for $\mu\rightarrow\mu_c$.}
\end{equation}
In general we expect that close to the critical point the exponential decay of $G(\mu,r)$
turns over into a power law. More precisely:
\begin{eqnarray}
  \label{di18}
  G(\mu,r) &\sim& e^{-m(\mu) r},~\text{for}~1\ll m(\mu)r,\\
  \label{di18a}
  G(\mu,r) &\sim& r^{1-\eta},~\text{for}~m(\mu)\ll m(\mu)r\ll 1.
\end{eqnarray}
The exponent $\eta$ is called the anomalous dimension in analogy to ordinary
statistical systems.
Furthermore we can define the susceptibility $\chi(\mu)$ in the discretized ensemble as
\begin{equation}
  \label{di19}
  \chi(\mu) = \sum_{r=1}^{\infty} G(\mu,r), 
\end{equation}
Close to the critical point, where triangulations with large $N_t$ dominate and symmetry
factors play no role, we have
\begin{equation}
  \label{di20}
  \chi(\mu) \sim \frac{\partial^2Z(\mu)}{\partial\mu^2} \sim (\mu-\mu_c)^{-\ga}.
\end{equation}
Similar to chapter~\sref{chap1} we can introduce several concepts of fractal dimensions
to characterize the geometrical structure of the two-dimensional quantum spacetime.
The Hausdorff dimension $d_H$ in the grand-canonical ensemble is defined by
\begin{equation}
  \label{di20a}
  \bra N\ket_r \sim r^{d_H}, \text{~for $r\rightarrow\infty$ and $m(\mu)r=\text{const}$,}
\end{equation}
where the average $\bra N\ket_r$ is given by
\begin{equation}
  \label{di20b}
  \bra N\ket_r  = \frac{1}{G(\mu,r)} \sum_{T\in\cT(2,r)} N_te^{-\mu N_t} =
  -\frac{\partial \log G(\mu,r)}{\partial \mu}.
\end{equation}
If we perform the derivative under the constraint in~\eref{di20a}
it follows that
\begin{equation}
  \label{di20c}
  \bra N\ket_r\sim m'(\mu)r \sim r^{\frac{1}{\nu}}.
\end{equation}
Thus the Hausdorff dimension $d_H$ is related to the scaling exponent $\nu$ by:
\begin{equation}
  \label{di20d}
  \nu = \frac{1}{d_H}. 
\end{equation}
With~\eref{di20} we can derive Fisher's scaling relation
\begin{equation}
  \label{di21}
  \ga = \nu (2-\eta),
\end{equation}
which relates the critical exponents defined above.
From the long distance behaviour of $G(\mu,r)$ we can compute the long distance behaviour
of $G(N,r)$ by a saddle point calculation:
\begin{equation}
  \label{di22}
  G(N,r) \sim e^{-c\left(\frac{r}{N^{\nu}}\right)^{\frac{1}{1-\nu}}},
\end{equation}
with $c=\left(\frac{1}{\nu}-1\right)\nu^{\frac{1}{1-\nu}}$. 

Analogously to the continuum formalism in chapter~\sref{chap1} we can define a Hausdorff
dimension $d_h$ by the short distance scaling of the two-point function in the canonical ensemble
with fixed number $N$ of triangles.
For $r=0$,  $G(N,0)$ is the one-point function  and behaves as
\begin{equation}
  \label{di22a}
  G(N,0) \sim e^{\mu_c N} N^{\ga-2}. 
\end{equation}
This is because for large $N$ the one-point function is proportional to $NZ(N)$ since it
counts triangulations with one marked link.

Now let $r\ll N^{1/d_h}$ and count the number $n(r)$ of triangles
which lie a geodesic distance $r$ away from a marked link. The average of $n(r)$ in the ensemble
of surfaces with one marked link defines the Hausdorff dimension $d_h$ by
\begin{equation}
  \label{di23}
  \bra n(r)\ket_N \sim \frac{G(N,r)}{G(N,0)} \sim r^{d_h-1}, ~\text{for $1\ll r\ll N^{1/d_h}$.}
\end{equation}
The first relation follows from the definition of $\bra n(r)\ket_N$. Together with~\eref{di22a}
we conclude:
\begin{equation}
  \label{di24}
  G(N,r) \sim r^{d_h-1} N^{\ga-2} e^{\mu_c N}, ~\text{for $1\ll r\ll N^{1/d_h}$.}
\end{equation}
Finally, the short distance behaviour of $G(\mu,r)$ follows by a discrete Laplace transformation.
Close to the critical point we get:
\begin{equation}
  \label{di25}
  G(\mu,r) \sim r^{\ga d_h-1}.
\end{equation}
Together with~\eref{di18a} this proves again Fisher's scaling relation.
Furthermore it turns out that both definitions of the intrinsic Hausdorff dimension
are equivalent in two-dimensional quantum gravity. However, counter examples exist,
as will be demonstrated in the next section.

The mass $m(\mu)$ determines the scaling of the discretized theory. To see how continuum
expressions can be approached in the discretized theory let us reintroduce dimensions. The
renormalized cosmological constant $\La$ in terms of the bare cosmological constant $\mu$ is defined by:
\begin{equation}
  \label{di26}
  \mu-\mu_c = \La a^2, ~\text{that means $a(\mu) \sim (\mu-\mu_c)^{1/2}$.}
\end{equation}
If the mass shall survive in the continuum limit, it has to be introduced as:
\begin{equation}
  \label{di27}
  M = m(\mu)\ a(\mu)^{-2\nu} = c \La^{\nu},
\end{equation}
with a constant $c$. If the two-point function shall survive, we must have
\begin{equation}
  \label{di17}
  m(\mu) r = M R,
\end{equation}
where the continuum parameter $M$ and $R$ are held fixed and the number of steps $r$ goes to infinity,
that means:
\begin{equation}
  \label{di28}
  R = r\ a(\mu)^{2\nu}.
\end{equation}
From~\eref{di18} and~\eref{di18a} we see that the continuum two-point function is given by:
\begin{equation}
  \label{di29}
  G(\La,R) = \lim_{\mu\rightarrow\mu_c}a(\mu)^{2\nu(1-\eta)} G(\mu,r), \quad m(\mu)r = MR.
\end{equation}
This scaling form can be explicitly verified by an exact calculation of $G(\mu,r)$ for
pure gravity~\cite{Kawai:1993cj,Watabiki:1995ym}, compare section~\sref{fs}.

\subsection{Branched polymers}
\label{sec:bpo}

An interesting simple example for the scenario described above is provided by the model
of branched polymers which is important in the understanding of the phases of two-dimensional
quantum gravity. The critical indices have been computed in detail in~\cite{Ambjorn:1986dn,Ambjorn:1990wp}.
The partition function is defined as
\begin{equation}
  \label{bp1}
  Z(\mu) = \sum_{BP} \frac{1}{C_{BP}}e^{-\mu N_{l}}\prod_{v\in BP}f(n_v),
\end{equation}
where the sum goes over all branched polymers, that means all connected planar tree graphs.
$N_l$ denotes the number of links in a branched polymer. The product goes over all vertices in the tree graphs
and $n_v$ is the number of links joining at the vertex $v$, called the order of $v$.
Usually, the (unnormalized) branching weight
$f(n)$ is a non-negative function of the order of the vertices. Finally the symmetry factor
$C_{BP}$ is chosen such that rooted branched polymers, that means with the first link marked, are counted only once.

The one-point function $G(\mu)$ is defined similarly as the  sum over the rooted branched polymers. In
\begin{figure}[htbp]
  \begin{center}
    \includegraphics[width=0.70\linewidth]{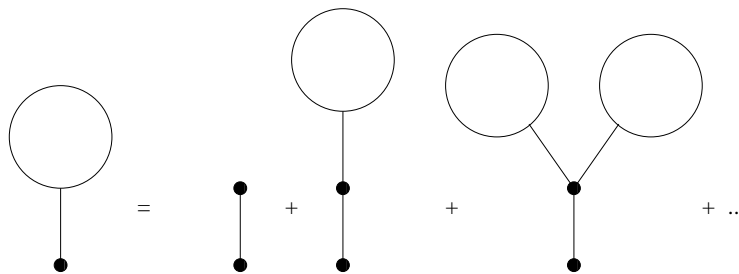}
    \parbox[t]{.85\textwidth}
    {
      \caption[bpfig]
      {
        \label{fig:bp1}
        \small
        Illustration of the self-consistent equation for rooted branched polymers. The first link
        gives a factor $e^{-\mu}$. At the vertices the branching weight has to be included.
        }
      }    
  \end{center}
\end{figure}
this case, the symmetry factor drops out. From figure \fref{bp1} we follow that $G(\mu)$ satisfies the equation
\begin{equation}
  \label{bp2}
  G(\mu) = e^{-\mu}\left(1 + f(2) G(\mu) + f(3) G(\mu)^2 + \ldots \right). 
\end{equation}
We can solve this relation for $e^{\mu}$ as a function of $G(\mu)$:
\begin{equation}
  \label{bp3}
  e^{\mu} = \frac{1 + f(2) G(\mu) + f(3) G(\mu)^2 + \ldots }{G(\mu)} \equiv \frac{\cF(G)}{G} \equiv F(G).
\end{equation}
\begin{figure}[htbp]
  \begin{center}
    \includegraphics[width=0.7\linewidth]{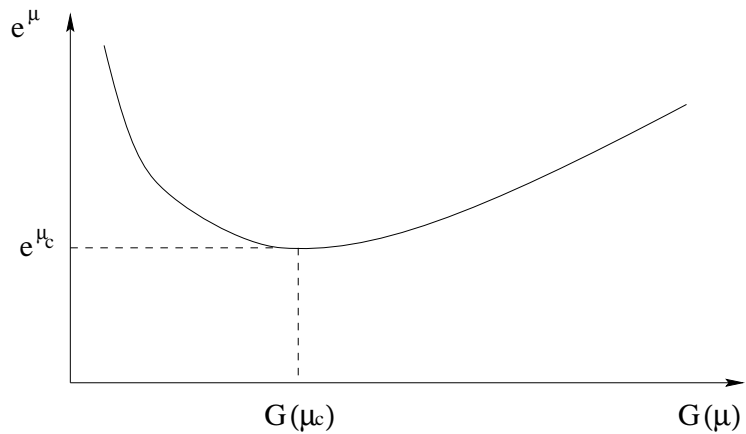}
    \parbox[t]{.85\textwidth}
    {
      \caption[solfig]
      {
        \label{fig:bp2}
        \small
        Graphical solution of~\eref{bp3}. The critical point is identified
        as the minimum of $e^{\mu}$ as a function of $G$.
        }
      }    
  \end{center}
\end{figure}
This equation is illustrated in figure \fref{bp2}. The critical point is at the minimum
of $F(G)$. Since all $f(n)$ are positive the minimum is unique and satisfies
$F'(G(\mu_c))=0$ and $F''(G(\mu_c))>0$. Therefore we have:
\begin{equation}
  \label{bp4}
  G(\mu) \sim G(\mu_c) - c (\mu-\mu_c)^{\frac{1}{2}}, ~\text{for $\mu\rightarrow\mu_c$,}
\end{equation}
where $c$ is some constant.
Since $G(\mu)$ is the one-point function we would have expected a behaviour
$G(\mu) \sim (\mu-\mu_c)^{1-\ga}$. Thus we conclude that the generic value
of $\ga$ for branched polymers is $\frac{1}{2}$.

If we allow some of the values $f(n)$ to be negative, we can fine tune the $f(n)$ such
that the minimum of $F(G)$ satisfies
\begin{equation}
  \label{bp5}
  F^{(k)}(G(\mu_c)) = 0, ~\text{for $k=1,\ldots,m-1$, and $F^{(m)}(G(\mu_c))\neq 0$.}
\end{equation}
This model is called $m$'th multicritical branched polymer model. For $m=2$
we simply recover the ordinary branched polymer model. The critical behaviour of the
one-point function is changed to
\begin{equation}
  \label{bp6}
  G(\mu) \sim G(\mu_c) - c (\mu-\mu_c)^{\frac{1}{m}}, ~\text{for $\mu\rightarrow\mu_c$,}
\end{equation}
and we get $\ga=\frac{m-1}{m}$.

To define the two-point function $G(\mu,r)$ for branched polymers, let the geodesic
distance between two marked vertices $x$ and $y$ be the minimal length of paths between the two points. This
is of course unique.
The two-point function is defined as the partition function of branched polymers with
the constraint that there are two marked points separated a geodesic distance $r$.
The path between the marked points can be viewed
as a random walk where at each vertex a rooted branched polymer can grow out.
If we think only in terms of intrinsic geometrical properties this leads to \cite{Ambjorn:1990wp}:
\begin{equation}
  \label{bp7}
  G(\mu,r) \sim \left(e^{-\mu}\cF'(G(\mu))\right)^r = e^{-m(\mu)r}, 
\end{equation}
from which we conclude, using $e^{-\mu}\cF'(G(\mu))= 1 - \frac{G(\mu)}{\chi(\mu)}$, that
\begin{equation}
  \label{bp8}
  m(\mu) = \ka (\mu-\mu_c)^{\frac{m-1}{m}},
\end{equation}
in the general case for $m\geq 2$, with a positive constant $\ka$.
Therefore it follows with~\eref{di20d} that the intrinsic Hausdorff dimension $d_H$ equals $\frac{m}{m-1}$.
For $m=2$ we have $d_H=2$ and the branched polymers have the same intrinsic dimension as
smooth two-dimensional manifolds. For $m\rightarrow\infty$, $d_H$ approaches one, in agreement
with the fact that in this limit the branched polymers approach ordinary random walks.
The two-point function $G(\mu,r)$ is given by
\begin{equation}
  \label{bp8a}
  G(\mu,r) = \text{const}\times e^{-\ka r (\mu-\mu_c)^{\frac{m-1}{m}}}.
\end{equation}

We can also compute the canonical intrinsic Hausdorff dimension $d_h$ for branched polymers
defined in the ensemble of branched polymers with a fixed number $N$ of links.
The volume $\bra n(r)\ket_N$ of a spherical shell of geodesic radius $r$ is defined in analogy
to~\eref{e17b} and~\eref{di23}. $Z(N)$ for branched polymers is given by an inverse Laplace transformation
of~\eref{bp1} and scales as $Z(N) \sim e^{\mu_c N}N^{\ga-3}$. Similarly the two-point function $G(N,r)$
for graphs with fixed volume $N$ is given by an inverse Laplace transformation of $G(\mu,r)$.
Inserting the scaling behaviour and taking only the leading orders we find for small $r$~\cite{Ambjorn:1997sy}:
\begin{equation}
  \label{bp9}
  \bra n(r)\ket_N \sim \frac{r}{N^{1-\frac{2}{m}}}, ~\text{for $N^{1-\frac{2}{m}}\ll r\ll N^{1-\frac{1}{m}}$.}
\end{equation}
This shows that $d_h=2$ for all values of $m$. Thus for the generic branched polymers ($m=2$)
the canonical and the grand-canonical definition of the intrinsic Hausdorff dimension
give the same result. However, this is not true for the $m$'th multicritical branched polymer
model with $m>2$, where $d_H=\frac{m}{m-1}$ and $d_h=2$.

The fractal structure of branched polymers embedded in $\mathbb{R}^D$ is described by the 
extrinsic Hausdorff dimension $D_H$. $D_H$ can be computed from the scaling of the embedded
two-point function. The path between two marked points $x$ and $y$ is an embedded random walk at which at each
point branched polymers can grow out. The integration over the embedding field is the same
as for the ordinary random walk in $D$ dimensions. The branching at the vertices corresponds
to a renormalization of $e^{-\mu}$. From this renormalization $D_H$ can be computed with the
result $D_H=\frac{2}{\ga} = \frac{2m}{m-1}$~\cite{Ambjorn:1990wp}. Thus $D_H=4$ for ordinary branched polymers and
$D_H\rightarrow 2$ for $m\rightarrow\infty$.

\section{Matrix models}
\label{sec:mm}

Major advances in the theory of two-dimensional quantum gravity have been made when it was
realized, that the theory is integrable in the discrete formulation, even for some types of matter
coupled to the
surfaces~\cite{Ambjorn:1985az,Boulatov:1987sb,David:1985tx,David:1985nj,David:1985et,Kazakov:1985ea,Kazakov:1986hu}. 

In the development of these results the formulation of dynamical triangulation
in terms of matrix models has been of some importance. 
The Hermitean one-matrix model
has a direct interpretation in terms of randomly triangulated surfaces. This is
demonstrated in the beginning of this section. Also the Hartle-Hawking wavefunctionals have a direct
natural formulation via matrix models. The Dyson-Schwinger equations for these models
allow the treatment of the theory by methods of complex analysis.
These so-called loop equations can alternatively be derived using only combinatorial methods~\cite{Ambjorn1997}.
The explicit solution of the loop equations provides deep insight into the properties
of the theory and allows the computation of the scaling limit of the Hartle-Hawking wavefunctionals.

We cannot review the vast literature on numerical simulations in quantum gravity in this work.
However, let us remark that all numerical simulations of two-dimensional dynamical triangulation
are consistent with the scaling hypotheses, with universality, and -- most remarkably -- with
the continuum theory of two-dimensional quantum gravity.

\subsection{Dynamical triangulation by matrix models}

The discretized partition function of two-dimensional quantum gravity can be represented as an integral over a
Hermitean $N\times N$-matrix $\phi$. Consider the Gaussian integral
\begin{equation}
  \label{mm1}
  \int\! d\phi\ e^{-\frac{1}{2}\tr\phi^2}\frac{1}{K!}\left(\frac{1}{3}\tr\phi^3\right)^K,
\end{equation}
where the measure $d\phi$ is defined as
\begin{equation}
  \label{mm2}
  d\phi = \prod_{i\leq j} d\Re \phi_{ij} \prod_{i<j} d\Im \phi_{ij}.
\end{equation}
The propagator of $\phi$ is given by
\begin{equation}
  \label{mm3}
  \bra\phi_{ij}\phi_{i'j'}\ket \equiv \frac{1}{Z(N,0)} \int\! d\phi\ e^{-\frac{1}{2}\tr\phi^2}
  \phi_{ij}\phi_{i'j'} = \de_{ij'}\de_{ji'},
\end{equation}
with $Z(N,0) = \int\! d\phi\ e^{-\frac{1}{2}\tr\phi^2}$.
Diagrammatically, the propagator can be represented as a double line where
the two lines are oriented in opposite directions.
The integral~\eref{mm1} is performed by doing all possible contractions of the
$K$ $\tr\phi^3$ vertices. The dual picture of this corresponds
to gluing $K$ triangles together to form all possible closed and not necessarily connected
surfaces of arbitrary topology. The contribution from a particular graph forming a closed surface is $N^{N_v}$,
since each vertex contributes with a factor $N$. 
If we make the substitution
\begin{equation}
  \label{mm4}
  \tr\phi^3 \rightarrow \frac{g}{\sqrt{N}}\tr\phi^3,
\end{equation}
each closed surface with Euler characteristic $\chi$
contributes with $g^KN^{N_v-K/2}=g^KN^{\chi}$. We get the sum over arbitrary closed
surfaces with any number of triangles if we sum~\eref{mm1} over $K$:
\begin{equation}
  \label{mm5}
  Z(N,g) = \int\! d\phi\ e^{-\frac{1}{2}\tr\phi^2 + \frac{g}{3\sqrt{N}}\tr\phi^3}.
\end{equation}
Taking the logarithm of this projects on the connected surfaces only.
Thus we see that 
\begin{equation}
  \label{mm6}
  Z(\mu,G) = \log{\frac{Z(N,g)}{Z(N,0)}},
\end{equation}
with
\begin{equation}
  \label{mm7}
  \frac{1}{G} = \log N, ~\text{and $\mu = -\log g$}
\end{equation}
is a formal definition of the discretized partition function of two-dimensional quantum gravity, including
the sum over topologies. Here $G$ is the gravitational coupling constant.
The matrix integral~\eref{mm5} is of course not convergent, but it has been suggested, that
a closed form like~\eref{mm5} might define a non-perturbative definition of the
sum over topologies after analytic continuation. In general however, the integral will be
complex~\cite{Ambjorn:1991pt,Ambjorn:1992km} and the problem of summing over topologies
has not yet been solved.

Equation~\eref{mm5} admits a $1/N^2$-expansion which is identical to an expansion
over topologies. Here we will restrict ourselves to spherical topology which corresponds to the
leading order in $1/N^2$.
It is convenient to generalize the matrix integral~\eref{mm5} to
\begin{eqnarray}
  \label{mm8}
  Z(N,g_i) &=& \int\! d\phi\ e^{-N\tr V(\phi)},\\
  V(\phi) &=& \sum_{n=1}^{\infty}\frac{g_n}{n}\phi^n,
\end{eqnarray}
where we have performed the rescaling $\phi\rightarrow \sqrt{N}\phi$.~\eref{mm8} is interpreted as
a perturbative expansion around a Gaussian integral. Thus the Gaussian coupling constant
is bigger than zero, while the others are chosen smaller than zero.
The expectation value of some observable $\cO(\phi)$ is defined as
\begin{equation}
  \label{mm8a}
  \bra\cO(\phi)\ket = \frac{1}{Z(N,g_i)} \int\! d\phi\ e^{-N\tr V(\phi)} \cO(\phi).
\end{equation}

Differentiating the logarithm of~\eref{mm8} with respect to the coupling constants $g_n$ defines
expectation values of observables like $\tr \phi^{k_1} \cdots \tr\phi^{k_b}$. They are interpreted
as the sum over all surfaces which have $b$ polygons with $k_i$ links as their boundary.
The generating function of the connected expectation values of these observables is given by
\begin{eqnarray}
  \label{mm9}
  W(z_1,\ldots,z_b) &=& N^{b-2} \sum_{k_1,\ldots,k_b=0}^{\infty}
  \frac{\bra\tr\phi^{k_1}\cdots\tr\phi^{k_b}\ket_{\text{conn}}}{z_1^{k_1+1}\cdots z_b^{k_b+1}}\nonumber\\
  &=& N^{b-2}\bra\tr\frac{1}{z_1-\phi}\cdots\tr\frac{1}{z_b-\phi}\ket_{\text{conn}}.
\end{eqnarray}
In the large-$N$ limit this is the generating function for the discretized
Hartle-Hawking wavefunctionals, whose continuum counterparts have been defined in section~\sref{hh}.
In fact, already the one-loop correlator $W(z)$ contains all necessary information, since
the higher correlators can be computed from it by differentiation:
\begin{equation}
  \label{mm10}
  W(z_1,\ldots,z_b)  = \frac{d}{dV(z_b)}\frac{d}{dV(z_{b-1})}\cdots\frac{d}{dV(z_2)} W(z_1),
\end{equation}
where the loop insertion operator $\frac{d}{dV(z)}$ is defined as
\begin{equation}
  \label{mm11}
  \frac{d}{dV(z)} \equiv -\sum_{n=1}^{\infty} \frac{n}{z^{n+1}} \frac{\partial}{\partial g_n}.
\end{equation}
Let us define the density $\rho(\la)$ of eigenvalues of the matrix integral~\eref{mm8} by
\begin{equation}
  \label{mm12}
  \rho(\la) = \bra \sum_{i=1}^N\de(\la_i-\la)\ket,
\end{equation}
where the $\la_i$ are the $N$ eigenvalues of the matrix $\phi$. Then the one-loop correlator
$W(z)$ can be written as:
\begin{equation}
  \label{mm13}
  W(z) = \int_{-\infty}^{\infty}\! d\la\ \frac{\rho(\la)}{z-\la}.
\end{equation}
For polynomial potentials the support of $\rho(\la)$ is confined to one or several intervals
on the real axes in the limit $N\rightarrow\infty$. The solution in the case of several
cuts has been given recently~\cite{Akemann:1996zr}. Here we will only deal with a single cut
$[\afr,\bfr]$ on the real axes.
Then $W(z)$ will be analytic in the complex plane with a cut at the support $[\afr,\bfr]$ of $\rho(\la)$.
It follows that
\begin{equation}
  \label{mm14}
  2\pi i\ \rho(\la) = \lim_{\eps\rightarrow 0} \left( W(\la+i\eps) - W(\la-i\eps)\right).
\end{equation}

\subsection{The loop equations}

The matrix model~\eref{mm8} can be solved by many methods. It is most conveniently and systematically
done by using  loop equations, the Dyson-Schwinger equations for the matrix models.
Let us  consider the transformation
\begin{equation}
  \label{le1}
  \phi \rightarrow \phi + \eps\frac{1}{z-\phi},
\end{equation}
which makes sense if $z$ is not an eigenvalue of $\phi$ and real, so that the new matrix
remains Hermitean. Then the measure and the action transform as
\begin{eqnarray}
  \label{le2}
  d\phi &\rightarrow& \left(1 + \eps \tr\frac{1}{z-\phi}\tr\frac{1}{z-\phi}\right)d\phi,\\
  \tr V(\phi) &\rightarrow& \tr V(\phi) + \eps \tr\frac{V'(\phi)}{z-\phi},
\end{eqnarray}
to first order in $\eps$.
Since the matrix integral~\eref{mm8} is invariant under such a definition of the
integration variable we get:
\begin{equation}
  \label{le3}
  \int\! d\phi\ e^{-N\tr V(\phi)}\ \left\lbrace \left(\tr\frac{1}{z-\phi}\right)^2
  - N\tr\frac{V'(\phi)}{z-\phi}\right\rbrace = 0.
\end{equation}
The second term can be rewritten as a contour integral involving
the one-loop correlator by using the eigenvalue density $\rho(\la)$, while the first
term is related to the two-loop correlator. This leads to the standard form of the loop
equations~\cite{David:1990ge}:
\begin{equation}
  \label{le4}
  \oint_{\cC}\! \frac{d\om}{2\pi i}\ \frac{V'(\om)}{z-\om}W(\om) = W(z)^2 + \frac{1}{N^2} W(z,z). 
\end{equation}
The contour $\cC$ goes around the cut $[\afr,\bfr]$ but does not enclose $z$. The loop equation
can be solved systematically to all orders in $1/N^2$, that means for all genera of the random
surfaces~\cite{Ambjorn:1992jf,Ambjorn:1993gw}. For spherical topology, that means in the
limit $N\rightarrow\infty$,~\eref{le4} simplifies to
\begin{equation}
  \label{le5}
  \oint_{\cC}\! \frac{d\om}{2\pi i}\ \frac{V'(\om)}{z-\om}W_0(\om) = W_0(z)^2,
\end{equation}
where the subscript $0$ denotes the genus of the surfaces.
By the definition we know that $W_0(z) = \frac{1}{z} + O(z^{-2})$.
Therefore~\eref{le5} can be solved by a deformation of the contour $\cC$ to infinity.
For a polynomial action $V$ of degree $n$ we get:
\begin{eqnarray}
  \label{le8}
  W_0(z) &=& \frac{1}{2}\left(V'(z) - \sqrt{(z-\afr)(z-\bfr)}\sum_{k=1}^{\infty}M_k[\afr,\bfr,g_i]
    (z-\bfr)^{k-1}\right)\nonumber \\
  &=& \oint_{\cC}\! \frac{d\om}{2\pi i}\
  \frac{V'(\om)}{z-\om}\frac{\sqrt{(z-\afr)(z-\bfr)}}{\sqrt{(\om-\afr)(\om-\bfr)}},
\end{eqnarray}
where the moments $M_k$ are defined as
\begin{equation}
  \label{le9}
  M_k[\afr,\bfr,g_i] = \oint_{\cC}\! \frac{d\om}{2\pi i}\
  \frac{V'(\om)}{(\om-\afr)^{\frac{1}{2}}(\om-\bfr)^{k+\frac{1}{2}}},
\end{equation}
and vanish for $k>n-1$. The endpoints $\afr$ and $\bfr$ of the cut are self-consistently determined by the equations
\begin{eqnarray}
  \label{le10}
  M_{-1}[\afr,\bfr,g_i] &=& 2,\\
  M_0[\afr,\bfr,g_i] &=& 0,
\end{eqnarray}
which are a consequence of~\eref{le8} together with $W_0(z) = \frac{1}{z} + O(z^{-2})$.

For the matrix model~\eref{mm5} which directly corresponds to triangulated surfaces,
the solution for $W_0(z)$ is:
\begin{eqnarray}
  \label{le10a}
  W_0(z) &=& \frac{1}{2}V'(z) + f(\mu,z),\nonumber \\
  f(\mu,z) &=& \frac{g}{2}(z-\cfr)\sqrt{(z-\afr)(z-\bfr)},
\end{eqnarray}
where $g=e^{-\mu}$ and $\cfr = \frac{1}{g}-\frac{\afr+\bfr}{2}$. 

In principle one could compute all higher loop correlators from~\eref{le8} by applying the loop
insertion operator~\eref{mm11}, which gives a complete solution of two-dimensional dynamical
triangulation for spherical topology.

In fact, a closed expression for the $b$-loop correlators can be obtained if the small scale
details of the theory are adjusted. Instead of the Hermitean matrix model one considers
the complex matrix model with a potential
\begin{equation}
  \label{le11}
  \tr V(\phi^+\phi) = \sum_{n=1}^{\infty}\frac{g_n}{n}\tr (\phi^+\phi)^n,
\end{equation}
and an integration measure
\begin{equation}
  \label{le12}
  d\phi = \prod_{i,j=1}^N d\Re \phi_{ij}d\Im \phi_{ij}.
\end{equation}
The one-loop correlator is now defined as
\begin{equation}
  \label{le13}
  W(z) = \frac{1}{N} \sum_{n=0}^{\infty}\frac{\bra\tr (\phi^+\phi)^n\ket}{z^{2n+1}},
\end{equation}
higher loop correlators are defined analogously. The term $\tr(\phi^+\phi)^n$ can be interpreted
as a $2n$-sided polygon whose links are alternately coloured black and white. These
polygons can be glued together as in the Hermitean matrix model with the additional
constraint that black links have to be glued to white links~\cite{Morris:1991cq}.
Such short distance details about the gluing should be unimportant in the continuum
limit.

The loop equations for this model have been derived in~\cite{Ambjorn:1990ji}.
Because of the symmetry $\phi\rightarrow -\phi$ we have $\afr=-\bfr$ for the cut of the
one-loop correlator. The solution for spherical topology is given by~\cite{Ambjorn:1990ji}:
\begin{eqnarray}
  \label{le14}
  W_0(z) &=& \frac{1}{2} \left(V'(z) - M(z) \sqrt{z^2-\bfr^2}\right),\nonumber \\
  M(z) &=& \oint_{\cC_{\infty}}\! \frac{d\om}{4\pi i}\ \frac{\om V'(\om)}{(\om^2-z^2)\sqrt{\om^2-\bfr^2}}
  = \sum_{k=1}^{\infty} M_k[\bfr,g_i](z^2-\bfr^2)^{k-1},\nonumber\\
  M_k[\bfr,g_i] &=& \oint_{\cC}\! \frac{d\om}{4\pi i}\ \frac{\om V'(\om)}{(\om^2-\bfr^2)^{k+\frac{1}{2}}}.
\end{eqnarray}
Here $\cC_{\infty}$ is a contour around the cut pushed to infinity. 
The position of the cut is determined by the equation
\begin{equation}
  \label{le14a}
  M_0[\bfr,g_i]=2.
\end{equation}
The higher loop
correlators are given by the following expressions~\cite{Ambjorn:1990ji}:
\begin{eqnarray}
  \label{le15}
  W_0(z_1,z_2) &=& \frac{1}{4(z_1^2-z_2^2)^2}\left(z_2^2\sqrt{\frac{z_1^2-\bfr^2}{z_2^2-\bfr^2}}
    + z_1^2\sqrt{\frac{z_2^2-\bfr^2}{z_1^2-\bfr^2}} -2z_1z_2\right),\nonumber \\
  W_0(z_1,z_2,z_3) &=& \frac{\bfr^4}{16M_1}\frac{1}{\sqrt{(z_1^2-\bfr^2)(z_2^2-\bfr^2)(z_3^2-\bfr^2)}},\nonumber \\
  W_0(z_1,\ldots,z_b) &=& \left(\frac{1}{M_1}\frac{d}{d\bfr^2}\right)^{b-3}
  \frac{1}{2\bfr^2M_1} \prod_{k=1}^b\frac{\bfr^2}{2(z_k^2-\bfr^2)^{3/2}}.
\end{eqnarray}
In these formulas all dependence on the coupling constants is hidden in $M_1$ and $\bfr$.
Similar and only slightly more complicated statements are valid also for the Hermitean matrix model.

\subsection{Scaling limit and computation of $\ga$}

It has been discussed in section~\sref{tpf} that the ensemble of triangulated random surfaces has
a critical point $\mu_c$. The continuum limit with a renormalized cosmological constant
$\La$ is approached by $\mu\rightarrow\mu_c$ with $\mu-\mu_c=\La a^2$, where $a$ is the lattice spacing.
In terms of the coupling constant $g$ for the model~\eref{mm5} this relation is
$g_c-g\sim g_c\La a^2$, close to the critical point. For the generalized matrix model~\eref{mm8}
with $n$ coupling constants we expect no qualitative changes except that the theory
will be critical on a $(n-1)$-dimensional hypersurface. This hypersurface is identified by
\begin{equation}
  \label{le16}
  M_1[\bfr(g_i),g_i] = 0,
\end{equation}
since the $b$-loop correlators $W_0(z_1,\ldots,z_b)$ are exactly divergent for $M_1=0$.
Let us denote a point in the critical hypersurface by $g_{c,i}, i=1,\ldots,n$ and the corresponding
endpoint of the eigenvalue distribution by $\bfr_c$.
If we move slightly away from the critical surface
\begin{equation}
  \label{le17}
  g_{i} = g_{c,i} (1-\La a^2)=g_{c,i}+\de g_i,
\end{equation}
there will be a corresponding change $\bfr_c^2\rightarrow \bfr_c^2+\de \bfr^2$
which can be computed from~\eref{le14a} to be
\begin{equation}
  \label{le18}
  (\de \bfr^2)^2 = -\frac{16}{3M_2[\bfr_c,g_{c,i}]}\La a^2 \sim \de g_i.
\end{equation}
We rescale the cosmological constant such that $\bfr^2 = \bfr_c^2 - a\sqrt{\La}$. Because
$z_i$ appears in~\eref{le15} always in the combination $(z_i^2-\bfr^2)$ it is natural
to introduce a scaling of $z_i$ by
\begin{equation}
  \label{le19}
  z_i^2  = \bfr_c^2 + aZ_i.
\end{equation}
Since $W_0(z_1,\ldots,z_b)$ is the generating function for $b$-loop correlators, we can
compute the transition amplitude for $b$ one-dimensional universes of lengths
$n_1,\ldots,n_b$ from it by a multiple contour integration. The physical lengths of the
loops will be $L_i = n_i a$. We see that in the continuum limit the number of links $n_i$
on the boundaries has to go to infinity as $a\rightarrow 0$ if the loops shall survive.
By following this procedure one gets the generating functional for
macroscopic $b$-loop amplitudes~\cite{Ambjorn:1990ji}:
\begin{eqnarray}
  \label{le20}
  W_0(z_1,\ldots,z_b) &\sim& a^{5-\frac{7b}{2}} W_0(\La,L_1,\ldots,L_b), \nonumber\\
  W_0(\La,Z_1,\ldots,Z_b) &=& \left(-\frac{d}{d\La}\right)^{b-3} \frac{1}{\sqrt{\La}}
  \prod_{k=1}^b (Z_k+\sqrt{\La})^{-\frac{3}{2}},
\end{eqnarray}
for $b\geq 3$. The inverse Laplace transform of $W_0(\La,Z_1,\ldots,Z_b)$
in the variables $Z_i$ gives the
Hartle-Hawking wavefunctionals of two-dimensional quantum gravity. It can be computed
from~\eref{le20} to be:
\begin{equation}
  \label{le21}
  W_0(\La,L_1,\ldots,L_b) = \left(-\frac{d}{d\La}\right)^{b-3} \frac{1}{\sqrt{\La}}
  \sqrt{L_1\cdots L_b}\ e^{-\sqrt{\La}(L_1 + \ldots + L_b)}.
\end{equation}
Since the $b$-point function in two-dimensional quantum gravity should
scale as $\La^{2-b-\ga}$, we can directly read off from this expression that
\begin{equation}
  \label{le22}
  \ga = -\frac{1}{2}
\end{equation}
for pure gravity, in agreement with the KPZ-formula~\eref{e3d}. These calculations can be
generalized to arbitrary topology~\cite{David:1990ge,Ambjorn:1992jf,Ambjorn:1992xu,Ambjorn:1993gw}
and to models with matter coupled to gravity~\cite{Kazakov:1989bc,Ambjorn:1990wg,Ambjorn1997}
by considering multicritical
models which are obtained by a fine-tuning of the critical coupling constants such that higher
moments vanish as well as $M_1$.

It is a major result that all values of the susceptibility exponent $\ga$ computed by these
or other methods in the model of dynamically triangulated quantum gravity agree with the
continuum formula~\eref{e3d} which has been derived in section~\sref{gd}. In fact, all
calculations which can be done by dynamical triangulation  and in the continuum approach have so far
yielded the same results. Therefore we identify both theories as the theory of
two-dimensional quantum gravity. Objects like the Hartle-Hawking wavefunctionals
and other correlation functions can much easier be obtained in the discretized approach
whose scaling limit yields the theory of two-dimensional quantum gravity.

\section{The fractal structure of pure gravity}
\label{sec:fs}

Another major advance in two-dimensional quantum gravity
was made by the explicit computation of the two-point function for pure quantum gravity
by constructing 
a transfer matrix \cite{Kawai:1993cj} and later by using a peeling decomposition \cite{Watabiki:1995ym}.
The two-point function is a natural object since it provides all details about the scaling
properties of two-dimensional quantum gravity. The astonishing result of these investigations is
that the intrinsic Hausdorff dimension of two-dimensional quantum gravity is {\em four}. That
means that even the dimensionality of spacetime is a dynamical quantity.

\subsection{The geodesic two-loop function}

Let us define with $\cT(l_1,l_2,r)$ the class of triangulations with an entrance loop
$l_1$ with one marked link and an exit loop $l_2$, separated a geodesic distance $r$.
We will also denote the number of links of $l_1$ and $l_2$ with the same symbols.
Then the geodesic two-loop function is defined as
\begin{equation}
  \label{tl1}
  G(\mu,r;l_1,l_2) = \sum_{T\in \cT(l_1,l_2,r)} e^{-\mu N_t}.
\end{equation}
For $r=0$ we have the initial condition:
\begin{equation}
  \label{tl1a}
  G(\mu,0;l_1,l_2) = \de_{l_1,l_2}.
\end{equation}
We introduce the generating function for $G(\mu,r;l_1,l_2)$ by
\begin{equation}
  \label{tl3}
  G(\mu,r;z_1,z_2) = \sum_{l_1,l_2=1}^{\infty} z_1^{-(l_1+1)}z_2^{-(l_2+1)} G(\mu,r;l_1,l_2),
\end{equation}
with the initial condition
\begin{equation}
  \label{tl4}
  G(\mu,0;z_1,z_2) = \frac{1}{z_1z_2}\frac{1}{z_1z_2-1}.
\end{equation}
By a two-fold contour integration the geodesic two-loop function can be reconstructed:
\begin{equation}
  \label{tl5}
  G(\mu,r;l_1,l_2) = \oint_{\cC_1}\! \frac{dz_1}{2\pi i}\ z_1^{l_1}
  \oint_{\cC_2}\! \frac{dz_2}{2\pi i}\ z_2^{l_2}\ G(\mu,r;z_1,z_2).
\end{equation}
It is an important observation that the two-point function $G(\mu,r)$ can be obtained
from the geodesic two-loop function. Consider the two-loop function with an entry
loop of length $l_1=1$ which is equivalent to a marked link,
and an exit loop of arbitrary length $l_2$ separated a geodesic distance $r$.
We can close this surface by gluing a disc to the boundary $l_2$.
The amplitude $W_0(l_2)$ of the disc can be computed from the one-loop correlator
$W_0(z)$ by a contour integration.
An additional factor of $l_2$ arises because we have
to mark one of the $l_2$ links of the exit loop. Thus we get:
\begin{eqnarray}
  \label{lt6}
  G(\mu,r) &=& \sum_{l_2=1}^{\infty}G(\mu,r;1,l_2)\ l_2 W_0(l_2) \nonumber\\
  &=& \oint_{\cC} \! \frac{d\om}{2\pi i}\frac{1}{\om} \left[z^2 G(\mu,r;z,\frac{1}{\om})\right]
  \left[-\frac{\partial}{\partial\om}\om W_0(\om)\right] \Big\vert_{z=\infty},
\end{eqnarray}
where $\cC$ is a contour around zero.
Thus we see that all information about the scaling of the theory can be obtained if the
generating function of geodesic two-loop amplitudes can be computed.

By a step-by-step decomposition of triangulations in $\cT(l_1,l_2,r)$
a differential equation for $G(\mu,r;z,\om)$ can be obtained . 
It is intuitive that any triangulation in $\cT(l_1,l_2,r)$ can be decomposed into $r$ rings
of thickness one \cite{Kawai:1993cj}. This leads to a transfer matrix
formalism for the two-point function. Alternatively, one can decompose the triangulations
by a peeling procedure~\cite{Watabiki:1995ym}. The differential equation one obtains is not
exact but should be valid close to the critical point of the theory:
\begin{equation}
  \label{lt7}
  \frac{\partial}{\partial r} G(\mu,r;z,\om) = -2 \frac{\partial}{\partial z}
  \left[f(\mu,z)G(\mu,r;z,\om)\right],
\end{equation}
where $f(\mu,z)$ is given by~\eref{le10a} for triangulated surfaces.

\subsection{Scaling of the two-point function}

The differential equation~\eref{lt7} can be solved by the method of characteristic
equations. The result is:
\begin{equation}
  \label{lt8}
  G(\mu,r;z,\om)  = \frac{f(\mu,\hat{z})}{f(\mu,z)} \frac{\hat{z}\om}{\hat{z}\om-1},
\end{equation}
where $\hat{z}$ is the solution to the characteristic equation
\begin{equation}
  \label{lt9}
  \frac{d\hat{z}(z,r)}{dr} = 2 f(\mu,\hat{z}),
\end{equation}
given by
\begin{equation}
  \label{lt10}
  \frac{1}{\hat{z}(z,r)} = \frac{1}{\cfr} - \frac{\de_1}{\cfr}
  \frac{1}{\sinh^2\left(-\de_0r+\sinh^{-1}\sqrt{\frac{\de_1}{1-c/z}-\de_2}\right)+\de_2}.
\end{equation}
Here the positive constants $\de_i$ scale as
\begin{eqnarray}
  \label{lt11}
  \de_0 &=& \frac{g}{2}\sqrt{(\cfr-\afr)(\cfr-\bfr)} = O((\mu-\mu_c)^{\frac{1}{4}}),\\
  \de_1 &=& \frac{(\cfr-\afr)(\cfr-\bfr)}{\cfr(\bfr-\afr)} = O((\mu-\mu_c)^{\frac{1}{2}}),\\
  \de_2 &=& -\frac{\afr(\cfr-\bfr)}{\cfr(\bfr-\afr)} = O((\mu-\mu_c)^{\frac{1}{2}}).
\end{eqnarray}
Now the two-point function $G(\mu,r)$ can be computed by inserting~\eref{lt8} into
\eref{lt6}. Close to the critical point one obtains~\cite{Ambjorn:1995dg}:
\begin{equation}
  \label{lt12}
  G(\mu,r) = \text{const}\times \de_0\de_1\frac{\cosh(\de_0 r)}{\sinh^3(\de_0 r)}(1+O(\de_0)).
\end{equation}
From~\eref{lt11} we get:
\begin{equation}
  \label{lt13}
  G(\mu,r) = \text{const}\times (\mu-\mu_c)^{\frac{3}{4}}
  \frac{\cosh \left(c(\mu-\mu_c)^{\frac{1}{4}}r\right)}{
    \sinh^3\left(c(\mu-\mu_c)^{\frac{1}{4}}r\right)},
\end{equation}
where $c=\sqrt{6}e^{-\mu_c}$ is a nonuniversal constant. To read off the critical behaviour of the theory we only
have to analyze the asymptotic  behaviour of the two-point function for large and for small distances.
We see that:
\begin{itemize}
\item $G(\mu,r)$ falls off exponentially as $e^{-2c(\mu-\mu_c)^{\frac{1}{4}}r}$, for $r\rightarrow\infty$.
  Thus the susceptibility exponent equals $\nu=\frac{1}{4}$ and the intrinsic Hausdorff dimension
  $d_H$ of pure two-dimensional quantum gravity equals {\em four}.
\item For $1\ll r\ll (\mu-\mu_c)^{-\frac{1}{4}}$ the two-point function behaves like $r^{-3}$,
  that means the anomalous scaling dimension is $\eta=4$. For ordinary statistical systems
  $\eta$ is smaller than two. It is remarkable, that the critical exponents of the theory still
  satisfy Fisher's scaling relation $\ga = \nu(2-\eta)$.
\item Any geodesic two-loop function $G(\mu,r;l_1,l_2)$,
  which can be computed from~\eref{tl5} and~\eref{lt8} fulfills
  the same scaling relations as the two-point function, provided that $l_1$ and $l_2$ stay finite
  in the limit $\mu\rightarrow\mu_c$.  
\item The continuum limit of the two-point function is given by
  \begin{equation}
    \label{lt14}
    G(\La,R) = C\La^{\frac{3}{4}} \frac{\cosh\left(c\La^{\frac{1}{4}}R\right)}{
      \sinh^3\left(c\La^{\frac{1}{4}}R\right)},
  \end{equation}
  where $C$ is a constant. 
\end{itemize}

\section{Conclusion}

In this chapter we have demonstrated, how to use discretized systems to describe
two-dimensional quantum gravity. In two dimensions, the continuum theory of
quantum gravity and the theory of dynamical triangulations agree with each other
in that region, where the continuum theory can be evaluated, that means for
matter with central charge $D\leq 1$ coupled to gravity.
The continuum limit of dynamical triangulations in two dimensions can therefore
be identified with the continuum theory.

In fact, the discretized approach is more powerful than the continuum approach
in the sense that natural observables like the Hartle-Hawking wavefunctionals can
be computed. Also the two-point function for pure quantum gravity 
can be obtained. This object is of central interest when
questions about the scaling of the theory are to be addressed.

On the technical side we have reviewed matrix models, which have developed
a status of interest on their own beyond dynamical triangulation in string theory
and in condensed matter theory.

Since there is at present no successful way to find a continuum theory of quantum
gravity in higher dimensions than two, it is natural to ask, whether such a theory
could be defined as the scaling limit of the corresponding higher dimensional
discretized theory. It turns out that dynamical triangulation can be defined
in three and four dimensions and that these theories have a phase transition
similar to the two-dimensional case.
While there has been some analytical progress in the understanding of entropy bounds
on the number of triangulations in higher dimensions~\cite{Carfora:1997} most work
in this field is numerical. Questions about the nature of the phase transition
in four dimensions are not settled yet and exciting research remains to be done.

\chapter{The failure of quantum Regge calculus}
\label{sec:regge}

In the continuum path integral formulation of quantum gravity
we are instructed to compute the integral
\begin{equation}
  \label{in1}
  \int\!\cD[\gmn] e^{-S_{\text{EH}}(g)}
  \equiv \int\!\frac{\cD\gmn}{\text{vol(Diff)}} e^{-S_{\text{EH}}(g)}
\end{equation}
over diffeomorphism classes of metrics weighted
with the exponential of the Einstein-Hilbert action, as has been
discussed in chapter \sref{chap1}.
Two different discretization schemes have been suggested. One,
dynamical triangulation, provides a regularization of the functional
integral~\eref{in1}. An explicit cutoff is introduced and one sums
over all (abstract) triangulations with
equilateral $d$-simplices of a $d$-dimensional manifold $M$,
compare chapter \sref{disc}.

A different attempt to discretize the functional integral~\eref{in1},
called quantum Regge calculus (QRC), has been suggested, see
\cite{Hamber:1984tm,Williams:1992cd,Williams:1996jb} for
reviews and references. This formalism is closely related to the classical
coordinate independent Regge discretization of general relativity:
One {\em fixes} a (suitably chosen) triangulation, while the dynamical
degrees of freedom are the $L$ link lengths. Thus the functional integration
in~\eref{in1} is replaced by
\begin{equation}
  \label{in2}
  \int\! d\mu(l_1,\ldots,l_L)\equiv
  \int_0^{\infty}\! \prod_{i=1}^L dl_i\ J(l_1,\ldots,l_L)\ \de(\De),
\end{equation}
where $J(l_1,\ldots,l_L)$ is the Jacobian of the transformation
$d\gmn(\xi)\rightarrow dl_i$. The integral in~\eref{in2} is over
all link length assignments consistent with the triangle inequalities,
as denoted by the delta function $\de(\De)$.
While it provides a discretization of the integration in~\eref{in1},
this replacement does not provide a regularization of~\eref{in1}.
Contrary to dynamical triangulation, no cutoff has to be
introduced~\cite{Hamber:1986gw}. One can of
course choose to work with a cutoff which has to be taken to zero at the
end of the computations. This will not make any difference in the
argumentation below.

The Jacobian in~\eref{in2} is very complicated and its form is
presently unknown.
For analytical and numerical investigations it is usually replaced by
local measures of the form $\prod_{i=1}^L f(l_i)$.

If a sensible continuum limit of this theory existed, it would be
approached by taking the number $L$ of links in the triangulation
to infinity at a critical point of the statistical ensemble defined
by~\eref{in2} and by the discretized Einstein Hilbert action. For this limit
to make sense, the resulting continuum theory should not
depend on details of the discretized theory like the chosen fixed triangulation
or the local measure $f(l_i)$.

It is one central objective of this chapter to show analytically
that such a continuum limit
of quantum Regge calculus in its present formulation cannot be defined in any
dimension $d>1$.

Such a result has been indicated by simulations in two-dimensional quantum
Regge calculus~\cite{Bock:1995mq,Holm:1995xr,Holm:1996fd}. These simulations revealed
that the KPZ-exponents~\eref{e3d} for the susceptibility of
conformal matter coupled to continuum quantum gravity could not be obtained.
This has often been called the failure of quantum Regge calculus. Most discussion
about this failure to reproduce continuum results in two dimensions has been
centered around the choice of the measure.

In the first part of this chapter some notation is introduced.
In the second part we discuss the measures which have
been suggested and used in the context of quantum Regge calculus.
In the third part we
show that in two dimensions all discussed measures for Regge calculus fail
to reproduce continuum results. Not even the concept of length can be
defined~\cite{Ambjorn:1997ub}.
In the conclusion we discuss the situation in higher dimensions.
We give a short review of an alternative approach, in
which the continuum functional integral over metrics is
restricted to
piecewise linear metrics~\cite{Menotti:1996de}.

This chapter is partly based on work presented in \cite{Ambjorn:1997ub}.



\section{Formulation of quantum Regge calculus}

Let us fix the connectivity of a triangulation with
$V$ vertices\footnote{We denote the volume of manifolds with the same
  symbol $V$. No confusion should arise.}, $T$ triangles and
$L$ links of lengths $l_1, \ldots, l_L$
of a two-dimensional closed manifold $M$ with
Euler-cha\-rac\-te\-ristic $\chi$.
The number of links can also be expressed as $L=3V-3\chi$.
We view the interior of the simplices as flat.
As explained in section \sref{dg}
the curvature of this piecewise linear space
is concentrated at the vertices and each assignment
of link lengths uniquely determines a metric on $M$.
For a triangle this metric can be written as 
\begin{equation}
  \label{r0c}
  \gmn = \left(
  \begin{array}[c]{cc}
    x_1 & \frac{1}{2}(x_1+x_2-x_3) \\
    \frac{1}{2}(x_1+x_2-x_3) & x_2
  \end{array}
  \right),
\end{equation}
where the squared edge lengths are defined as 
$x_i=l_i^2,\ i=1, \ldots, L$. The area $A$ of a single triangle
can be expressed as
\begin{equation}
  \label{r0d}
  A = \int\! d^2\xi \sqrt{g(\xi)} = \frac{1}{2}\sqrt{g} =
  \frac{1}{2}\sqrt{x_1x_2-\frac{1}{4}(x_1+x_2-x_3)^2}.
\end{equation}

Since in quantum Regge calculus the link lengths are the dynamical
variables one attempts to replace the functional integration over
equivalence classes of metrics by an integration
over all link length assignments which correspond to different
geometries. Clearly, this replacement
\begin{equation}
  \label{r1}
  \int\! \cD[\gmn] = \int\! \frac{\cD \gmn}{\text{Vol(Diff)}}
  \rightarrow
  \int\!\prod_{i=1}^L dl_i\ J(l_1, \ldots, l_L) \de(\De)
  \equiv \int\! d\mu(l_1,\ldots,l_L)
\end{equation}
involves a highly non-trivial Jacobian $J(l_1,\ldots,l_L)$. The integration is over
all link lengths compatible with the triangle inequalities. 
But not all assignments of link lengths define independent geometries as can
be seen in the flat case. All vertices can be moved around in the plane,
changing the link lengths without changing the flat geometry.
The Jacobian has to be such that this additional invariance is divided out of the
integral.

Thus, we obtain an $L=3V-3\chi$-dimensional subspace of the infinite dimensional
space of equivalence classes of metrics. Quantum Regge calculus replaces the
functional integration in~\eref{e1} by an integral over this finite dimensional
subspace. It is hoped that in the limit $L\rightarrow\infty$ expectation values
of observables converge in some suitable way to their continuum values.

\section{Regge integration measures}

Presently, the form of the Jacobian in~\eref{r1} is not known. However, by appealing
to universality one might assume that the choice of this measure is not very
important. A large number of measures have been suggested and tried out in numerical
experiments to study whether quantum Regge calculus in its present form agrees
with the results of continuum calculations and dynamical triangulations in
two dimensions. The result of these investigations is negative.
In this section we will define the most important and common of
these measures and study some of their
properties.

\subsection{DeWitt-like measures}

One possible way to construct a measure for quantum Regge calculus
in terms of the link lengths is to repeat the construction in section
\sref{fm} with the constraint that the deformations of the metric
are allowed by quantum Regge calculus. For an alternative discussion
see~\cite{Williams:1996jb} where
the corresponding metric is called Lund-Regge metric, refering to an
unpublished work by Lund and Regge. However, since the starting point
of our discussion is the DeWitt metric
we rather call the measure a DeWitt-like measure.
Note however that it 
is not the DeWitt measure restricted to piecewise linear metrics.
This will be given in section \sref{menotti}.

For a single triangle the variation of the metric~\eref{r0c} in terms
of the $\de x_i$ is given by
\begin{equation}
  \label{dw1}
  \de\gmn = \left(
  \begin{array}[c]{cc}
    \de x_1 & \frac{1}{2}(\de x_1+\de x_2-\de x_3) \\
   \frac{1}{2}(\de x_1+\de x_2-\de x_3) & \de x_2
  \end{array}
  \right).
\end{equation}
Using this and the DeWitt metric, the scalar product
$\langle\de g,\de g\rangle_T$, which has been defined in section \sref{fm}
can be computed for a single triangle with area $A$.
It turns out that it simplifies considerably when we use
the canonical value $-2$ for the parameter $C$ in the DeWitt metric:
\begin{eqnarray}
  \label{dw2}
  \langle\de g,\de g\rangle_T &=& \int\! d^2\xi \sqrt{g(\xi)}\
  \de\gmn(\xi) \GMNAB \de\gab(\xi)\nonumber\\
  &=& \frac{A}{2}\left(2\de\gmn g^{\mu\al} g^{\nu\be} \de\gab
    + C \de\gmn\gMN\gAB\de\gab\right)\nonumber\\
  &=& -2A\det(\de\gmn) \det(\gMN),~\text{for $C=-2$}.
\end{eqnarray}
Using~\eref{r0c} and~\eref{dw1} leads to
\begin{eqnarray}
  \label{dw3}
  \langle\de g,\de g\rangle_T 
&=& [\,\delta x_1, \delta x_2, \delta x_3\,]\:\frac{1}{16A}\!\left[
  \begin{array}{rrr} 1 & -1 & -1 \\ -1 & 1 & -1 \\ -1 & -1 & 1
  \end{array} \right] \left[ \begin{array}{c} \delta x_1 \\ \delta x_2
  \\ \delta x_3 \end{array} \right].
\end{eqnarray}
For a general two-dimensional triangulation this line element is given
by the sum of~\eref{dw3} over all triangles:
\begin{eqnarray}
  \label{dw4}
 \langle\de g,\de g\rangle_T &=& \sum_t\int d^2\xi \sqrt{g^t(\xi)}\,\delta
g^t_{\mu\nu}(\xi)(G^t)^{\mu\nu\alpha\beta}\delta g^t_{\alpha\beta}(\xi)
\nonumber \\
&=& [\,\delta x_1,\ldots,\delta x_L\,]\:{M}\left[
  \begin{array}{c} \delta x_1 \\ \vdots \\ \delta x_L \end{array} \right].
\end{eqnarray}
Here ${M}$ is an $L\times L$ matrix. ${M}_{ij}$ is the element
corresponding to the links $l_i$ and $l_j$. If both $l_i$ and $l_j$
belong to the same triangle $t$ with area $A_t$ the corresponding non-diagonal
entry is
\begin{equation}
  \label{dw5}
  {M}_{ij} = {M}_{ji} = -\frac{1}{A_t}.
\end{equation}
All other off-diagonal entries are zero.
It follows that ${M}$ has four nonvanishing off-diagonal entries in each row.
Diagonal entries are given by
\begin{equation}
  \label{dw6}
  M_{ii} =  \frac{1}{A_{t_1}} + \frac{1}{A_{t_2}}, 
\end{equation}
where $t_1$ and $t_2$ are the two triangles which contain $l_i$.
From the i'th row of ${M}$ the product $(A_{t_1}A_{t_2})^{-1}$ can
be factorized. Since each triangle has three sides, this amounts
to factorizing $\prod_{k=1}^T A_k^{-3}$ from the determinant of ${M}$:
\begin{eqnarray}
  \label{dw7}
   \det M &=& \prod_{k=1}^T A_k^{-3}
  \left\vert
    \begin{array}{ccccccc}
      A_1+A_2 & -A_2 & -A_2 & -A_1 & -A_1 & 0 & \ldots \\
      \mc{7}{c}{\ldots}
    \end{array}
    \right\vert \nonumber \\
    &=:&  P(A_1,\ldots,A_T) \prod_{k=1}^T A_k^{-3}.
\end{eqnarray}
$P(A_1 , \ldots ,A_T)$ is a polynomial in the areas of the triangles
which vanishes, whenever two adjacent triangle areas vanish. It follows
directly that $P$ is a highly nonlocal function of the areas, since each
monomial of $P$ must contain at least half of the triangles.

Therefore the DeWitt-like integration measure for quantum Regge calculus
is given by
\begin{equation}
  \label{dw8}
   d\mu(l_1,\ldots,l_L) = \text{const}\times
  \frac{\sqrt{P(A_1 , \ldots ,A_T)}}{\prod_{k=1}^T A_k^{3/2}}
  \prod_{j=1}^L l_jdl_j\ \de(\De).
\end{equation}
On a first glance there are similarities with the continuum measure
\eref{meas1.15}. Using $g(\xi) \sim A^2$ we see that the power
of the ``local areas'' in both measures equals $-3/2$. However, reparametrization
invariance fixes the continuum measure completely, while in the discretized
version~\eref{dw3} might be multiplied by the diffeomorphism invariant
area factor $A^{1-2\be/3}$, leading to the measure
\begin{equation}
  \label{dw9}
    d\mu(l_1,\ldots,l_L) = \text{const}\times
  \frac{\sqrt{P(A_1^{\frac{2\be}{3}} ,
      \ldots ,A_T^{\frac{2\be}{3}})}}{\prod_{k=1}^T A_k^{\be}}
  \prod_{j=1}^L l_jdl_j\ \de(\De). 
\end{equation}
This generalization stresses the fact that the power of the areas is not
fixed in quantum Regge calculus by any obvious principle
since the areas are diffeomorphism invariant
quantities with no direct local interpretation in the continuum.

The measure~\eref{dw9} is highly nonlocal and thus
not suited for numerical simulations.
Below we will show that it does not reproduce continuum physics.

\subsection{The DeWitt-like measure in other dimensions than two}
\label{sec:dwone}

We can perform the derivation of the DeWitt-like measure in one and
in higher dimensions. A one-dimensional piecewise linear
manifold consists of $L$ links of lengths $l_i~(x_i=l_i^2), i=1,\ldots,L$,
which are glued together at their ends. 
The canonical metric is thus given by $\gmn^i = (x_i)$.
Therefore the DeWitt metric gives:
\begin{equation}
  \label{od1}
  \langle\de g,\de g\rangle_T = (1+\frac{C}{2})\sum_{i=1}^L x_i^{-\frac{3}{2}} (\de x_i)^2.
\end{equation}
Thus the DeWitt-like measure in one dimension is
\begin{equation}
  \label{od2}
  d\mu(l_1,\ldots,l_L) = \text{const}\times \prod_{i=1}^L \frac{dl_i}{l_i^{\frac{1}{2}}}.
\end{equation}
In this case the measure is local. The factor $(1+\frac{Cd}{2})$ factorizes as in the
continuum. Note that in $d=1$ we cannot use the canonical value $C=-2$ for $C$ since
the DeWitt metric would be singular.

In higher dimensions the situation is more complicated. 
For a $d$-simplex the measure is derived from the scalar product
\begin{equation}
  \label{hd1}
  \langle\de g,\de g\rangle_T = \de x {M}_d \de x,
\end{equation}
where $\de x$ is the $\frac{d(d+1)}{2}$-dimensional vector associated
with link length deformations of the $d$-simplex. As in two dimensions,
the complete norm for the triangulation can be written as a straightforward
superposition of the matrices ${M}_d$. However, in dimensions larger than two,
the entries of this matrix ${M}_d^{\text{total}}$ depend on the $x_i$'s.
In general, the measure is given by
\begin{equation}
  \label{hd1a}
  d\mu(l_1,\ldots,l_L) = \text{const}\times\sqrt{\det{{M}_D^{\text{total}}}}
  \prod_{i=1}^L l_i dl_i\ \de(\De),
\end{equation}
where $\de(\De)$ now stands for the generalized triangle inequalities.
As an example we have computed the measure for the simplest closed three-dimensional
manifold which consists of two tetrahedra glued together along the six links.
The result is
\begin{equation}
  \label{hd2}
  d\mu(l_1,\ldots,l_6) = \text{const}\times \frac{1}{V}
  \prod_{i=1}^6 l_idl_i\ \de(\De),
\end{equation}
where $V$ is the $3$-volume of the manifold.

Note that also in four dimensions,
where the DeWitt metric is simply $\prod_{\xi\in M}\prod_{\mu\leq\nu}d\gmn$,
the determinant of ${M}_d^{\text{total}}$ does not evaluate to a constant,
not even for the simplest $4$-geometries.

\subsection{The DeWitt-like measures for special geometries}
\label{sec:DWspecial}

We have not been able to obtain
a general closed expression for the polynomial $P$. However, in a number of special
cases it is possible to find explicit expressions for the measure~\eref{dw8}
which can immediately be generalized for the measure~\eref{dw9}~\cite{Ambjorn:1997ub}.

The easiest case of a closed piecewise linear manifold are two triangles
of area $A$ glued together along the three links.
The measure is
\begin{equation}
  \label{dw10}
  d\mu(l_1,l_2,l_3) = \text{const}\times\frac{1}{A^{\frac{3}{2}}}
  \prod_{j=1}^3 l_jdl_j\ \de(\De).
\end{equation}

A tetrahedron has six links and four faces. The determinant of the
$6\times 6$ matrix ${M}$ can be computed. The measure is given by
\begin{equation}
  \label{dw11}
  d\mu(l_1,\ldots,l_6) = \text{const}\times
  \frac{\sum_{i=1}^4 A_i}{\prod_{i=1}^4 A_i}
  \prod_{j=1}^6 l_jdl_j\ \de(\De).
\end{equation}

\begin{figure}[htbp]
  \begin{center}
    \includegraphics[width=0.4\linewidth]{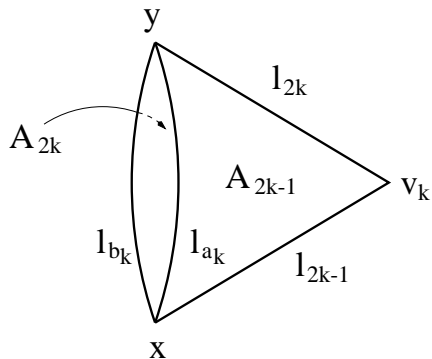}
    \parbox[t]{.85\textwidth}
    {
      \caption[sing]
      {
        \label{fig:hat}
        \small
        Building blocks for the hedgehog geometry which
        consists of $K$ of these blocks glued together along the links
        $l_{a_k}$ and $l_{b_k}$.
        }
      }    
  \end{center}
\end{figure}
Several other more complicated examples are possible. As a last example
we display the measure for a hedgehog geometry which consists of $K$
hat-like building blocks such as the one shown in figure \fref{hat}. This block
consists of two triangles with areas $A_{2k-1}$ and $A_{2k}$ respectively,
glued together along the links $l_{2k-1}$ and $l_{2k}$.
The hedgehog geometry is built by gluing these elements together along
the links $l_{b_k}$ and $l_{a_{k+1}}$ successively. After this gluing
the vertices $x$ and $y$ have the order $2K$, while the other vertices
are of order two.
Although this geometry might
seem a bit artificial it plays an important role in the analysis of
the phase
transitions of two dimensional quantum gravity coupled to
matter fields~\cite{Ambjorn:1986dn}. 
The matrix ${M}$ for this geometry can be written in a block diagonal
form. Its determinant can be computed, resulting in the measure
\begin{eqnarray}
  \label{dw12}
   d\mu(l_1,\ldots,l_{3K}) & = & {\rm const}\times\frac{\prod_{j=1}^{3K} l_j
    dl_j}{\prod_{k=1}^{2K}A_k^{3/2}}\prod_{k=1}^K\left(A_{2k-1}+A_{2k}
      \right)^{1/2} \nonumber \\
      & & \times \left|\,\prod_{k=1}^K A_{2k}+(-1)^{K-1}\prod_{k=1}^K
        A_{2k-1}\,\right|. 
\end{eqnarray}
For an even number of elements this measure vanishes whenever the product
over the even areas equals the product over the odd areas. For odd $K$, the
measure is always positive. Clearly, this points to a severe sickness of
the measure~\eref{dw8}.

\subsection{Commonly used measures}
\label{sec:cum}

As has been mentioned above, the DeWitt-like measures~\eref{dw9} are
not suited for numerical simulations. Also, we have demonstrated, that
the measure in quantum Regge calculus is not fixed as in the continuum
case. 
Therefore it is natural to search for other, local measures for which
a sensible continuum limit can be defined. The principle of universality
should ensure that all reasonable choices of the measure result
in the same continuum theory. 

For numerical simulations in two-dimensional
quantum Regge calculus, a number of local measures
have been suggested~\cite{Hamber:1984tm,Bander:1986kd,Gross:1990fq}.
In all cases these are
of the form
\begin{equation}
  \label{a1}
  d\mu(l_1,\ldots,l_L) = \frac{1}{\prod_{j=1}^T A_j^{\be}}
  \prod_{i=1}^L l_i^{-\al}dl_i\ \de(\De),
\end{equation}
for some parameter $\al$ or $\be$, or derived from this\label{p1}\footnote{Sometimes the product
  over all triangle areas is replaced by the product over all areas assigned to the links.
  In our arguments this is not important and corresponds to choosing another value for
  the parameters $\al$ and $\be$.}. 
These measures are arrived at by intuitively translating the continuum measure
$\prod_{\xi\in M}\prod_{\mu\leq\nu}d\gmn$ as $\prod_{i=1}^L l_i dl_i$ and
by replacing the product $\prod_{\xi\in M} g(\xi)^{\sig}$ by the product
of the ``local'' areas $\prod_{k=1}^T A_k^{2\sig}$. The most popular
choices are the scale invariant measure $\prod_{i=1}^L\frac{dl_i}{l_i}$
and the uniform measure $\prod_{i=1}^Ldx_i\sim \prod_{i=1}^Ll_idl_i$.

Motivation for the first of these measures is often suggested by an analogy to the
scale invariant continuum-measure
\begin{equation}
  \label{a3}
  \cD\gmn = \prod_{x\in M} g(x)^{-\frac{d+1}{2}}\prod_{\mu\leq\nu} d\gmn(x),
\end{equation}
which has been advocated for four-dimensional quantum gravity in~\cite{Misner,FP}.
However, this measure is not diffeomorphism invariant, as has been stressed in
section \sref{fm}, although the opposite has been claimed and ``proved''
by wrong formal arguments~\cite{FP,Hamber:1997ut}.
The DeWitt measure is the
only diffeomorphism invariant measure for continuum quantum
gravity.
Therefore the measure~\eref{a3}
cannot replace the DeWitt measure in continuum quantum gravity and
any translations to discretized versions should be avoided.

Also the DeWitt-like measure~\eref{dw8} cannot be used to motivate~\eref{a1}
in dimensions higher than one, since~\eref{dw8} is highly nonlocal while
the measures~\eref{a1} are local. Note that also in four dimensions, where the
DeWitt measure is simply
\begin{equation}
  \label{a4}
  \prod_{x\in M}\prod_{\mu\leq\nu} d\gmn(x),
\end{equation}
the discrete DeWitt-like measure is very complicated and highly nonlocal and does
not equal the uniform measure~\eref{a1} with $\al=-1$ and $\be=0$, not even for the
simplest 4-geometries. This clearly shows that translations like
\begin{equation}
  \label{a5}
  \prod_{x\in M}g(x)^{\sig}\prod_{\mu\leq\nu} d\gmn(x)
  \rightarrow \prod_{j=1}^T A_j^{2\sig}\prod_{i=1}^L l_idl_i\ \de(\De)
\end{equation}
or similar
are far too simple minded, since the Jacobian of this variable transformation
is not taken into account. The correct treatment of the diffeomorphisms
and the Jacobian leads to the results in section \sref{menotti}.

However, it has been argued that, appealing to universality, the
precise form of the measure need not to be known.
In the continuum limit, averages of observables should result in their continuum
values independent of details of the discretization like
the exponents $\al$ and $\be$ in~\eref{a1} or the underlying connectivity of the
piecewise linear space. But numerical simulations revealed
that averages of observables in the context of quantum Regge calculus
\begin{itemize}
\item do not reproduce results of continuum quantum gravity. In two dimensions
  not even the KPZ exponents~\eref{e3d} could be obtained.
\item show a significant dependence on the measure.
\item show a significant dependence on the underlying triangulation.
\end{itemize}
These are serious flaws in the Regge approach to quantum gravity. 

\subsection{Numerical simulations in two-dimensional quantum Regge calculus}
\label{sec:sim}

Quantum Regge calculus has repeatedly been used for
numerical simulations in two, three and four dimensions.
However, its status as a theory of quantum gravity remained unclear even in two dimensions.
To clarify this situation, two groups have recently performed
large scale numerical investigations in two dimensions where it is possible to
compare with continuum results.

Bock and Vink~\cite{Bock:1995mq} have performed a Monte Carlo simulation of
two-dimensional Regge calculus applied to pure gravity. They added the term
\begin{equation}
  \label{bv1}
  \be \int_{M}\! d^2\xi\sqrt{g}\ \cR^2
\end{equation}
to the action. Such curvature square terms have been introduced in the
context of four-dimensional quantum gravity
to overcome  the unboundedness
of the Einstein-Hilbert action from below, which is induced by conformal
fluctuations~\cite{Gibbons:1978ab}. In two dimensions this term is the only nontrivial
part of the action. Its discretized version can be written as
\begin{equation}
  \label{bv2}
  \be \sum_{i=1}^V \frac{\de_i^2}{A_{(i)}},
\end{equation}
where $\de_i$ is the deficit angle at the vertex $i$ and $A_{(i)}$ is the area
assigned to the vertex $i$ by a barycentric division of the triangle areas
around $i$. The addition of this term to the action allows a scaling
analysis which gives the susceptibility exponent $\ga$.
This exponent can for example be defined as an entropy exponent for the
partition function
\begin{equation}
  \label{bv2a}
  Z(V) \propto V^{3-\ga} e^{\La_c V},
\end{equation}
where the dominant exponential is nonuniversal and the universal part is given
by the subleading power correction.
Continuum calculations
show, that for pure gravity $\ga$ is given by
\begin{equation}
  \label{bv3}
  \ga = 2 - \frac{5}{2}(1-h),
\end{equation}
compare~\eref{e3d}.
It has been shown in~\cite{David:1985nj,Boulatov:1986jd}, that the addition of a small
curvature square term does not change the value of $\ga$ in dynamical triangulation.

To be conservative, Bock and Vink used the scale invariant measure~\eref{a1} with
$\al=1$ and $\be=0$ for their simulations. This enabled them to compare their results
with previous calculations~\cite{Gross:1990fq}, where agreement between
quantum Regge calculus and the formula~\eref{bv3} has been reported.
They computed the susceptibility exponent for the topology of a sphere ($h=0$),
a torus ($h=1$) and a bitorus, that means a sphere with two handles ($h=2$).
\begin{table}
\renewcommand{\baselinestretch}{1.2}
  \normalsize
  \begin{center}
    \begin{tabular}{llll}
      \hline\hline\hline
      $h$ & $0$ (sphere) & $1$ (torus) & $2$ (bitorus)\\ \hline
      $\ga$ in Liouville theory& $-0.5$ & $2$ & $4.5$ \\
      $\ga$ in QRC from~\cite{Bock:1995mq} & $\gtrapprox 5.5$ & $\approxeq 2.0$ &
      $\gtrapprox 5.5$\\ \hline\hline\hline      
    \end{tabular}
    \renewcommand{\baselinestretch}{1.0}
    \normalsize
    \parbox[t]{0.85\textwidth}
    {
      \caption[bv1]
      {\label{tab:bv1}
        \small
        Values for the susceptibility exponent $\ga$ for pure two-dimensional
        gravity at different topologies as computed in Liouville theory compared
        to the values from the Monte Carlo simulation in~\cite{Bock:1995mq}.
        Bock and Vink do not give error bars for their values. However,
        the deviation from the Liouville values is significant.
        }
      }
  \end{center}
\end{table}
The results are compared with the continuum values in table \tref{bv1}.
For spherical and bitoroidal topology they found that $\ga$ in quantum Regge
calculus differs significantly from Liouville theory. That means that quantum
Regge calculus does not reproduce the fundamental KPZ-result~\eref{bv3}.
Note that the torus topology is not well suited to test the KPZ formula~\eref{e3d}.
For $h=1$ $\ga$ is independent of the conformal charge of the matter coupled to gravity.
Furthermore $\ga$ takes the same value $2$ in the classical case without
metrical fluctuations ($c=-\infty$) and in the quantum case.

The results of Bock and Vink have first been confirmed~\cite{Holm:1994un} and later
been questioned~\cite{Holm:1995kq,Holm:1996kw} by Holm and Janke. Using a very
careful finite size scaling ansatz for pure quantum gravity
on a fixed random triangulation of the sphere, they arrive
at the value $\ga=-10(2)$ for the susceptibility exponent~\cite{Holm:1996fd}, in the same
setup. This is in clear disagreement with the prediction from Liouville theory and with
the computations in~\cite{Bock:1995mq,Gross:1990fq}.

Using an alternative measure similar to~\eref{a1} with $\al=0$ and
$\be=\frac{1}{2}$ it was found in~\cite{Holm:1994un} that a scaling can no longer
be observed and one encounters numerical difficulties. The observables
did not reach equilibrium values. A similar change in the scaling behaviour
was observed in~\cite{Bock:1995mq}. The value of $\ga$ is changed when the
scale invariant measure is replaced by $\prod_{i=1}^L \frac{dl_i}{l_i}l_i^{\zeta}$,
for $\zeta\neq 0$.
Furthermore it has been demonstrated in~\cite{Holm:1996fd},
that the results of quantum Regge calculus depend on the chosen
fixed triangulation.

\section{The appearance of spikes}
\label{sec:spikes}

In this section we will show, that two-dimensional quantum Regge calculus
does not reproduce results of continuum quantum gravity with any of the
proposed measures mentioned above~\cite{Ambjorn:1997ub}.

Let us recall that the partition function $G(V,R)$ for closed two-dimensional universes
with fixed volume $V$ and with two marked points separated a geodesic
distance $R$ has the asymptotic behaviour~\eref{e16g}
\begin{equation}
  \label{as1}
  G(V,R) \sim V^{-1/4} e^{-\left(\frac{R}{V^{1/4}}\right)^{\frac{4}{3}}},~\text{
    for}~\frac{R}{V^{1/4}}\rightarrow\infty.
\end{equation}
Thus the expectation value of any power of the radius $R$ can be calculated as
\begin{equation}
  \label{as2}
  \langle R^n \rangle = \int_0^{\infty}\! dr\ r^n G(V,r) \sim
  V^{n/4}.
\end{equation}
We can show that~\eref{as2} is not fulfilled in quantum Regge calculus.

The physical reason is, that contrary to continuum quantum gravity in two
dimensions, the volume, or equivalently the cosmological constant, does
{\it not} set an intrinsic scale for the theory.
It follows that the concept of length has no natural definition in this
formalism and that every generic manifold degenerates into spikes.

\subsection{General proof of appearance}

\begin{theorem}
  \label{th:1}
  For any value of $\al$ or $\be$ in~\eref{dw9} and~\eref{a1}
  there exists an $n$ such that for any link $l$ in a given arbitrary
  triangulation
  \begin{equation}
    \label{as3}
    \langle l^n \rangle_V = \infty
  \end{equation}
  for any fixed value $V$ of the spacetime volume. Thus the average radius
  $\langle R\rangle$, or some suitable power $\langle R^n\rangle$, does
  not exist~\cite{Ambjorn:1997ub}.
\end{theorem}
\vspace{2mm}
\noindent
{\bf Proof:}~In the situation of figure \fref{supplfig},
\begin{figure}[htbp]
  \begin{center}
    \includegraphics[width=0.45\linewidth]{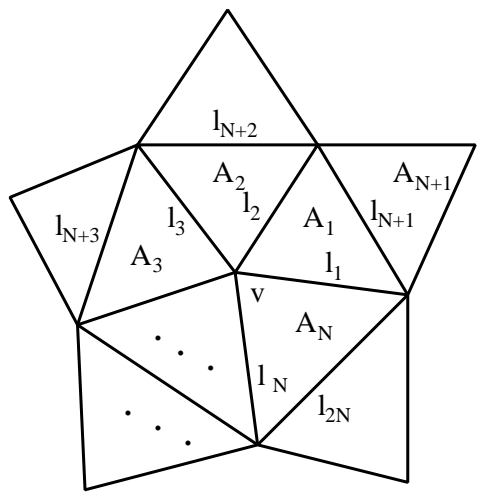}
    \parbox[t]{.85\textwidth}
    {
      \caption[supplfig]
      {
        \label{fig:supplfig}
        \small
        Parametrization of the link lengths and the triangle areas in a part of
        the piecewise linear manifold $M$. The vertex $v$ is
        of order $N$. A spike can be formed by making the links
        $l_1,\ldots,l_N$ arbitrary large. When the area of the
        surface is fixed, the link lengths $l_{N+1},\ldots,l_{2N}$
        have to be small of order $1/l_1$.
        }
      }    
  \end{center}
\end{figure}
which displays a small part of the piecewise linear
manifold $M$,
the vertex $v$ has the arbitrary order $N$. The links connected to
$v$ are labelled as $l_1,\ldots,l_N$, while the links opposite to $v$ are
labelled as $l_{N+1},\ldots,l_{2N}$. We want to analyze the Regge integral
in that part of the phase space where the vertex $v$ is pulled to
infinity and forms a spike while the volume of $M$ is held bounded.
In this situation the link lengths $l_1,\ldots,l_N$ are large, while
the link lengths $l_{N+1},\ldots,l_{2N}$ have to be of the order $l_1^{-1}$.

First we analyze the measure~\eref{a1} with $\be=0$.
Let $\La$ be a large number. Then $l_1$ can be integrated freely between
$\La$ and infinity, while the integration over $l_2,\ldots,l_N$ is restricted
by triangle inequalities. The integral over $l_2,\ldots,l_N$ gives a factor
$l_1^{(1-N)\al}(l_{N+1}\ldots l_{2N})^{\frac{N-1}{N}}$ after symmetrizing over
$l_{N+1},\ldots,l_{2N}$. An additional factor $l_{N+1}\ldots l_{2N}$ comes
from the triangle inequalities for the adjacent triangles which contain the
links $l_{N+1},\ldots,l_{2N}$. Taking these factors together results in
\begin{equation}
  \label{as}
  \int_{\La}^{\infty}\! dl_1\ l_1^{-N\al} \int_0^{\frac{\La}{l_1}}\!
  dl_{N+1}\ldots dl_{2N}\ (l_{N+1}\ldots l_{2N})^{\frac{2N-1}{N} - \al},
\end{equation}
which exists if
\begin{equation}
  \label{as5}
  \al < 3 - \frac{1}{N}.
\end{equation}
If that is fulfilled, the integral over $l_{N+1}\ldots l_{2N}$ can be performed,
which gives
\begin{equation}
  \label{as6}
  \int_{\La}^{\infty}\! dl_1\ l_1^{-N\al}
  \left(\frac{\La}{l_1}\right)^{N\left(\frac{3N-1}{N}-\al\right)}
  \sim \int_{\La}^{\infty}\! dl_1\ l_1^{1-3N}.
\end{equation}
This means that $\langle l_1^{n}\rangle = \infty$ for all $n\geq 3N-2$, which
proves the theorem~\ref{th:1} for the measure~\eref{a1} with $\be=0$..

To discuss the general form of~\eref{a1} we have to parametrize
the areas $A_1,\ldots,A_{2N}$.
In our situation these are given as the products of a small
and a large link length if $v$ is pulled to infinity. Thus the product
${\prod_{j=1}^T A_j^{-\be}}$  gives an extra factor
$(l_1\ldots l_N)^{-\be}(l_{N+1}\ldots l_{2N})^{-2\be}$. After integrating
over $l_2,\ldots,l_N$ and after taking the triangle inequalities into account we get:
\begin{equation}
  \label{as7}
  \int_{\La}^{\infty}\! dl_1\ l_1^{-N(\al+\be)}
  \int_0^{\frac{\La}{l_1}}\! dl_{N+1}\ldots dl_{2N}\
  (l_{N+1}\ldots l_{2N})^{\frac{2N-1}{N}-\al-2\be},
\end{equation}
which exists if $\al+2\be<3-\frac{1}{N}$. Then the integrals can be computed to give
\begin{equation}
  \label{as8}
  \int_{\La}^{\infty}\! dl_1\ l_1^{1+N(\be-3)}.
\end{equation}
Thus for all $n\geq N(3-\be)-2$ the expectation value of $l_1^n$ is infinite.

For the measure~\eref{dw9} the same analysis can be repeated if one notes,
that from the nonlocal factor $\sqrt{P(A_1^{2\be/3},\ldots,A_T^{2\be/3})}$
the product $\prod_{i=1}^L l_i^{\be/3}$ can be split of. It turns out that this
doesn't change the result and we get again~\eref{as8}.
This completes the proof. We denote $\langle l^n\rangle=\infty$ as the
appearance of spikes.

\subsection{Spikes in special geometries}

Clearly, if a discretized theory shall make any sense,
the results should not depend on details like the chosen triangulation.
In fact however, the proof of theorem \ref{th:1} shows,
that $\langle l^n\rangle$ depends
crucially on the order of the vertices, to which the link $l$ is attached.
To illustrate this dependence of quantum Regge calculus on the
underlying triangulation further, we analyze some special geometries in the
same way as in the proof to theorem \ref{th:1}. For some geometries even
the partition function is ill defined for some measures, while for other measures or
geometries $\langle l^n\rangle$ might be finite for some values
of $n$.

\subsubsection{Spikes in a hexagonal geometry}

\begin{figure}[htbp]
  \begin{center}
     \includegraphics[width=0.7\linewidth]{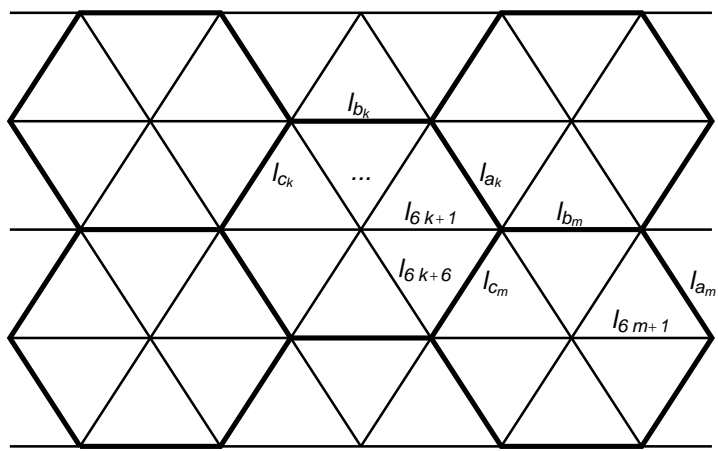}
    \parbox[t]{.85\textwidth}
    {
      \caption[hex]
      {
        \label{fig:hex}
        \small        
        Parametrization of a hexagonal geometry; $k\neq m$. All centers
        of the cells, which are marked by thicker lines form spikes.
        }
      }    
  \end{center}
\end{figure}
We analyze a regular triangulation with torus topology
in which all vertices have the
order $6$, consisting of $K$ hexagonal cells and parametrized as in
figure \fref{hex}.
We want to study the measure~\eref{a1} in that part of the configuration space, where
all centers of cells form spikes while the total area
of the surface is held bounded to prevent exponential damping from
the action. That means that the $6K$ link
lengths $l_i, i=1,\ldots 6K$ are very large while the $3K$ link lengths
$l_{a_k},l_{b_k},l_{c_k}, k=0,\ldots,K-1$ are very small, of order $l_1^{-1}$.
In cell number $k$ one link, say $l_{6k+1}$ can be integrated freely
from a large number $\La$ to infinity, while the integration
over $l_{6k+2},\ldots,l_{6k+6}$ is then constrained by the triangle
inequalities.
Integrating these out and symmetrizing over the small links
$l_{a_k},l_{b_k},l_{c_k}, k=0,\ldots,K-1$ gives
\begin{equation}
  \label{sg1}
  \int_{\La}^{\infty}\! \prod_{k=1}^K l_{6k+1}^{-6(\al+\be)}
  \int_0^{\frac{\La}{l_{6k+1}}}\! \prod_{k=1}^K
  (l_{a_k}l_{b_k}l_{c_k})^{\frac{5}{3}-\al-2\be} dl_{a_k}dl_{b_k}dl_{c_k}.
\end{equation}
Thus the integral over $l_{a_k},l_{b_k},l_{c_k}$ only exists if $\al+2\be<\frac{8}{3}$.
Furthermore $\langle l_1^n\rangle=\infty$ if $n\geq 7+3\al$.

For the measure~\eref{dw9} we extract a factor $\prod_{i=1}^L l_i^{\frac{\be}{3}}$
from the nonlocal factor. Performing the analysis gives the bound
$\be<\frac{11}{5}$ on $\be$. $\langle l_1^{n}\rangle$ is infinite
if $n\geq 4-\be$. These results are summarized in table \tref{bounds}.

\subsubsection{Spikes in a $12$-$3$ geometry}

Similarly we can analyze the integration measures for quantum Regge calculus
for a triangulation of the torus where two thirds of the vertices have order
\begin{figure}[htbp]
  \begin{center}
    \includegraphics[width=0.7\linewidth]{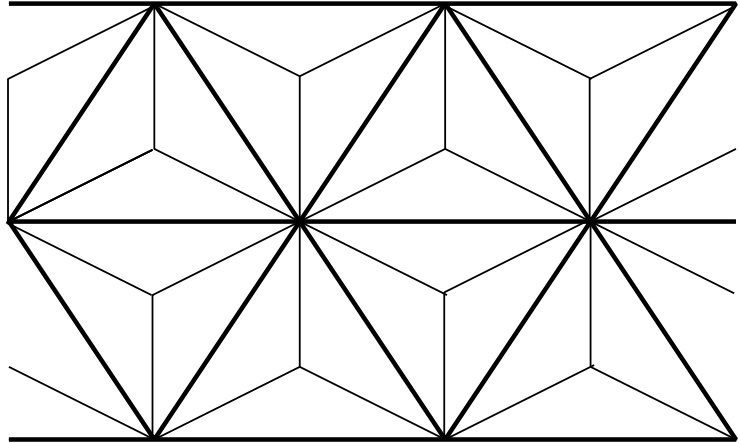}
    \parbox[t]{.85\textwidth}
    {
      \caption[12-3]
      {
        \label{fig:12-3}
        \small
        Illustration of the $12$-$3$ geometry. All vertices
        of order $3$ form spikes such that the total area of
        the surface is held bounded from above, to prevent
        exponential damping from the action.
        }
      }    
  \end{center}
\end{figure}
$3$ and one third of the vertices have order $12$, see figure \fref{12-3}.
We consider that part of the configuration space where all vertices
of order $3$ form spikes. The bounds on the exponents $\al$ and $\be$
in~\eref{a1} and~\eref{dw9} are depicted in table \tref{bounds}.
The resulting bounds are sharper because the order of the vertices
which form spikes is lower which leaves less geometrical restriction.
For $\al\leq -1$ the the average of a link length cannot be defined
with the measure~\eref{a1} (independent of $\be$). The same holds true for
the measure~\eref{dw9} if $\be>0$.

\subsubsection{Spikes in degenerate geometries}

One can get sharper bounds for geometries with vertices of
order two, like the hedgehog geometry introduced
in section \sref{DWspecial}. As explained in that section, these
geometries do occur in phases of simplicial quantum gravity in the
framework of dynamical triangulation and it
would be unnatural to forbid their occurrence as basically local
parts of very large triangulations.

As a first example we analyze the situation for the geometry shown
\begin{figure}[htbp]
  \begin{center}
    \includegraphics[width=0.7\linewidth]{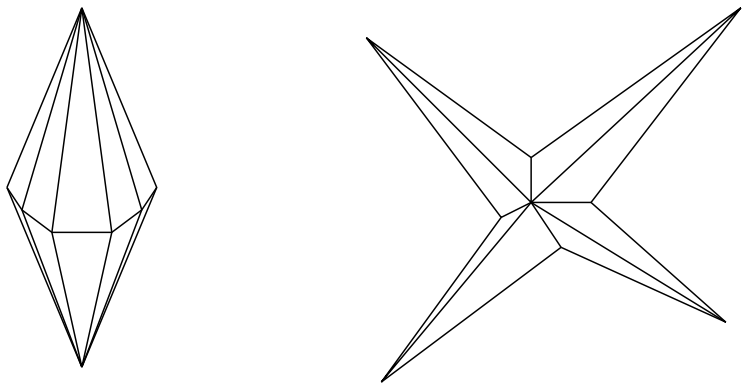}
    \parbox[t]{0.85\textwidth}
    {
      \caption[brill]
      {
        \label{fig:degenerate}
        \small
        A degenerate geometry which has two vertices
        of high order while all other vertices are of order $4$.
        Spikes can be formed in several ways. In the text we analyze
        the behaviour in that part of the phase space where every second
        vertex of order $4$ forms a spike.
        }
      }    
  \end{center}
\end{figure}
in figure \fref{degenerate} which has two vertices of high order while
all other vertices are of order $4$. This is slightly more regular
than the hedgehog geometry associated with figure \fref{hat}.
In table \tref{bounds} we present the results from our analysis of
the situation, where every second vertex of order $4$ forms a spike.
While with measure~\eref{a1} $\langle l^2\rangle$ is undefined for $\al\leq -1$,
the DeWitt-like measure~\eref{dw8} does not even
admit the computation of the average $\langle l\rangle$.

For the hedgehog geometry these measures themselves are illdefined,
see table \tref{bounds}.
Measure~\eref{a1} is illdefined for $\al=-1$ (the uniform measure),
while for instance for the scale invariant measure~\eref{a1} with $\al=1$ and $\be=0$
expectation values of $l^2$ and higher are infinite.
\begin{table}
\renewcommand{\baselinestretch}{1.2}
  \normalsize
  \begin{center}
    \begin{tabular}{lcll}
      \hline\hline\hline
      Geometry & Measure & Bounds &  Values of $n$ \\
               &         &        &  s.t.~$\langle l^n\rangle=\infty$\\ \hline
      hexagonal, &\eref{a1}&$\al+2\be<\frac{8}{3}$, $-\frac{7}{3}<\al$&$n\geq 7+3\al$\\
      see figure \fref{hex} &\eref{dw9}&$\be<\frac{11}{5}$&$n\geq 4-\be$\\ 
                                                                   &&&\\
      $12$-$3$, see &\eref{a1}&$\al+2\be<\frac{7}{3}$, $-\frac{5}{3}<\al$&$n\geq \frac{5}{2}+\frac{3}{2}\al$\\
      figure \fref{12-3} &\eref{dw9}&$\be<2$&$n\geq 1-\frac{\be}{2}$\\ 
                          &&&\\
      degenerate,&\eref{a1}&$\al+2\be<\frac{5}{2}$, $-2<\al$&$n\geq 4+2\al$\\
      see figure \fref{degenerate}&\eref{dw9}&$\be<\frac{21}{10}$&$n\geq 2-\frac{2\be}{3}$\\ 
                           &&&\\
      hedgehog, see&\eref{a1}&$\al+2\be<2$, $-1<\al$&$n\geq 1+\al$\\
      section \sref{DWspecial}&\eref{dw9}&$\be<0$&$n\geq -\frac{\be}{3}$\\ \hline\hline\hline
    \end{tabular}
    \renewcommand{\baselinestretch}{1.0}
    \normalsize
    \parbox[t]{0.85\textwidth}
    {
      \caption[bounds]
      {\label{tab:bounds}
        \small
        Bounds for the exponents $\al$ and $\be$ for the measures
       ~\eref{a1} and~\eref{dw9} and values of $n$
        such that $\langle l^n\rangle=\infty$ for various geometries.
        }
      }
  \end{center}
\end{table}

\section{Concluding remarks}

\subsection{The appearance of spikes in higher dimensions}

Although the derivation in the last section has been given only for the
two-dimensional case, we expect, that one has to deal with the same problems
in higher-dimensional quantum Regge calculus.
The measures \eref{a1} have obvious generalizations in higher dimensions.
An extra problem in
higher dimensions is the unboundedness of the action which can be
overcome by adding a cutoff involving for instance a curvature squared term.
However, if the cutoff is taken to zero, we expect the spikes to
reemerge. Actually, this has been observed in numerical simulations
\cite{Beirl:1992ap,Beirl:1994st}. In \cite{Beirl:1994st} four-dimensional
quantum Regge calculus is explored on a fixed but not regular triangulation
of the $4$-torus. The triangulation is constructed from a regular triangulation
with $31$ vertices following \cite{KuehnelLassmann} by inserting
three additional vertices with low order. While in the conventional
regular triangulation of the $4$-torus all vertices are of order $30$,
the additional vertices are of order $5$. The Einstein-Hilbert action plus
a cosmological constant term is discretized in the usual way. A cutoff is
introduced by the requirement that the local fatness
\begin{equation}
  \label{c1}
  \phi_s = 576 \frac{V_s^2}{\text{max}_{l_i\in s}(l_i^8)}
\end{equation}
for every $4$-simplex $s$ is held bounded away from zero,
$\phi_s\geq f=\text{const}>0$.  Here we have introduced the
$4$-volume $V_s$ of the simplex $s$. The fatness is maximal for
equilateral simplices and vanishes for collapsing ones.

For the integration over the link lengths $l_i$ the measure
\begin{equation}
  \label{c2}
  d\mu(l_1,\ldots,l_L) = \prod_{i=1}^L l_i^{2\sig-1} dl_i \de(\De)
\end{equation}
has been chosen for $\sig$ between $0$ and $1$. In figure 1 of \cite{Beirl:1994st}
the expectation value of $l_i^2$ is depicted as a function of the cutoff
$f$ at the values $f=2^m10^{-6},~m=9,\ldots,0$ for $\sig=1$, $\sig=\frac{1}{2}$
and $\sig=\frac{1}{10}$. For $\sig=1$ and $\sig=\frac{1}{2}$
this expectation value tends to infinity if the cutoff is removed while
for $\sig=\frac{1}{10}$ the situation cannot be decided from the numerical
data in \cite{Beirl:1994st}. From the figure we deduce that the behaviour
of $\langle l_i^2\rangle$ is a power of $f$:
\begin{eqnarray}
  \label{c3}
  \langle l_i^2\rangle &\sim& f^{-0.26(2)},~\text{for $\sig=1$,}\\
  \langle l_i^2\rangle &\sim& f^{-0.16(1)},~\text{for $\sig=\frac{1}{2}$}.
\end{eqnarray}
This clearly demonstrates the appearance of spikes in four dimensions
and shows that one has to expect the dependence of the physical results
on the choice of the measure and on the choice of the fixed triangulation
in four dimensions as well as in two dimensions.

\subsection{The role of diffeomorphisms in two dimensions}
\label{sec:menotti}

While we have shown that quantum Regge calculus in its present
form is not a suitable candidate for a theory of quantum gravity
it is an interesting question, whether one can develop a theory
of quantum gravity in the spirit of quantum Regge calculus.

A promising alternative approach to the functional integration~\eref{e1}
was studied in~\cite{Menotti:1995ih,Menotti:1996de,Menotti:1997tm}.
The integration over all metrics modulo diffeomorphisms is replaced
by the continuum functional integral over piecewise linear metrics
with a finite number of singularities modulo diffeomorphisms.
The singularities are the $V$ vertices of the piecewise linear space
at which the curvature is located.

This approach is remarkable since at each stage in the calculations
diffeomorphisms are treated exactly. Thus it includes all metrics
which are related to a piecewise linear metric by diffeomorphisms.
It turns out, that this is sufficient to reproduce a discretized form
of the Liouville action. Therefore it is believed, that for
$V\rightarrow\infty$ this discretized version approaches the continuum
expressions in some sense. 
No ad hoc assumption about the form of the measure has to be made.
Note that the connectivity of the
piecewise linear space is not fixed. In particular, all metrics
of dynamical triangulation are contained in this approach.

We will give a short review of these ideas for the two-dimensional
case.
As in the continuum calculations, the starting point is the DeWitt metric.
Since the reduction of the degrees of freedom involves only geometries
but not diffeomorphisms, the gauge fixing which leads to~\eref{z8}
can be performed in the same way as in the continuum. Now the integration
$\cD\phi$ over the conformal factor has to be restricted to those conformal
factors which describe Regge geometries. For spherical topology, to which we
confine the discussion here, there are no Teichm\"uller parameters and one can choose
a unique background metric $\hat{g}$. Menotti et al.~adopt the usual way of a
stereographic projection of the piecewise linear surface on the plane
with $\hat{g}_{\mu\nu} = \de_{\mu\nu}$~\cite{Forster:1987bw}. Then the conformal factor for
piecewise linear geometries can be parametrized as
\begin{equation}
  \label{m1}
  e^{\phi} = e^{2\la_0}\prod_{i=1}^V \vert z-z_i\vert^{2(\al_i-1)}.
\end{equation}
Here $2\pi\al_i$ is the angular aperture at the vertex $i$ while
the $z_i$ are the coordinates of the singularities in the complex plane.
The sum of the deficit angles $1-\al_i$ has to equal the Euler characteristic,
\begin{equation}
  \label{m2}
  \sum_{i=1}^V (1-\al_i) = 2,
\end{equation}
that means that $\al_V$ can be expressed by the other angles. $\la_0$ is
an overall scale factor. Note that this parametrization simply stems from
a generalization of the common Schwarz-Christoffel mappings which map the
complex upper half plane to an arbitrary polygon~\cite{Smirnov}.
One advantage of this parametrization over a parametrization via link lengths
is that there are no triangle inequalities.

Due to the presence of six conformal Killing vectors, compare section \sref{liouville},
the gauge fixing is
not complete and the conformal factor $\phi$ is invariant under an
$SL(2,\mathbb{C})$ transformation given by
\begin{equation}
  \label{m3}
  z' = \frac{az+b}{cz+d},~ad-bc=1,~a,b,c,d\in\mathbb{C}.
\end{equation}
A short calculation gives
\begin{equation}
  \label{m4}
  \phi'(z',\la_0, z_i, \al_i) = \phi(z',\la_0', z_i', \al_i),
\end{equation}
with
\begin{equation}
  \label{m5}
  \la_0' = \la_0 - \sum_{i=1}^V (1-\al_i)\log\vert cz_i+d\vert.
\end{equation}
Geometrical invariants as the angles $\al_i$ and the area
\begin{equation}
  \label{m6}
  A=e^{2\la_0}\int\! d^2z\ \prod_{i=1}^V \vert z-z_i\vert^{2(\al_i-1)}
\end{equation}
are invariant under the transformation~\eref{m3}. A counting
shows, that as in the usual parametrization
in Regge calculus there are $3V-6$ dynamical degrees of freedom.

To compute the Fadeev-Popov determinant for these conformal factors
one employs the same techniques as in the continuum. Namely one
computes the change of
\begin{equation}
  \label{m7}
  \log\frac{\det'(P^+P)}{\det\langle\om_a,\om_b\rangle_V}
\end{equation}
under a variation of the parameters $\la_0$, $z_i$ and $\al_i$ and
integrates the result back. One gets~\cite{Menotti:1996de}:
\begin{eqnarray}
  \label{m8}
  \lefteqn{
    \log\sqrt{\frac{\det'(P^+P)}{\det\langle\om_a,
      \om_b\rangle_V}}
  =}\\
  &&
  {\frac{26}{12}\left\lbrace
  \sum_{i, j\neq i} \frac{(1-\al_i)(1-\al_j)}{\al_i}
  \log\vert z_i-z_j\vert
  + \la_0\sum_{i}(\al_i-\frac{1}{\al_i}) - \sum_i F(\al_i)
  \right\rbrace}.\nonumber
\end{eqnarray}
This is the discretized form of the Liouville action. In the limit
$V\rightarrow\infty$ it 
approaches the continuum Liouville action.
The function $F$ depends only on the angles $\al_i$. For
$V\rightarrow\infty$, $\sum_{i=1}^VF(\al_i)$ goes to a topological
invariant.
The discretized Liouville action is invariant under the
$SL(2,\mathbb{C})$ transformation~\eref{m3}. Indeed, it has been shown
in~\cite{Menotti:1997xz} that this invariance induced by the conformal
Killing vectors is sufficient to derive the form~\eref{m8} of the
Liouville action up to the sum over the $F(\al_i)$. 
Note that the discretized
Liouville action is strongly unlocal.
Overcounting of equivalent locally flat geometries is avoided by
expression~\eref{m8}. Indeed, an angle $\al_i=1$ simply does not
contribute to~\eref{m8}.

The integration measure for the conformal factor is defined
by the scalar product $ds^2 = \int\! d^2z\ e^{\phi}\ \de\phi\de\phi$.
The change to the variables $(q_1,\ldots,q_{3V})=
((z_1)_x,\ldots,(z_V)_y, \la_0, \al_1,\ldots,\al_{V-1})$ involves a Jacobian:
\begin{equation}
  \label{m9}
  \cD_{e^{\phi}\de_{\mu\nu}}\phi = \prod_{i=1}^V d^2z_i
  \prod_{j=1}^{V-1}d\al_j d\la_0\ \sqrt{\det J},
\end{equation}
which is given by the determinant of 
\begin{equation}
  \label{m10}
  J_{ij} = \int\! d^2z\ e^{\phi} \frac{\partial\phi}{\partial q_i}
  \frac{\partial\phi}{\partial q_j},
\end{equation}
with the conformal factor $\phi$ defined by~\eref{m1}.
The diagonal entries $J_{ii}$ for $i=1,\ldots,2V$ of this matrix
require some  regularization. 
The explicit form of the measure is unfortunately unknown even
for the simplest geometries which makes the approach presently
inaccessible for numerical tests.
While it has been possible to extract the dependence of
the Jacobian on the scale factor $\la_0$, almost nothing is known about
the other variables.
It is an interesting open problem to show that the conformal anomaly
arises from this measure.

\subsection{Conclusion}

We have shown that in two-dimensional quantum Regge calculus the expectation value
of a suitable power of a single link length diverges. 
No natural scale is set by the measures \eref{a1} or \eref{dw9}.
Even if the total area is kept bounded from above, the probability
of having arbitrary large link lengths is not exponentially suppressed.
In two-dimensional quantum gravity however, the probability of having two points
separated a distance $R$ is exponentially suppressed as
$e^{-R\sqrt{\La}}$ for a given cosmological constant $\La$ and
suppressed as $e^{-R^{4/3}/V^{1/3}}$ if the area $V$ of spacetime
is fixed. That means, that quantum Regge calculus in two dimensions
does not provide a formulation of quantum gravity. In addition,
the mere existence of the partition function depends crucially on
the chosen measure as well as on the chosen fixed triangulation. Clearly,
this is a severe problem: if quantum Regge calculus was a realistic
theory, these microscopic details should be unimportant and not affect
physical expectation values. 

Furthermore, we have shown that in four dimensions spikes have been
observed in numerical studies of quantum Regge calculus. Numerical
investigations demonstrate that also in four dimensions the results
do depend crucially on the underlying triangulation and on the
measure. Thus our conclusion in four dimensions is the same as in
two dimensions: Quantum Regge calculus does not provide a theory
of quantum gravity. A sensible continuum limit cannot be defined. Not
even the concept of length can be defined unambiguously.

\chapter*{Conclusion}
\addtocontents{toc}{\protect\contentsline {chapter}{\protect\numberline {\ }Conclusion}{82}}
\enlargethispage*{\baselineskip}

In this work we have analyzed the fractal structure of two-dimensional quantum gravity.
A central result is that the spectral dimension equals two for all types
of conformal matter with central charge smaller than one coupled to gravity.
Other aspects of the fractal nature of the spacetime have been introduced and discussed
such as the extrinsic and intrinsic Hausdorff dimension, the branching ratio into
minimal bottleneck baby universes and the scaling of the two-point function.
We have compared our results with the latest numerical data. 
Numerical simulations are performed in the framework of dynamical triangulation.
We have discussed this method and illustrated its power by a computation of an exact
expression of the two-point function in two-dimensional quantum gravity.
Furthermore we have discussed quantum Regge calculus which has been suggested as
an alternative discretization of quantum gravity. We have shown that this
theory does not provide a suitable discretization of quantum gravity since
it disagrees with the continuum results. This point constitutes another central result of this work.

There are several open questions which have to be addressed in future work.
It would be very interesting to develop a method which would allow the solution
of Liouville theory. More modestly, the computation of the geodesic distance
or similar geometrical objects would be interesting. The derivation of the intrinsic
Hausdorff dimension is not very well understood. An alternative derivation of this
result would provide further insight into the theory. Related to this is the question, why
the fractal dimension seems to equal four for unitary matter coupled to quantum
gravity in two dimensions.
Of course the final triumph would be an analogous analysis of the fractal
properties of the quantum spacetime of four dimensional quantum gravity in the continuum approach
and in the discretized approach.

While this work settles the question about the viability of quantum Regge calculus
in its present formulation it would be interesting to develop different schemes to discretize
quantum gravity. We have discussed one possibility in chapter \sref{regge}.
It would be interesting to study further properties of this approach.

\small
\bibliographystyle{utphys}
\addtocontents{toc}{\protect\contentsline {chapter}{\protect\numberline {\ }Bibliography}{83}}
\bibliography{thesis}
\end{document}